\documentclass[twocolumn,english,aps,prb,floats,showpacs]{revtex4}
\usepackage[T1]{fontenc}
\usepackage[latin9]{inputenc}
\usepackage{babel}
\usepackage{amsmath}
\usepackage{amssymb}
\usepackage{graphicx}
\usepackage{esint}

\makeatletter

\newcommand{\lyxdot}{.}

\@ifundefined{textcolor}{}
{%
 \definecolor{BLACK}{gray}{0}
 \definecolor{WHITE}{gray}{1}
 \definecolor{RED}{rgb}{1,0,0}
 \definecolor{GREEN}{rgb}{0,1,0}
 \definecolor{BLUE}{rgb}{0,0,1}
 \definecolor{CYAN}{cmyk}{1,0,0,0}
 \definecolor{MAGENTA}{cmyk}{0,1,0,0}
 \definecolor{YELLOW}{cmyk}{0,0,1,0}
}

\usepackage{babel}

\usepackage{babel}

\makeatother

\begin{document}
\bibliographystyle{prsty}

\title{Random Field XY Model in Three Dimensions}

\author{D. A. Garanin, E. M. Chudnovsky, and T. Proctor}

\affiliation{Physics Department, Lehman College, City University of New York \\
 250 Bedford Park Boulevard West, Bronx, New York 10468-1589, USA}

\date{\today}
\begin{abstract}
We study random-field $xy$ spin model at $T=0$ numerically on lattices
of up to $1000\times1000\times1000$ spins with the accent on the
weak random field. Our numerical method is physically equivalent to
slow cooling in which the system is gradually losing the energy and
relaxing to an energy minimum. The system shows glass properties,
the resulting spin states depending strongly on the initial conditions.
Random initial condition for the spins leads to the vortex glass (VG) state with short-range spin-spin
correlations defined by the average
distance between vortex lines. Collinear and some other vortex-free
initial conditions result in the vortex-free ferromagnetic (F) states
that have a lower energy. The energy difference between the F and
VG states correlates with vorticity of the VG state. Correlation functions
in the F states agree with the Larkin-Imry-Ma theory at short distances.
Hysteresis curves for weak random field are dominated by topologically
stable spin walls raptured by vortex loops. We find no relaxation
paths from the F, VG, or any other states to the hypothetical vortex-free
state with zero magnetization.
\end{abstract}


\maketitle

\section{Introduction}

\label{introduction}

Studies of the effect of static random field on the long-range order
in system with continuous order parameter have a long history. Larkin
\cite{Larkin-JETP1970} argued that weak random pinning, no matter
how weak, destroys the long-range translational order in the Abrikosov
vortex lattice. Later similar ideas were applied to spin- and charge-density
waves,\cite{Efetov-Larkin-JETP1977} magnets,\cite{CSS} Josephson
junction arrays, and even cosmology. The question of interest for
superconductors is the distortion of the vortex lattice due to the
collective pinning of vortex lines by randomly distributed point defects.
In magnets it is a question of long-range behavior of ferromagnetic
correlations in the presence of torques applied to individual spins
by randomly distributed static local fields. The analytical results obtained
for the magnetic and superconducting systems are similar.

In 1975 Imry and Ma \cite{Imry-Ma-PRL1975} made a landmark observation
known as the Imry-Ma argument. It states that a static random field,
no matter how weak, destroys the long-range order in a system with
any continuous-symmetry order parameter in less than four dimensions.
The Imry-Ma correlation length $R_{f}$ (speaking of ferromagnetic
models) in $d$ dimensions scales as $R_{f}\propto h^{2/(d-4)}$ with
the strength of the random field $h$. Aizenman and Wehr \cite{Aizenman-Wehr-PRL,Aizenman-Wehr-CMP}
provided a mathematical scheme that is considered to be a rigorous
proof of the Imry-Ma argument. The effects of random magnetic anisotropy
relevant to the properties of amorphous and sintered ferromagnets
have been shown to resemble those of a random field. \cite{Pelcovits,Aharony,CSS}
It was demonstrated that random fields grow naturally out of magnetic
anisotropy in disordered antiferromagnets. \cite{Fishman} Early results
on magnets and superconductors with quenched randomness have been
summarized in Refs. \onlinecite{Fisher-PRB1985} and \onlinecite{Blatter-RMP1994},
and also discussed in the context of spin-glasses. \cite{Binder}
Larkin-Imry-Ma (LIM) approach leads to exponential decay of correlations
at large distances.\cite{Larkin-JETP1970,Efetov-Larkin-JETP1977,CSS,EC-PRB1991}
Recently, this approach has been employed to describe superconductor-insulator
transition in disordered films. \cite{SI}

Despite the appealing simplicity of the Imry-Ma argument the renormalization
group treatments of the problem by Cardy and Ostlund \cite{Cardy-PRB1982}
and by Villain and Fernandez in early 1980s \cite{Villain-ZPB1984}
had questioned the validity of that argument for distances $R\gtrsim R_{f}$.
The application of scaling and replica-symmetry breaking arguments to statistical mechanics of flux lattices,
\cite{Nattermann,Kierfield,Korshunov-PRB1993,Giamarchi-94,Giamarchi-95,LeDoussal-Wiese-PRL2006,LeDoussal-PRL2006,LeDoussal-PRL07,Bogner}
as well as variational approach, \cite{Orland-EPL,Garel-PRB} yielded
power-law decay of correlations at large distances, that suggested that
ordering could be more robust against weak static randomness than expected
from the LIM theory. Such a {\em quasiordered} phase, presumed
to be vortex-free in spin systems and dislocation-free in Abrikosov
lattices, received the name of a Bragg glass.

In parallel with analytical studies, the effect of static disorder
has been investigated by numerical methods. Early results on $1d$
(Ref. \onlinecite{DC-PRB1991}) and $2d$ (Ref. \onlinecite{Dieny-PRB1990})
spin systems with quenched randomness have established strong non-equilibrium
effects, such as magnetic hysteresis and dependence on initial conditions,
as well as significant departure of the correlation functions from
the prediction of the LIM theory. Gingras and Huse \cite{Gingras-Huse-PRB1996}
attempted to test numerically the existence of the vortex-free Bragg
glass phase in $2d$ and $3d$ random-field $xy$ spin model. While
they found some evidence of the expulsion of vortices below the critical
temperature, rapid freezing of spin dynamics prevented them from making
a definitive comparison with the Bragg glass theory. For the interested
reader, Ref. \onlinecite{Gingras-Huse-PRB1996} also provides a
discussion of similarities and differences of the $xy$ random-field spin model
and flux-lattice model in the background of the random pinning potential,
see also review Ref. \onlinecite{Nat-review}. In the absence of topological defects, numerical evidence
of the logarithmic growth of misalignment with the size of the system
has been found in $2d$ Monte Carlo studies of a crystal layer on
a disordered substrate and for pinned flux lattices. \cite{Zeng,Rieger}
Power-law decay of spin-spin correlations has been also reported in
Monte Carlo studies of the random-field Heisenberg model, \cite{Itakura-03}
as well as for the $xy$ model. \cite{Itakura-05} In a follow-up
on Ref. \onlinecite{Gingras-Huse-PRB1996}, further argument in
favor of the Bragg glass phase in the $xy$ model was presented by Fisher \cite{Fisher-PRL1997}
who analyzed energies of randomly pinned dislocation loops. Defect-free
spin models with relatively large random field and random anisotropy have
been studied numerically on small lattices by Fisch. \cite{Fisch}
At elevated temperatures the numerical evidence of the power-law decay
of correlations in a $2d$ random-field $xy$ model has been recently
obtained by Perret et al. \cite{Perret-PRL2012}

In spite of the large body of work, no agreement currently exists
on ordering and correlations in systems with quenched randomness.
The complexity of such systems appears to be in the same class as
the complexity of a spin-glass, even in the limit of a vanishingly
small random field. This contributed to the decline of the analytical
effort on random-field models after intensive work in 1980s and 1990s.
Numerical work on this problem has been hampered by the fact that
the ordered regions grow as the random field goes to zero. One can
easily come up with a wrong statement on the long-range behavior if
the size of the system is not sufficiently large. Numerical calculations
on large systems require large computation times, which, to a large
degree, contributed to the decline of the numerical effort in this
field. Nowadays the increased computational capabilities allow one
to re-address the question of the long-range behavior of the random-field
model in three dimensions. This has been the main motivation of our
work on the magnetic model.

Our main finding is that arguments of analytical theories about the behavior of
systems with quenched randomness, while having undisputable conceptual value
and serving as reference points for numerical studies, are, probably, oversimplified.
Properties of such systems
are dominated by pinned topological defects and metastability due to large energy
barriers that are practically unsurpassable at any temperature below the temperature of local ordering. We do not
find any relaxation path from typical initial states, such as random
and collinear initial orientation of spins, toward a completely disordered vortex-free
state.

The paper is organized as follows. The model is formulated in Section
\ref{model}. Some analytical results of the LIM theory are presented
in Section \ref{sec:Analytical}, as the reference frame for comparison
with numerical results. The effect of disorder on correlations of
spin directions (angular correlations) is discussed in Section \ref{sub:short-range-corr}.
Spin-spin correlation functions are discussed in Section \ref{sub:long-range}.
The leading contribution to the energy is derived in Section \ref{SR-E}.
Analytical formulas for the approach to saturation in the external
field are obtained in Section \ref{saturation}. Zero-field susceptibility
of the Imry-Ma state is obtained in Section \ref{sub:susceptibility}.
Expressions for the average magnetization of a finite system due to
statistical fluctuations are derived in Section \ref{sub:fluctuations}.
Correlated random-field is considered in Section \ref{sub:corr-RF}.
The details of our numerical method are discussed in Section \ref{sub:method}.
Section \ref{overview} gives overview of our numerical findings.
Relaxation of the spin system from different initial conditions is
studied in Section \ref{sub:relaxation}. The resulting spin structures
are reviewed in Section \ref{sub:spin structures}. Relation between
magnetization and vorticity is discussed in Section \ref{sub:mag-vorticity}.
Energies of different equilibrium states are compared in Section \ref{sub:energy}.
Numerical results on the hysteresis curves are presented in Section
\ref{sub:saturation}. Spin-spin correlation functions are computed
in Section \ref{sub:corr-functions}. Section \ref{discussion} contains
discussion of the results and possible interpretations.

\section{The model}

\label{model}

We study the $xy$ model described by the Hamiltonian
\begin{equation}
{\cal H}=-\frac{1}{2}\sum_{ij}J_{ij}{\bf s}_{i}\cdot{\bf s}_{j}-\sum_{i}{\bf h}_{i}\cdot{\bf s}_{i}-{\bf H}\cdot\sum_{i}{\bf s}_{i},\label{eq:ham-discrete}
\end{equation}
where ${\bf s}_{i}$ is a two-component constant-length ($\left|{\bf s}_{i}\right|=s$)
spin at the site $i$ of a cubic lattice, ${\bf h}_{i}$ is a quenched
random field (RF) at that site, and ${\bf H}$ is the external field.
The summation is over the nearest neighbors. The factor 1/2 in the
first term is compensating for the double counting of the exchange
bonds. In what follows we assume isotropic exchange, ($J_{ij}\equiv J$).
Below we present numerical results of the energy minimization in Eq.
(\ref{eq:ham-discrete}) for the uncorrelated RF,
\begin{equation}
\left\langle h_{i\alpha}h_{j\beta}\right\rangle =\frac{1}{2}h^{2}\delta_{\alpha\beta}\delta_{ij},\label{eq:h-corr-ij}
\end{equation}
(Greek indices being the Cartesian components of the vectors) although
computations for a correlated RF have been performed as well. The
correlator above has the same form for the fixed-length RF, $|\mathbf{h}_{i}|=h=\mathrm{const}$,
our main choice, as well as for models with a distributed RF strength
$h$, such as Gaussian distribution. No difference between the fixed-length
and Gaussian models has been found in numerical calculations.

The continuous counterpart of this model in $d$ dimensions is
\begin{equation}
{\cal H}=\int\frac{d^{d}r}{a^{d}}\left[\frac{Ja^{2}}{2}\left(\frac{\partial s_{\alpha}(\mathbf{r})}{\partial r_{\beta}}\right)^{2}-\mathbf{h}(\mathbf{r})\cdot\mathbf{s}(\mathbf{r})-{\bf H}\cdot\mathbf{s}(\mathbf{r})\right],\label{continuous}
\end{equation}
where ${\bf s}({\bf r})$ is dimensionless spin-density field, $\left|{\bf s}({\bf r})\right|=s$,
and ${\bf h}({\bf r})$ is the random field density, ${\bf r}=(x,y,z)$.
In the continuous model Eq. (\ref{eq:h-corr-ij}) corresponds to
\begin{equation}
\langle h_{\alpha}({\bf r'})h_{\beta}({\bf r}'')\rangle=\frac{1}{2}h^{2}a^{d}\delta_{\alpha\beta}\delta({\bf r}'-{\bf r}'').\label{h-corr}
\end{equation}
Another possible choice could be random field that is correlated on
a short scale $\rho$. This would correspond to
\begin{equation}
\langle h_{\alpha}({\bf r}')h_{\beta}({\bf r}'')\rangle=\frac{1}{2}h^{2}\delta_{\alpha\beta}\Gamma({\bf r}'-{\bf r}'')\label{h-Gamma}
\end{equation}
with $\Gamma(r)$ rapidly going to zero at $r\gg\rho$, e.g., $\Gamma(r)=\exp(-r/\rho)$
or $\Gamma(r)=\exp(-r^{2}/\rho^{2})$. Eq. (\ref{h-corr}) can be
generated by the Gaussian distribution \cite{Ziman} of the realizations
of the random field ${\bf h}({\bf r})$,
\begin{equation}
P[{\bf h}({\bf r})]\propto\exp\left[-\frac{1}{h^{2}}\int\frac{d^{d}r}{a^{d}}{\bf h}^{2}({\bf r})\right].\label{eq:h-Gaussian}
\end{equation}

It is convenient to express the continuous field model in terms of
the angles $\phi({\bf r})$ and $\varphi({\bf r})$ that determine
orientation of ${\bf s}({\bf r)}$ and ${\bf h}({\bf r)}$ in the
$xy$ plane. Writing
\begin{eqnarray}
 &  & {\bf s}({\bf r})=s[\sin\phi({\bf r}),\cos\phi({\bf r})]\\
 &  & {\bf h}({\bf r})=h[\sin\varphi({\bf r}),\cos\varphi({\bf r})]
\end{eqnarray}
and assuming that ${\bf H}$ is directed along the $x$ axis, ${\bf H}=(H,0)$,
one obtains from Eq.\ (\ref{continuous})
\begin{equation}
{\cal H}=s\int\frac{d^{d}r}{a^{d}}\left[\frac{Jsa^{2}}{2}(\nabla\phi)^{2}-h\cos(\phi-\varphi)-H\cos\phi\right].\label{Ham-continuous}
\end{equation}

For analytical calculations, the above $xy$ random-field spin model
is simpler than the Heisenberg spin model that, in general, would
have more than two components of ${\bf s}$ and ${\bf h}$. The reason
is that $xy$ spins can be described by one angle per spin, as dynamic
variables.The generalization is straightforward, though. Both models
can be modified to study the effects of random anisotropy, which applies,
e.g., to amorphous magnets. This would require the replacement of
the $-{\bf h}_{i}\cdot{\bf s}_{i}$ interaction with $-D_{i}({\bf n}_{i}\cdot{\bf s}_{i})^{2}$,
where $D_{i}$ and ${\bf n}_{i}$ describe strength and direction
of the local magnetic anisotropy. These other models will be be studied
separately. In this paper we focus on $3d$ $xy$ random-field model.
We will calculate analytically and numerically the correlation function
(CF) defined by
\begin{equation}
C(R)=\frac{1}{N}\sum_{i}\left\langle {\bf s}({\bf r}_{i})\cdot{\bf s}({\bf r}_{i}+\mathbf{R})\right\rangle ,\label{eq:C(R)-Def}
\end{equation}
where $N$ is the total number of spins. In analytical calculations,
there is no averaging over $i$ and $\left\langle \ldots\right\rangle $
mean averaging over realizations of the random field. In numerical
work, $\left\langle \ldots\right\rangle $ can be dropped for large
enough system sizes where there is a sufficient self-averaging.

\section{Qualitative analysis}

\label{sec:Qualitative}

The idea of a whatever weak random field destroying LRO has been proposed
by Imry and Ma. \cite{Imry-Ma-PRL1975} According to it, spins in
different regions will order in different directions dictated by the
random field, so that the magnetization of the bulk will be zero.
Since spins are coupled by a strong exchange, they cannot follow the
RF at each lattice site. Instead, they adjust to the direction of
the RF averaged over large correlated volumes of linear size $R_{f}$,
determined self-consistently. The energy per spin of the Imry-Ma (IM)
state can be estimated as
\begin{equation}
E-E_{0}\sim-sh\left(\frac{a}{R_{f}}\right)^{d/2}+s^{2}J\left(\frac{a}{R_{f}}\right)^{2},\label{eq:Imry-Ma-1}
\end{equation}
 where $E_{0}$ is the energy per spin of a collinear state. Minimization
of the energy with respect to $R_{f}$ yields
\begin{equation}
R_{f}\sim a\left(\frac{sJ}{h}\right)^{2/(4-d)}.\label{eq:Rf-IM-d-1}
\end{equation}
Finiteness of $R_{f}$ for any $d<4$ supports the initial assumption
that spins follow the averaged RF and thus the state is disordered,
$m=0$. The resulting energy of the IM state is
\begin{equation}
E-E_{0}\sim-s^{2}J\left(\frac{h}{sJ}\right)^{4/(4-d)}\label{eq:E-IM-d-1}
\end{equation}
that yields $E-E_{0}\sim-h^{4}/J^{3}$ in $3d$. However, the main
contribution to the adjustment energy arizes at the atomic scale and
is given by $E-E_{0}\sim-h^{2}/J$ in all dimensions.

One can modify the IM argument by taking into account adjustment of
spins to the RF at all length scales. For this purpose, consider a
reference state perfectly ordered in some direction. Spins will turn
away from this state under the influence of the RF. More precisely,
groups of spins of linear size $R$ will rotate by an adjustment angle
$\phi$ (considered as small to begin with) under the influence of
the RF averaged over this region. The corresponding energy per spin
is given by the generalization of Eq. (\ref{eq:Imry-Ma-1})
\begin{equation}
E-E_{0}\sim-sh\left(\frac{a}{R}\right)^{d/2}\phi+s^{2}J\left(\frac{a}{R}\right)^{2}\phi^{2}.\label{eq:Imry-Ma-phi}
\end{equation}
Minimizing this expression with respect to $\phi$, one obtains
\begin{equation}
\phi\sim\frac{h}{sJ}\left(\frac{R}{a}\right)^{(4-d)/2}\label{eq:phi-IM}
\end{equation}
that grows with the distance $R$, as expected. The square of the
angular deviation increases as $\phi^{2}\sim\left(R/R_{f}\right)^{4-d}$,
where $R_{f}$ is given by Eq. (\ref{eq:Rf-IM-d-1}). This defines
the spin CF at small distances
\begin{equation}
C(R)=s^{2}\cos\phi\cong s^{2}\left(1-\frac{1}{2}\phi^{2}\right)=s^{2}\left[1-A\left(\frac{R}{R_{f}}\right)^{4-d}\right],\label{eq:CF-IM}
\end{equation}
where $A$ is a number.

The energy per spin corresponding to spin adjustment at the distance
$R$ is
\begin{equation}
E-E_{0}\sim-\frac{h^{2}}{J}\left(\frac{a}{R}\right)^{d-2}.\label{eq:Energy-IM-d-R-1}
\end{equation}
One can see that the highest energy gain is provided by spin adjustments
at the atomic scale, $R\sim a$. In this case one obtains
\begin{equation}
E-E_{0}\sim-h^{2}/J.\label{eq:E-atomic}
\end{equation}
Spin misalignments grow large, $\phi\sim1$, at $R\sim R_{f}$. Substituting
$R_{f}$ into Eq. (\ref{eq:Energy-IM-d-R-1}), one recovers the IM
energy of Eq. (\ref{eq:E-IM-d-1}). It should be stressed that the
IM energy is much smaller than the main short-distance energy contribution
and it is not accessible numerically.

It has been speculated \cite{Nat-review} that at $R>R_{f}$, when
$\phi$ becomes large, it is distributed with a Gaussian probability,
making the energy associated with the random field scale as $sh(a/R)^{d/2}\exp(-\phi^{2}/2)$
instead of $-sh(a/R)^{d/2}\phi$ for small $\phi$. Then the minimum
of the total energy that includes the exchange energy $s^{2}J(a/R)^{2}\phi^{2}$,
would correspond to $\phi^{2}\sim(4-d)\ln(R/R_{f})$ in accordance
with the Bragg glass result. \cite{Nattermann,Korshunov-PRB1993,Giamarchi-94}

\section{Analytical results}

\label{sec:Analytical}

If the random field is sufficiently strong, then in the absence of
a strong external field, a strong local Zeeman interaction should
align the spins with the random field at each site independently.
The case of a weak random field is less straightforward. On one hand,
such a field cannot destroy the parallel alignment of neighboring
spins created by the strong ferromagnetic exchange. On the other hand,
neither the exchange nor the local random field can determine the
direction of the local magnetization. The latter can, therefore, wander
around the sample, with some characteristic ferromagnetic correlation
length that can be, in principle, either finite or infinite. This
non-obvious effect of the weak random field will be the main focus
of our investigation.

\subsection{Angular correlations}

\label{sub:short-range-corr}

At $H=0$ the correlation function of the spin angles $\phi$ can
be computed by noticing that the extremal configurations of $\phi({\bf r})$
with the Hamiltonian (\ref{Ham-continuous}) satisfy
\begin{equation}
Jsa^{2}\nabla^{2}\phi=h\sin(\phi-\varphi)=h_{x}\sin\phi-h_{y}\cos\phi
\end{equation}
where $h_{x}=h\cos\varphi$ and $h_{y}=h\sin\varphi$. This equation
has an implicit solution
\begin{eqnarray}
 &  & \phi({\bf r})=\frac{1}{Jsa^{2}}\int d^{d}r'G_{d}({\bf r}-{\bf r}')\times\nonumber \\
 &  & [h_{x}({\bf r}')\sin\phi({\bf r}')-h_{y}({\bf r}')\cos\phi({\bf r}')],\label{eq:angles}
\end{eqnarray}
where $G_{d}({\bf r})$ is the Green function of the Laplace equation
in $d$ dimensions: $G_{2}({\bf r})=-(2\pi)^{-1}\ln|{\bf r}|$ and
$G_{3}({\bf r})=-1/(4\pi|{\bf r}|)$. Its Fourier transform is $G_{d}({\bf q})=-1/{\bf q}^{2}$
for all $d$. Eq.\ (\ref{eq:angles}) then gives
\begin{eqnarray}
 &  & \langle[\phi({\bf r}_{1})-\phi({\bf r}_{2})]^{2}\rangle=\frac{1}{J^{2}s^{2}a^{4}}\int d^{d}r'\int d^{d}r''\times\nonumber \\
 &  & [G_{d}({\bf r}_{1}-{\bf r}')-G_{d}({\bf r}_{2}-{\bf r}')][G_{d}({\bf r}_{1}-{\bf r}'')-G_{d}({\bf r}_{2}-{\bf r}'')]\nonumber \\
 &  & \times[\langle h_{x}({\bf r}')h_{x}({\bf r}'')\rangle\langle\sin\phi({\bf r}')\sin\phi({\bf r}'')\rangle\nonumber \\
 &  & +\langle h_{y}({\bf r}')h_{y}({\bf r}'')\rangle\langle\cos\phi({\bf r}')\cos\phi({\bf r}'')\rangle\nonumber \\
 &  & -\langle h_{x}({\bf r}')h_{y}({\bf r}'')\rangle\langle\sin\phi({\bf r}')\cos\phi({\bf r}'')\rangle\nonumber \\
 &  & -\langle h_{y}({\bf r}')h_{x}({\bf r}'')\rangle\langle\cos\phi({\bf r}')\sin\phi({\bf r}'')\rangle]
\end{eqnarray}
Here we used the fact that for a weak random field the direction of
the spin at a particular site must have very weak correlation with
the direction of the random field at that site, leading to $\langle h_{x}({\bf r}')h_{x}({\bf r}'')\sin\phi({\bf r}')\sin\phi({\bf r}'')\rangle\approx\langle h_{x}({\bf r}')h_{x}({\bf r}'')\rangle\langle\sin\phi({\bf r}')\sin\phi({\bf r}'')\rangle$
and so on.

With the help of Eq.\ (\ref{h-corr}), one obtains in three dimensions
at $H=0$
\begin{eqnarray}
 &  & \langle[\phi({\bf r}_{1})-\phi({\bf r}_{2})]^{2}\rangle=\nonumber \\
 &  & \frac{h^{2}}{2J^{2}s^{2}a}\int d^{3}r[G_{3}({\bf r}_{1}-{\bf r})-G_{3}({\bf r}_{2}-{\bf r})]^{2}=\nonumber \\
 &  & \frac{h^{2}}{J^{2}s^{2}a}\int\frac{d^{3}q}{(2\pi)^{3}}\frac{1-\cos[{\bf q}\cdot({\bf r}_{1}-{\bf r}_{2})]}{q^{4}}=\nonumber \\
 &  & \frac{h^{2}}{8\pi J^{2}s^{2}a}|{\bf r}_{1}-{\bf r}_{2}|\label{eq:phiDifferenceCalculation}
\end{eqnarray}
and, finally,
\begin{equation}
\langle[\phi({\bf r}_{1})-\phi({\bf r}_{2})]^{2}\rangle=2\frac{|{\bf r}_{1}-{\bf r}_{2}|}{R_{f}},\qquad\frac{R_{f}}{a}=16\pi\left(\frac{Js}{h}\right)^{2},\label{eq:Rf-Def}
\end{equation}
where $R_{f}$ is the ferromagnetic correlation length. As we shall
see later, this formula is in excellent agreement with numerical results.
The linear decay of short-range correlations due to the random field
was first obtained by Larkin in the application to translational correlations
in flux lattices. \cite{Larkin-JETP1970} Extrapolating Eq.\ (\ref{eq:Rf-Def})
to greater distances, one should expect that the spin field would
rotate significantly at distances $|{\bf r}_{1}-{\bf r}_{2}|\sim R_{f}$.
The long-range behavior of spin-spin correlations has been, however,
subject of a significant controversy in the last forty years.

\subsection{Spin correlations}

\label{sub:long-range}

At short distances the spin correlation function directly follows
from the angular-deviation correlator computed above:
\begin{eqnarray}
\langle{\bf s}({\bf r}_{1})\cdot{\bf s}({\bf r}_{2})\rangle & = & s^{2}\langle\cos[\phi({\bf r}_{1})-\phi({\bf r}_{2})]\rangle\nonumber \\
 & = & s^{2}\left(1-\frac{1}{2}\langle[\phi({\bf r}_{1})-\phi({\bf r}_{2})]^{2}\rangle\right)\nonumber \\
 & = & s^{2}\left(1-\frac{|{\bf r}_{1}-{\bf r}_{2}|}{R_{f}}\right),\label{eq:CF-short}
\end{eqnarray}
in accordance with Eq. (\ref{eq:CF-IM}) in $3d$. More generally,
one can write
\begin{eqnarray}
 &  & \langle{\bf s}({\bf r}_{1})\cdot{\bf s}({\bf r}_{2})\rangle=s^{2}\langle\cos[\phi({\bf r}_{1})-\phi({\bf r}_{2})]\rangle\nonumber \\
 &  & =s^{2}\exp\left\{ -\frac{1}{2}\langle[\phi({\bf r}_{1})-\phi({\bf r}_{2})]^{2}\rangle\right\} .\label{eq:Exp-CF-derivation}
\end{eqnarray}
Substituting here Eq.\ (\ref{eq:Rf-Def}), in $3d$ one obtains
\begin{equation}
\langle{\bf s}({\bf r}_{1})\cdot{\bf s}({\bf r}_{2})\rangle=s^{2}\exp\left(-\frac{|{\bf r}_{1}-{\bf r}_{2}|}{R_{f}}\right).\label{Exp-CF}
\end{equation}

Equation (\ref{Exp-CF}) can be obtained in the whole range of distances
by the functional integration over the distribution of the random
field given by Eq. (\ref{eq:h-Gaussian}). The calculation in $3d$
proceeds as follows
\begin{eqnarray}
 &  & \langle{\bf s}({\bf r}_{1})\cdot{\bf s}({\bf r}_{2})\rangle=s^{2}\langle\exp{i[\phi({\bf r}_{1})-\phi({\bf r}_{2})]}\rangle\nonumber \\
 &  & =s^{2}\left[\int D\{h_{x}\}D\{h_{y}\}\exp\left\{ -\frac{\int d^{3}r\,(h_{x}^{2}+h_{y}^{2})}{h^{2}a^{3}}\right\} \right]^{-1}\nonumber \\
 &  & \times\int D\{h_{x}\}D\{h_{y}\}\exp\big\{ i\int d^{3}r[\frac{1}{Jsa^{2}}[G_{3}({\bf r}-{\bf r}_{1})-\nonumber \\
 &  & G_{3}({\bf r}-{\bf r}_{2})][h_{x}\sin\phi({\bf r})-h_{y}\cos\phi({\bf r})]-\frac{h_{x}^{2}+h_{y}^{2}}{h^{2}a^{3}}]\big\}\nonumber \\
 &  & =s^{2}\exp\left\{ -\frac{h^{2}}{4J^{2}s^{2}a}\int d^{3}r\left[G_{3}({\bf r}-{\bf r}_{1})-G_{3}({\bf r}-{\bf r}_{2})\right]^{2}\right\} \nonumber \\
 &  & =s^{2}\exp\left\{ -\frac{h^{2}}{2J^{2}s^{2}a}\int\frac{d^{3}q}{(2\pi)^{3}}\frac{1-\cos[{\bf q}\cdot({\bf r}_{1}-{\bf r}_{2})]}{q^{4}}\right\} \nonumber \\
 &  & =s^{2}\exp\left(-\frac{|{\bf r}_{1}-{\bf r}_{2}|}{R_{f}}\right),\label{eq:pathint}
\end{eqnarray}
where we have used Eq.\ (\ref{eq:angles}).

The increase of spin misalignments with distance according to Eq.
(\ref{eq:Rf-Def}) is unquestionable and it is also true that at some
distance misalignments become large. It was questioned by many researchers,
however, whether the averaging employed to obtain Eq.\ (\ref{eq:pathint})
provides correct description of the behavior at large distances. Theory
based upon scaling and replica-symmetry breaking arguments \cite{Nattermann,Giamarchi-94}
yielded $\langle[\phi({\bf r}_{1})-\phi({\bf r}_{2})]^{2}\rangle=A\ln|{\bf r}_{1}-{\bf r}_{2}|$
at $R\gg R_{f}$, with $A$ depending on the dimensionality only.
While this theory was initially developed for flux lattices, it was
later argued that the result must be relevant for the $xy$ random-field
spin model as well. \cite{Gingras-Huse-PRB1996,Garel-PRB,Fisher-PRL1997,Perret-PRL2012}
This would imply universal power law decay of long-range correlations,
\begin{equation}
\langle{\bf s}({\bf r}_{1})\cdot{\bf s}({\bf r}_{2})\rangle\sim\frac{1}{|{\bf r}_{1}-{\bf r}_{2}|}\label{NGL}
\end{equation}
in $3d$ according to Eq. (\ref{Exp-CF}). Such a quasiordered phase,
presumed to be vortex-free in spin systems and dislocation-free in
flux lattices, received the name of Bragg glass. As we shall see below
neither Imry-Ma argument nor the Bragg glass argument provides the
correct description of the random-field system that would agree with
numerical results. Crucial for its behavior is magnetic hysteresis,
which implies that energy barriers and metastable states play an important
role regardless of the strength of the random field. We shall also
demonstrate that the behavior of the random-field system cannot be
understood without invoking topological defects.

\subsection{Short-range energy due to random field}

\label{SR-E}

The random field contributes to the energy of the system through Zeeman
interaction with the spin field and through the exchange energy associated
with the non-uniformity of the spin field. The latter can be computed
as
\begin{equation}
\langle{\cal H}_{\mathrm{ex}}\rangle=\frac{1}{2}J\sum_{ij}\langle s^{2}-{\bf s}_{i}\cdot{\bf s}_{j}\rangle=\frac{1}{4}Js^{2}\sum_{ij}\langle(\phi_{i}-\phi_{j})^{2}\rangle,\label{micro}
\end{equation}
where the summation is over $N$ sites $i$ and the nearest neighbors
$j$ of each $i$-site, with six such neighbors in a $3d$ cubic lattice,
separated by $|{\bf r}_{i}-{\bf r}_{j}|=a$. According to Eq.\ (\ref{eq:Rf-Def}),
for the nearest neighbors $\langle(\phi_{i}-\phi_{j})^{2}\rangle=h^{2}/(8\pi J^{2}s^{2})$,
so that per spin
\begin{equation}
\frac{\langle{\cal H}_{\mathrm{ex}}\rangle}{N}=\frac{1}{4}Js^{2}6\frac{h^{2}}{8\pi J^{2}s^{2}}=\frac{3h^{2}}{16\pi J}.\label{ex-SR}
\end{equation}

The total energy is a sum of the exchange energy and Zeeman energy,
given by Eq.\ (\ref{Ham-continuous}). Let us consider the case of
$H=0$. The contribution of the weak random field to the energy is
a sum of almost independent contributions from small volumes inside
which the deviation, $\delta\phi({\bf r})$, from the local ferromagnetic
alignment of spins is small. Thus, to obtain the main part of the
energy due to random field, one can replace $\phi$ in Eq.\ (\ref{Ham-continuous})
with $\delta\phi({\bf r})\ll1$,
\begin{equation}
{\cal H}_{\mathrm{SR}}=s\int\frac{d^{3}r}{a^{3}}\left[\frac{1}{2}Jsa^{2}(\nabla\delta\phi)^{2}-h\delta\phi\sin\varphi\right].\label{H-SR}
\end{equation}
Low temperature behavior is dominated by the extremal configurations
satisfying
\begin{equation}
Jsa^{2}\nabla^{2}\delta\phi=-h\sin\varphi
\end{equation}
Substituting $\sin\varphi$ from this equation into Eq. (\ref{H-SR})
and integrating by parts one obtains
\begin{eqnarray}
 &  & {\cal H}_{\mathrm{SR}}=s\int\frac{d^{3}r}{a^{3}}\left\{ \frac{1}{2}Jsa^{2}(\nabla\delta\phi)^{2}+Jsa^{2}\delta\phi\nabla^{2}\phi\right\} \nonumber \\
 &  & =s\int\frac{d^{3}r}{a^{3}}\left\{ \frac{1}{2}Jsa^{2}(\nabla\delta\phi)^{2}-Jsa^{2}(\nabla\delta\phi)^{2}\right\} .
\end{eqnarray}
It is clear from this expression that the short-range Zeeman energy
is twice the short-range exchange energy with a minus sign,
\begin{equation}
\frac{\langle{\cal H}_{Z}\rangle}{N}=-2\frac{\langle{\cal H}_{\mathrm{ex}}\rangle}{N}=-\frac{3h^{2}}{8\pi J}.\label{Z-SR}
\end{equation}

The total short-range energy per spin is
\begin{equation}
\frac{\langle{\cal H}\rangle}{N}=\frac{\langle{\cal H}_{\mathrm{ex}}\rangle+\langle{\cal H}_{Z}\rangle}{N}=-\frac{3h^{2}}{16\pi J},\label{total-SR}
\end{equation}
in accordance with Eq. (\ref{eq:E-atomic}). It is insensitive to
the long-range behavior of the spin field, that is, to the spatial
scale of the rotation of the direction of the local magnetization
over the sample. This is because for a weak random field such rotations
involve large distances, and therefore they contribute much less to
the exchange energy then the weak misalignment of the neighboring
spins due to the random field. As we shall see below, equations (\ref{ex-SR}),
(\ref{Z-SR}), and (\ref{total-SR}) are in excellent agreement with
numerical results. Small deviations are due to the contribution of
vortices to the short-range behavior.

\subsection{Approach to saturation}

\label{saturation}

In the presence of the external magnetic field the extremal configurations
satisfy
\begin{equation}
Jsa^{2}\nabla^{2}\phi-H\sin\phi=h\sin(\phi-\varphi).\label{phi-H}
\end{equation}
Let the field $H$ be sufficiently large to ensure a small deviation
of spins from the $x$ axis, that is, small angle $\phi({\bf r})$.
Then Eq.\ (\ref{phi-H}) can be approximately written as
\begin{equation}
\nabla^{2}\phi-k_{H}^{2}\phi=-\frac{h}{Jsa^{2}}\sin\varphi,\label{Lap-H}
\end{equation}
where
\begin{equation}
\frac{1}{k_{H}^{2}}=R_{H}^{2}=\left(\frac{Js}{H}\right)a^{2}.\label{RH}
\end{equation}
The solution of Eq.\ (\ref{Lap-H}) is
\begin{equation}
\phi({\bf r})=\frac{h}{Jsa^{2}}\int d^{3}r'\frac{e^{-k_{H}|{\bf r}-{\bf r}'|)}}{4\pi|{\bf r}-{\bf r}'|}\sin\varphi({\bf r}').
\end{equation}
Consequently,
\begin{eqnarray}
\langle\phi^{2}\rangle & = & \left(\frac{h}{Jsa^{2}}\right)^{2}\int d^{3}r'\int d^{3}r''\frac{e^{-k_{H}|{\bf r}-{\bf r}'|}e^{-k_{H}|{\bf r}-{\bf r}''|}}{16\pi^{2}|{\bf r}-{\bf r}'||{\bf r}-{\bf r}''|}\nonumber \\
 & \times & \langle\sin\varphi({\bf r}')\sin\varphi({\bf r}'')\rangle.
\end{eqnarray}
With the help of Eq.\ (\ref{h-corr}) one obtains for $k_{H}a\ll1$
($R_{H}\gg a$)
\begin{eqnarray}
\langle\phi^{2}\rangle & = & \frac{a^{3}}{32\pi^{2}}\left(\frac{h}{Jsa^{2}}\right)^{2}\int d^{3}r\frac{e^{-2k_{H}r}}{r^{2}}\nonumber \\
 & = & \frac{1}{16\pi}\left(\frac{h}{Js}\right)^{3/2}\left(\frac{h}{H}\right)^{1/2}.
\end{eqnarray}
The above formulas describe the approach to saturation on increasing
the field:
\begin{equation}
1-\frac{m}{s}=\langle1-\cos\phi\rangle=\frac{1}{2}\langle\phi^{2}\rangle=\frac{1}{32\pi}\left(\frac{h}{Js}\right)^{3/2}\left(\frac{h}{H}\right)^{1/2}\label{sqrt}
\end{equation}

The square root dependence on $H$, Eq.\ (\ref{sqrt}), must hold
as long as the field satisfies $R_{H}>a$, which translates into $H<Js$.
At $H>Js$ the length $R_{H}$ becomes small compared to $a$ and
the exchange-generated Laplacian in Eq.\ (\ref{Lap-H}) is no longer
relevant because the $r$ in the Green function of that equation cannot
be smaller than $a$. In this case the approach to saturation is dominated
by the spin torque of the external field $H$ against the local field
$h({\bf r})$. The Laplacian in Eq. (\ref{Lap-H}) can be safely dropped
and one ends up with $\phi=(h/H)\sin{\varphi}$. This gives
\begin{equation}
1-\frac{m}{s}=\frac{1}{2}\langle\phi^{2}\rangle=\frac{h^{2}}{2H^{2}}\langle\sin^{2}\varphi\rangle=\frac{h^{2}}{4H^{2}}.\label{square}
\end{equation}
Eqs. (\ref{sqrt}) and (\ref{square}) are confirmed by numerical
results with high accuracy, see below.

\subsection{Zero-field susceptibility}

\label{sub:susceptibility}

To have some reference point for comparison with numerical results,
it is important to have the zero-field susceptibility of the Imry-Ma
state. Application of a small field $H\rightarrow0$ in the $x$ direction
slightly perturbs $\phi({\bf r})$ created by the random field,
\begin{equation}
\phi({\bf r})\rightarrow\phi({\bf r})+\delta\phi({\bf r}).
\end{equation}
Linearization of Eq.\ (\ref{phi-H}) gives
\begin{equation}
Jsa^{2}\nabla^{2}\delta\phi-H\sin\phi=h\delta\phi\cos(\phi-\varphi).
\end{equation}
Neglecting the rapidly oscillating small term in the right-hand-side
of this equation we obtain in $3d$
\begin{equation}
\delta\phi({\bf r})=-\frac{H}{Jsa^{2}}\int d^{3}r'\frac{\sin({\bf r}')}{4\pi|{\bf r}-{\bf r}'|}.
\end{equation}
The magnetization per spin in the direction of the field is given
by
\begin{eqnarray}
\frac{\langle m\rangle}{s} & = & \langle\cos\phi\rangle=-\langle\delta\phi\sin\phi\rangle\nonumber \\
 & = & \frac{H}{Jsa^{2}}\int d^{3}r'\frac{\langle\sin\phi({\bf r})\sin\phi({\bf r}')\rangle}{4\pi|{\bf r}-{\bf r}'|}.
\end{eqnarray}
This can be related to
\begin{eqnarray}
\langle{\bf s}({\bf r})\cdot{\bf s}({\bf r}')\rangle & = & s^{2}\langle\cos\phi({\bf r})\cos\phi({\bf r}')+\sin\phi({\bf r})\sin\phi({\bf r}')\rangle\nonumber \\
 & = & 2s^{2}\langle\sin\phi({\bf r})\sin\phi({\bf r}')\rangle.
\end{eqnarray}
Consequently,
\begin{eqnarray}
\frac{m}{s} & = & \frac{H}{2Js^{3}a^{2}}\int d^{3}r'\frac{\langle{\bf s}({\bf r})\cdot{\bf s}({\bf r}')\rangle}{4\pi|{\bf r}-{\bf r}'|}\nonumber \\
 & = & \frac{H}{2Jsa^{2}}\int d^{3}r'\frac{\exp(|{\bf r}-{\bf r}'|/R_{f})}{4\pi|{\bf r}-{\bf r}'|}.
\end{eqnarray}
Integration gives
\begin{equation}
\frac{m}{s}=\frac{H}{2Js}\left(\frac{R_{f}}{a}\right)^{2}.
\end{equation}
Zero-field susceptibility defined through
\begin{equation}
\frac{m}{s}=\chi\frac{H}{Js}
\end{equation}
is given by
\begin{equation}
\chi=\frac{1}{2}\left(\frac{R_{f}}{a}\right)^{2}=128\pi^{2}\left(\frac{Js}{h}\right)^{4}.\label{chi}
\end{equation}
In the limit of small $h$ it is very large, which may have prompted
some statements in the past about infinite susceptibility of the Imry-Ma
state. \cite{Aharony} As we shall see below, the actual zero-field
susceptibility in a zero magnetization state is dominated by the dynamics
of vortices and is much smaller.

Note that the initial magnetization of the Imry-Ma state in the limit
of a very weak field and the approach to saturation at a higher field
can be presented as
\begin{equation}
\frac{m}{s}=\frac{1}{2}\left(\frac{R_{f}}{R_{H}}\right)^{2},\quad R_{H}\gg R_{f}\label{ZF-mag}
\end{equation}
\begin{equation}
1-\frac{m}{s}=\frac{1}{2}\left(\frac{R_{H}}{R_{f}}\right),\quad R_{H}\ll R_{f},\label{sat}
\end{equation}
where $R_{H}$ and $R_{f}$ are given by equations (\ref{RH}) and
(\ref{Exp-CF}) respectively. Both formulas provide $m\sim s$ at
$R_{H}\sim R_{f}$, which translates into
\begin{equation}
\frac{H}{Js}\sim\frac{1}{256\pi^{2}}\left(\frac{h}{Js}\right)^{4}.\label{H-sat}
\end{equation}
For a weak random field, $h<Js$, this gives a very small value of
$H$. It has the following physical meaning. If one studies the full
hysteresis loop of the random-field system then the state close to
saturation must have no vortices. In this case Eq.\
(\ref{sat}) is exact. Thus, the field in Eq.\ (\ref{H-sat}) provides
the estimate of the maximal width of the hysteresis loop. In the limit
of small $h$ the loop must be very narrow, which is confirmed by
our numerical results. For small $h$ one should use a very small
field step in order not to confuse a very steep magnetization curve
with a discontinuity in the magnetization curve. When the width of
the hysteresis loop is so small that it cannot be resolved in either
real or numerical experiment, its slope may well be described by Eq.\ (\ref{chi}).

\subsection{Average magnetization of a finite system}

\label{sub:fluctuations}

As we have seen, at $h<Js$, the regions that are ferromagnetically
ordered can be quite large. A system of size $L<R_{f}$ will always
exhibit ferromagnetic order. Thus, it may be difficult to numerically
test the Imry-Ma statement that a random field, however weak it may
be, destroys the long-range order in three dimensions. Even when $R_{f}$
is small compared to $L$ it may not be easy to distinguish between
spontaneously magnetized states and zero-magnetization states because
of the magnetization arising from statistical fluctuations. The problem
is similar to that of a finite-size paramagnet: $N$ spins randomly
distributed between spin-up and spin-down states will have an average
total magnetization proportional to $\sqrt{N}$ and thus average magnetization
per spin proportional to $1/\sqrt{N}$.

The magnetization of the system is given by
\begin{equation}
\mathbf{m}=\frac{1}{N}\sum_{i}\mathbf{s}_{i},
\end{equation}
where $N$ is the total number of spins. The absolute value of $m$
is related to the spin correlation function of Eq. (\ref{eq:C(R)-Def})
as
\begin{equation}
m^{2}=C(\infty)+\frac{1}{V}\int d^{d}R\left[C(\mathbf{R})-C(\infty)\right],\label{eq:m2_via_CF}
\end{equation}
where $C(\infty)$ describes long-range order (LRO) and $V=L^{3}$
is the system volume. Plotting $m^{2}$ vs $1/V$ shows if there is
a LRO in the system in the limit $V\rightarrow\infty$.

Substituting here $C(\mathbf{R})=s^{2}\exp(-R/R_{f})$ (no long-range
order) in $3d$, one obtains
\begin{equation}
m=s\left(\frac{8\pi R_{f}^{3}}{V}\right)^{1/2}=\sqrt{8\pi}s\left(\frac{R_{f}}{L}\right)^{3/2},\label{m_fluct}
\end{equation}
where $L$ is the size of the system, $N=(L/a)^{3}$. At, e.g., $h=0.5Js$,
the Imry-Ma correlation length is $R_{f}\approx200a$. For $L=1000a$
this gives $m\approx0.45s$. Such a large value of $m$ for a not
very weak random field in a system of the maximum size that we can
access numerically suggests that any evidence of the long-range ferromagnetic
order based upon finite $m$ should be dealt with care. However, LRO,
if it is present, reduces the value of the $1/V$ term in Eq. (\ref{eq:m2_via_CF}).

\subsection{Correlated random field}

\label{sub:corr-RF}

All the above formulas have been written for uncorrelated random field
described by Eq.\ (\ref{h-corr}). Meanwhile, in physical problems
involving flux lattices and random magnets, the static randomness
can be correlated over a certain distance $\rho>a$. Such situation
is described by Eq.\ (\ref{h-Gamma}). It is easy to see that it
leads to the following modification of $R_{f}$ in Eq. (\ref{eq:Rf-Def}):
\begin{equation}
\frac{R_{f}}{a}=\frac{16\pi a^{3}}{\Omega}\left(\frac{Js}{h}\right)^{2},\label{Rf-Omega}
\end{equation}
where
\begin{equation}
\Omega=\int d^{3}r\Gamma(r)
\end{equation}
is the correlated volume, with $\Gamma(r)$ describing the short-range
correlations of the random field.

For uncorrelated random field one has $\Omega=a^{3}$ and Eq. (\ref{Rf-Omega})
goes back to Eq.\ (\ref{Exp-CF}). In the case of a correlated random
field $\Omega>a^{3}$ and $R_{f}$ is reduced. For, e.g., $\Gamma=\exp(-r/\rho)$,
one obtains $\Omega=8\pi\rho^{3}$ and
\begin{equation}
\frac{R_{f}}{a}=2\left(\frac{a}{\rho}\right)^{3}\left(\frac{Js}{h}\right)^{2}.
\end{equation}
Notice that the reduction in $R_{f}$ is by a factor $8\pi(\rho/a)^{3}$
which can be quite significant. This, in principle, may allow one
to test the effect of a very small $h$ in a finite-size system. When
the above formulas produce $R_{f}<\rho$, this means that $R_{f}=\rho$.

\section{Numerical results}

\label{numerical}

\subsection{Numerical method}

\label{sub:method}
\begin{figure}
\centering\includegraphics[width=8cm]{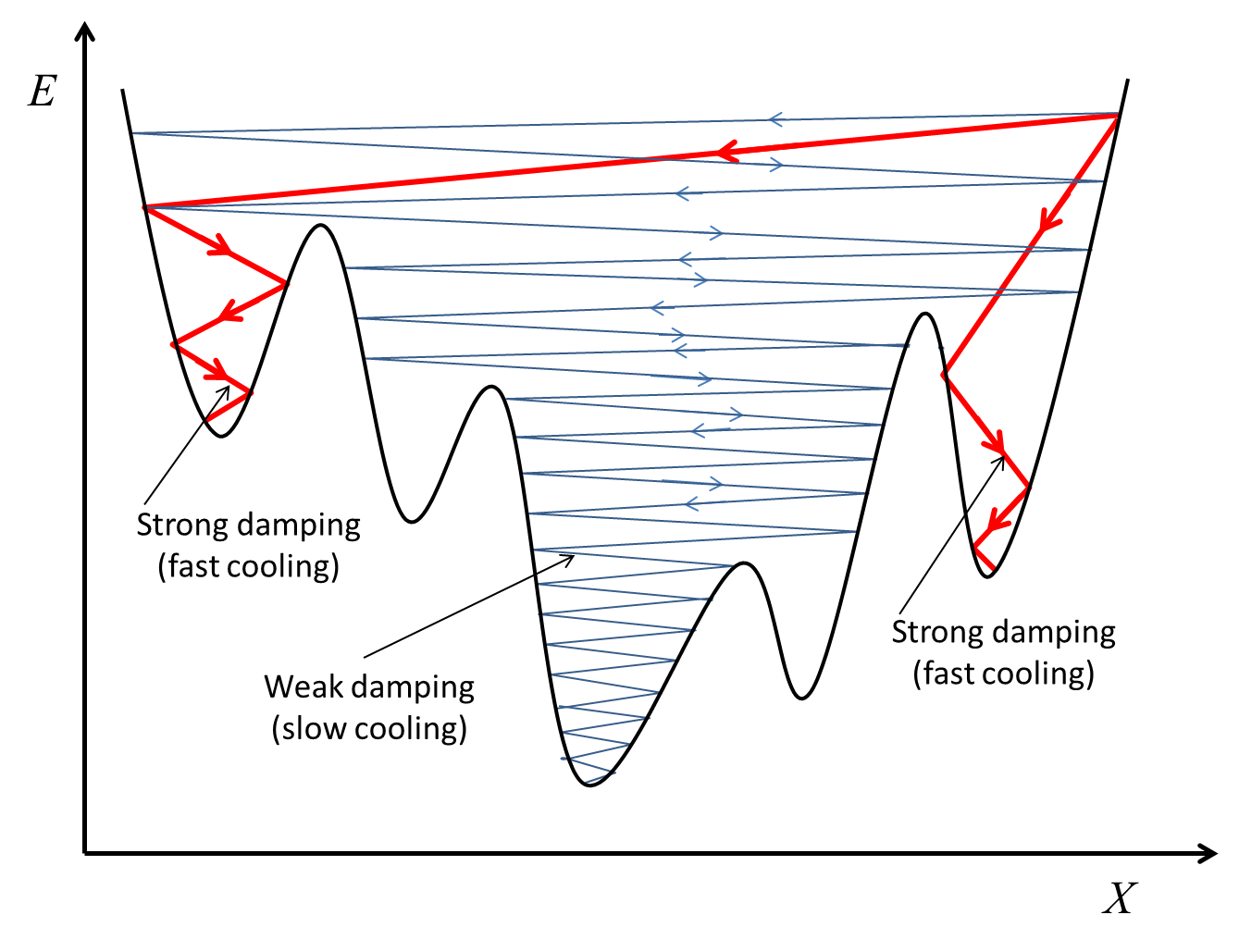}\caption{Efficiency of the weak-damping (slow-cooling) method for glassy systems. }

\label{Fig:Energy_minimization}
\end{figure}

The task is to find energy minima of Eq. (\ref{eq:ham-discrete})
by a numerical algorithm starting from an initial state (IC) and using
some relaxation protocol. It turns out that there are multitudes of
local energy minima and the situation resembles that of a spin glass.
At the end of relaxation the system ends up in one of them. We do
not attempt to search for the ground state of the system, which would
require different numerical methods. Rather, we are interested in
representative local minima obtained by relaxation from typical IC
such as random and collinear initial conditions. This corresponds
to experimental situations, and the results for physical quantities
in the final state are reproducible up to statistical noise due to
different realizations of the random field and different realizations
of the relaxation protocol that may have a stochastic part. The larger
the system size, the smaller the fluctuations. For smaller sizes,
averaging of the results over realizations of the random field is
necessary.

One could use the Landau-Lifshitz equation of motion with damping
(with no precession term for the $xy$ model) to find local energy
minima. One can expect an even faster relaxation if one rotates every
spin, sequentially, straight in the direction of the effective magnetic
field
\begin{equation}
\mathbf{H}_{i,\mathrm{eff}}=\sum_{j}J_{ij}\mathbf{s}_{j}+\mathbf{h}_{j}+\mathbf{H}\label{eq:Heff_Def}
\end{equation}
that is \cite{Dieny-PRB1990}
\begin{equation}
\mathbf{s}_{i,\mathrm{new}}=\mathbf{H}_{i,\mathrm{eff}}/\left|\mathbf{H}_{i,\mathrm{eff}}.\right|\label{eq:FR_Def}
\end{equation}
We call this the finite rotation (FR) method. Although this method
works very well in cases when there is only one energy minimum (such
as the collinear state for pure ferromagnetic models), it leads to
slow relaxation in the case of glassy behavior characterized by many
local minima. The problem is that the relaxation described by Eq.
(\ref{eq:FR_Def}) is initially too fast and the system falls into
the nearest local minimum that is not the deepest and not the most
representative. As in the multi-dimensional space of our model there
are narrow valleys rather than simple local minima, the system quickly
falls into one of these valleys and then begins a long travel along
it.

To counter this slow relaxation, it is convenient to combine the FR
method with so-called over-relaxation \cite{Adler-PRD1981} that is, in fact, a conservative
pseudo-dynamics described by
\begin{equation}
\mathbf{s}_{i,\mathrm{new}}=\frac{2\left(\mathbf{s}_{i,\mathrm{old}}\cdot\mathbf{H}_{i,\mathrm{eff}}\right)\mathbf{H}_{i,\mathrm{eff}}}{H_{i,\mathrm{eff}}^{2}}-\mathbf{s}_{i,\mathrm{old}}.\label{eq:Overrel_Def}
\end{equation}
Here spins are sequentially flipped onto the other side of the effective
field (half of the precession period for the Heisenberg model) and
the energy is conserved. This method is very convenient to quickly
explore the hypersurface of constant energy of the system. Whereas
the FR method searches for a minimal energy, the over-relaxation method
searches for the maximal entropy. It is a standard numerical method for classical spin systems,
usually combined with Monte Carlo updates (see, e.g., Ref. \onlinecite{Chen-Landau-PRB1994}).

For instance, starting from the collinear state and using the over-relaxation
method, one can describe FR-induced transition of the system from
the initial state that has the minimal statistical weight to a more
disordered state having the same energy but a much higher statistical
weight. This process describes an irreversible relaxation in which
the magnetization value $m$ decreases from 1 to a smaller value.
The resulting final state is above the ground state, so it can be
interpreted as a thermal state with some small temperature. Adding
the energy-lowering evolution described by Eq. (\ref{eq:FR_Def})
one can find the lowest-energy state in this particular region of
phase space.

Practically it is convenient to combine both methods. In the main
method we used, Eq. (\ref{eq:FR_Def}) is applied with the probability
$\alpha$ while Eq. (\ref{eq:Overrel_Def}) is applied with the probability
$1-\alpha$. The optimal value of $\alpha$ that plays the role of
a relaxation constant is in the range $0.1-0.01,$ typically 0.03.
Physically this corresponds to slow cooling the system. Such a choice
results in convergence acceleration by factors greater than 10 in
comparison to $\alpha=1$. The efficiency of the combined weak damping
method for glassy systems is shown in Fig. \ref{Fig:Energy_minimization},
assuming that deeper minima have broader basins of attraction.

Starting from the collinear state, we also used a two-stage relaxation
method. The first stage, which we call ``chaotization'', is the
conservative pseudo-dynamics given by Eq. (\ref{eq:Overrel_Def}).
The second stage is the combined relaxation process described above.
In some cases during chaotization damped oscillating behavior was
observed. In this case suppression of oscillations and a faster convergence
can be achieved by performing Eq. ($\ref{eq:Overrel_Def}$) with a
probability $1-\eta$ and leaving the spin unchanged with the probability
$\eta$. The constant $\eta$ that has the optimal value about 0.01
plays the role of a decoherence constant in the numerical method.

To check the predicted absence of ordering in RF magnets, one has
to numerically solve models of a size $L\gg R_{f}$ that must be strongly
fulfilled in accordance with the discussion in Section {\ref{sub:fluctuations}.
This sets a lower bound on the numerically accessible $H_{R}\equiv h$.
With a Mac Pro with 96 GB RAM running Wolfram Mathematica, we can
compute $3d$ models up to the size $L=800$, i.e., half a billion
spins, including correlation functions. The memory usage during relaxation
to the energy minima is about 30 GB while computation of the correlation
function takes 85 GB. This means that we cannot further increase the
size while computing correlation functions, although we can compute
the relaxation of a system of $L=1000,$ a billion spins. Our Mathematica
program that uses compilation and parallelization is comparable in
speed with programs written in Fortran and C. Relaxation of a $3d$
system of $L=800$ for moderately small $H_{R}$ takes 1-2 days.

We also compute the vorticity by analyzing rotation of spin vectors
along any unitary square plaquette in $xy$ planes. If spins rotate
by 0 angle along the plaquette, there is no singularity of the spin
field at this plaquette. If spins rotate by $\pm2\pi$, there is a
vortex or antivortex. For initial states that have no global vorticity,
such as collinear and random initial state, the numbers of vortices
and antivortices are always the same. Thus we just count them as ``vortices''
and define vorticity $f_{V}$ as the fraction of plaquettes that contain
singularities. In the random state one has $f_{V}=1/3$ while in the
energy minima that we find $f_{V}$ is zero or a small number. In
the latter case there are vortex loops in the system.

In the numerical work we use $J=a=s=1$.

Our lowest value of $H_{R}$ in $3d$ is 0.7, which corresponds to
$R_{f}=103$ and is still much smaller than our largest size, $L=800$,
and convergence of our method is still fast enough. For $H_{R}=0.5$
one has $R_{f}=201$, so that that the ratio $R_{f}/L$ is not small
enough even for $L=800$ and here convergence of our method is noticeably
slower. Although we can reach an energy minimum in this case spending
more time, the resulting correlation functions depend on the realization
of the random field and are bumpy. This is the consequence of insufficient
self-averaging for $R_{f}/L$ not small enough. In this case additional
averaging over random-field realizations is needed, which for $L=800$
would take too much time. To the contrary, if $R_{f}/L\ll1$ is strongly
satisfied, self-averaging is sufficient and correlation functions
have a smooth shape. Also in this case convergence of our method is
pretty fast. For instance, relaxation of the system of $L=800$ out
of a collinear state takes only about 5 hours. For smaller sizes we
do averaging over realization of the random field to achieve a better
precision.

Computations in $2d$ are numerically less challenging, in particular,
because of the much shorter $R_{f}$. Finding a local minimum of the
energy at $T=0$ for $L=10000$ (100 millions of spins) and $H_{R}=0.1$
does not present problems.

\subsection{General results}

Our main finding is that for a weak random field the state of the
system is always a glassy state with many local energy minima, so
that the final state that we find depends on the initial state or
initial condition, as well as somewhat on the details of the relaxation
protocol. This to a some degree disqualifies earlier attempts to describe
the random-field system by a unique magnetic state. Instead, the system exhibits magnetic hysteresis similar
to that in conventional ferromagnets with pinning of the domain walls.

Starting from random initial condition we find states having small
values of $m$ (decreasing to zero in the large-size limit) and substantial
vorticity. For $R_{f}\gg1$ there is a strong short-range order everywhere
except the vicinity of vortex loops. The correlation function in this
state decays to zero but the correlation length is defined by the
average distance between the vortices rather then by $R_{f}$, the
former being much shorter. We call this state a vortex glass (VG).

\begin{figure}
\centering\includegraphics[width=8cm]{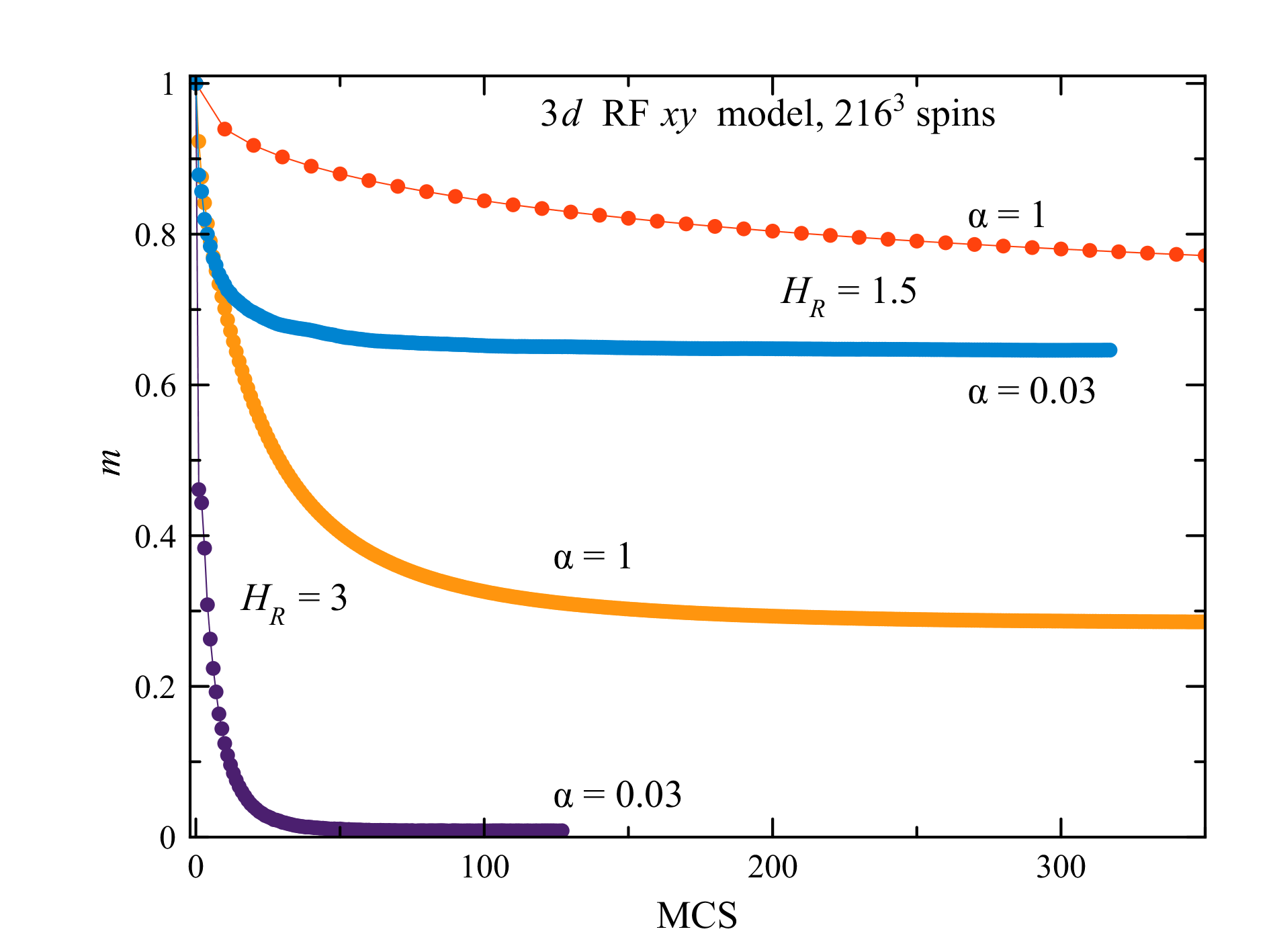}

\caption{Magnetization relaxation curves starting from a collinear initial
condition. The method with a small damping constant, $\alpha=0.01-0.03$ is most efficient.}

\label{Fig-m_vs_MCS_L=00003D216_HR=00003D1.5_pbc_coll_IC_alp=00003D1_0.03}
\end{figure}

\begin{figure}
\centering\includegraphics[width=8cm]{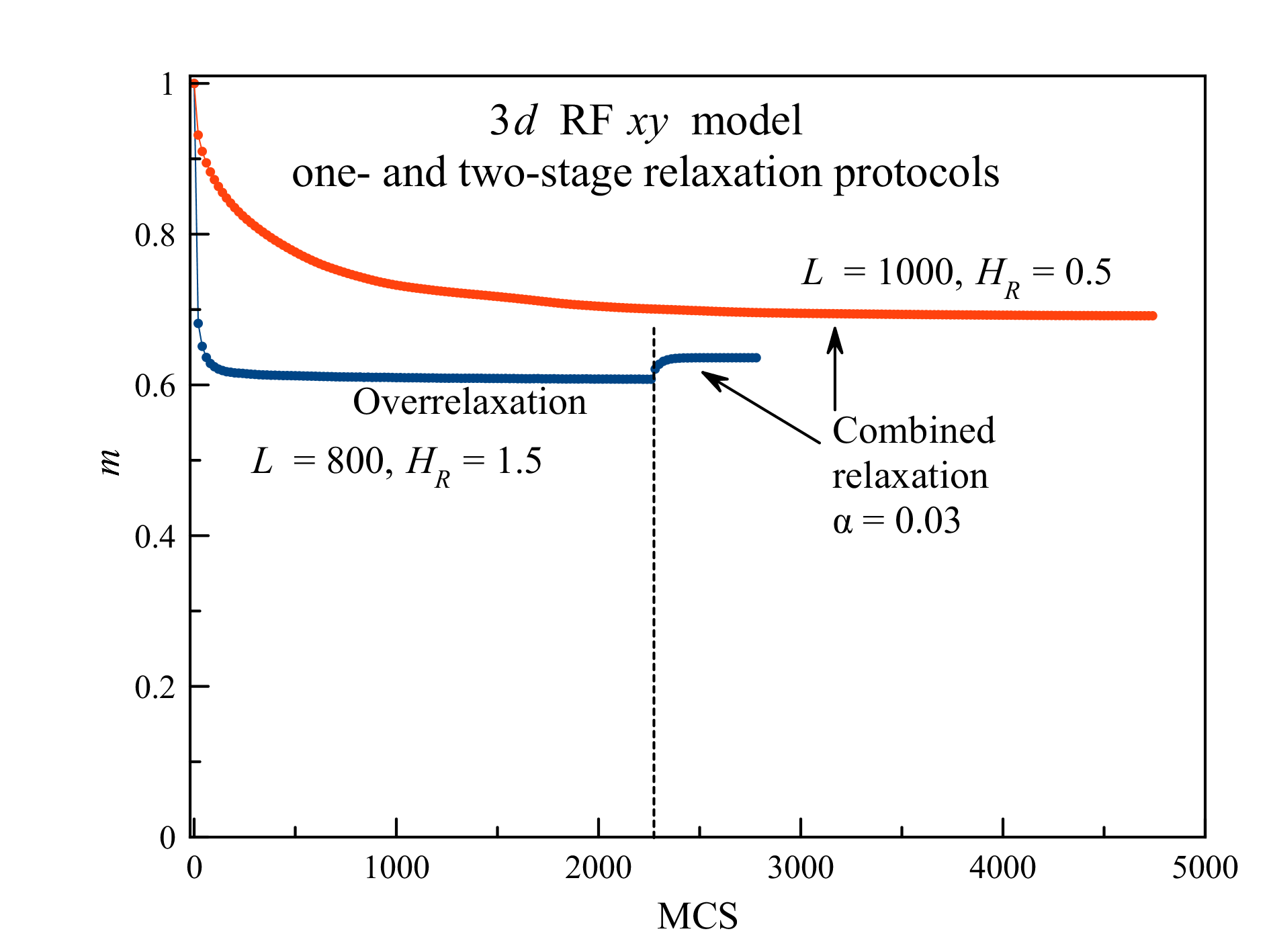}

\caption{Two-stage relaxation starting from a collinear state for $H_{R}=1.5$
and $L=800$ and one-stage relaxation for $H_{R}=0.5$ and $L=1000$,
our largest system size. Note a slow relaxation for $H_{R}=0.5$.
\label{Fig-m_vs_MCS_L=00003D800_1000_HR=00003D1.5_0.5_pbc_coll_IC_alp=00003D0_0.03}}
\end{figure}

Starting from the collinear initial condition, for $H_{R}\lesssim2$
we find only partially disordered states having $m$ still of order
1 (stable in the large-size limit) and zero or extremely small vorticity.
In this state, the correlation function follows Eq. (\ref{Exp-CF})
at short distances but reaches a plateau at longer distances, thus
showing a long-range order. We call this state ferromagnetic, although
it should be stressed that the system does not order spontaneously
on lowering temperature but freezes into the vortex glass state instead.

For $H_{R}\gtrsim2,$ starting from the collinear initial state, vortex
loops are spontaneously generated and magnetization is strongly reduced.

The energy of the VG state is always higher than the energy of the
ferromagnetic state. (This holds for both $xy$ and Heisenberg models
in $1d$, $2d$, and $3d$, as well as for random-anisotropy models.)
Thus the vortex glass state is a metastable state that could, in principle,
relax to the ferromagnetic state by eliminating vortex loops that
cost energy. However, this does not happen because vortex loops are
pinned by the random field. It is possible that the ferromagnetic
state is also a metastable state, while there is a true ground state
with $m=0$, in accordance with the implicit theorem by Aizenman and
Wehr. \cite{Aizenman-Wehr-PRL,Aizenman-Wehr-CMP} However, we were
unable to find this state by relaxation from typical states. To the
contrary, sampling local energy minima shows that starting with a
low $m$ state it is easier to find lower energy states with higher
$m$ than with lower $m$.

\subsection{Relaxation from the collinear state leading to a ferromagnetic state}

\label{overview}

\begin{figure}
\centering\includegraphics[width=8cm]{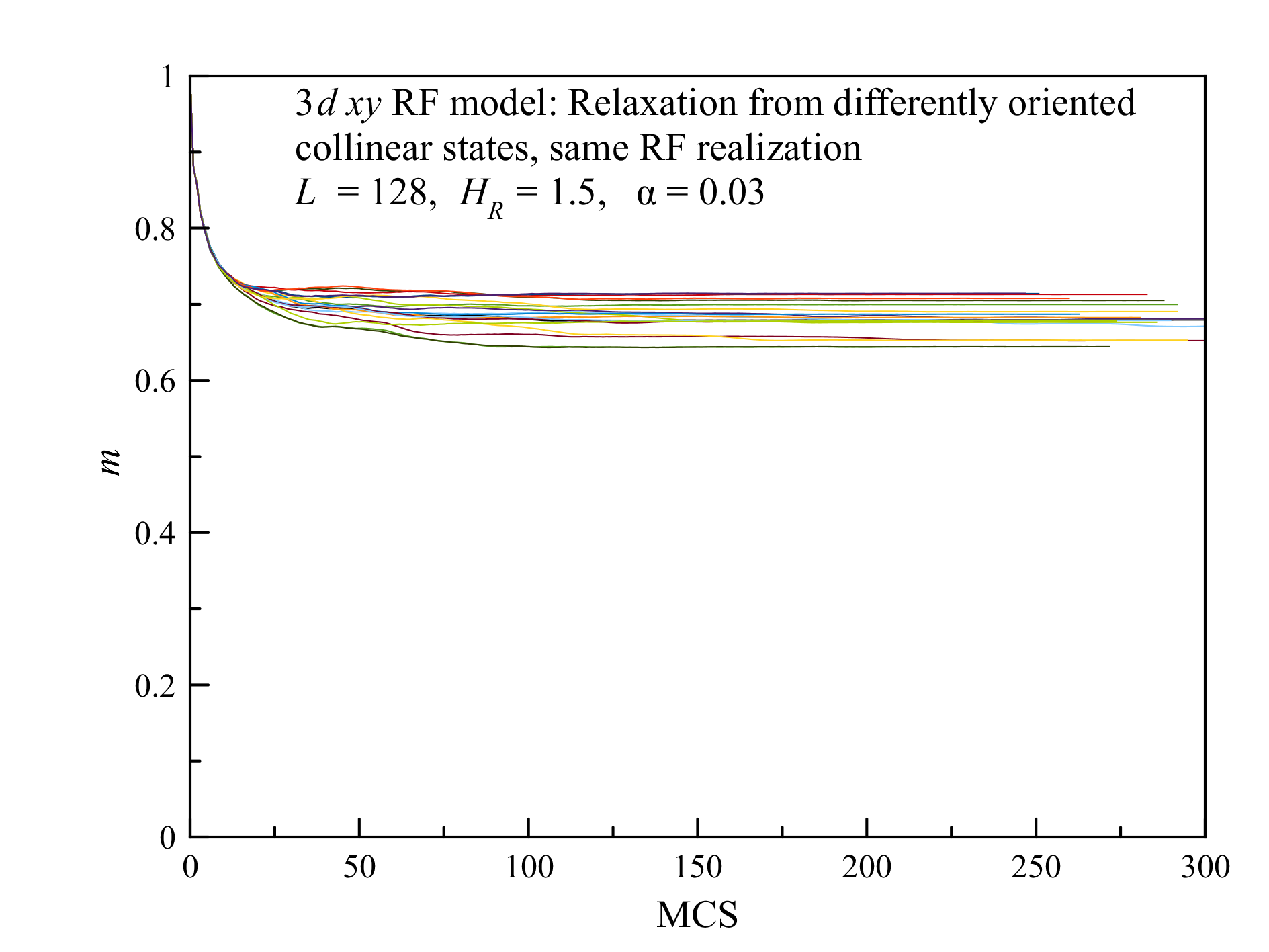}

\caption{Relaxation starting from differently oriented collinear states for the same realization of the random field,
showing glassy nature of the RF magnet.
\label{Fig-m_vs_MCS_many_coll_IC_same_RF_Nalp=2_Nx=Ny=Nz=128_HR=1.5}}
\end{figure}

\begin{figure}
\centering\includegraphics[width=8cm]{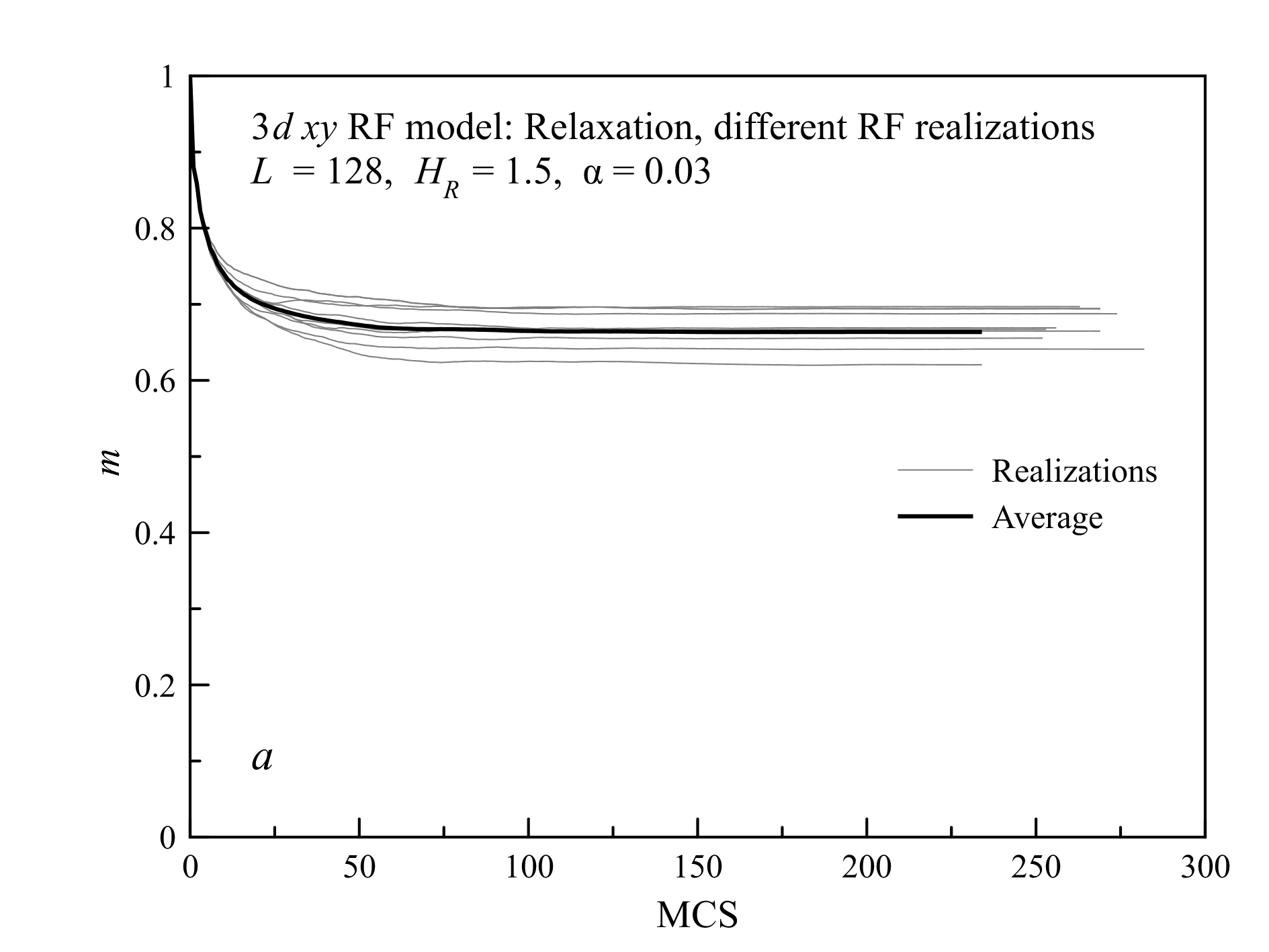}
\centering\includegraphics[width=8cm]{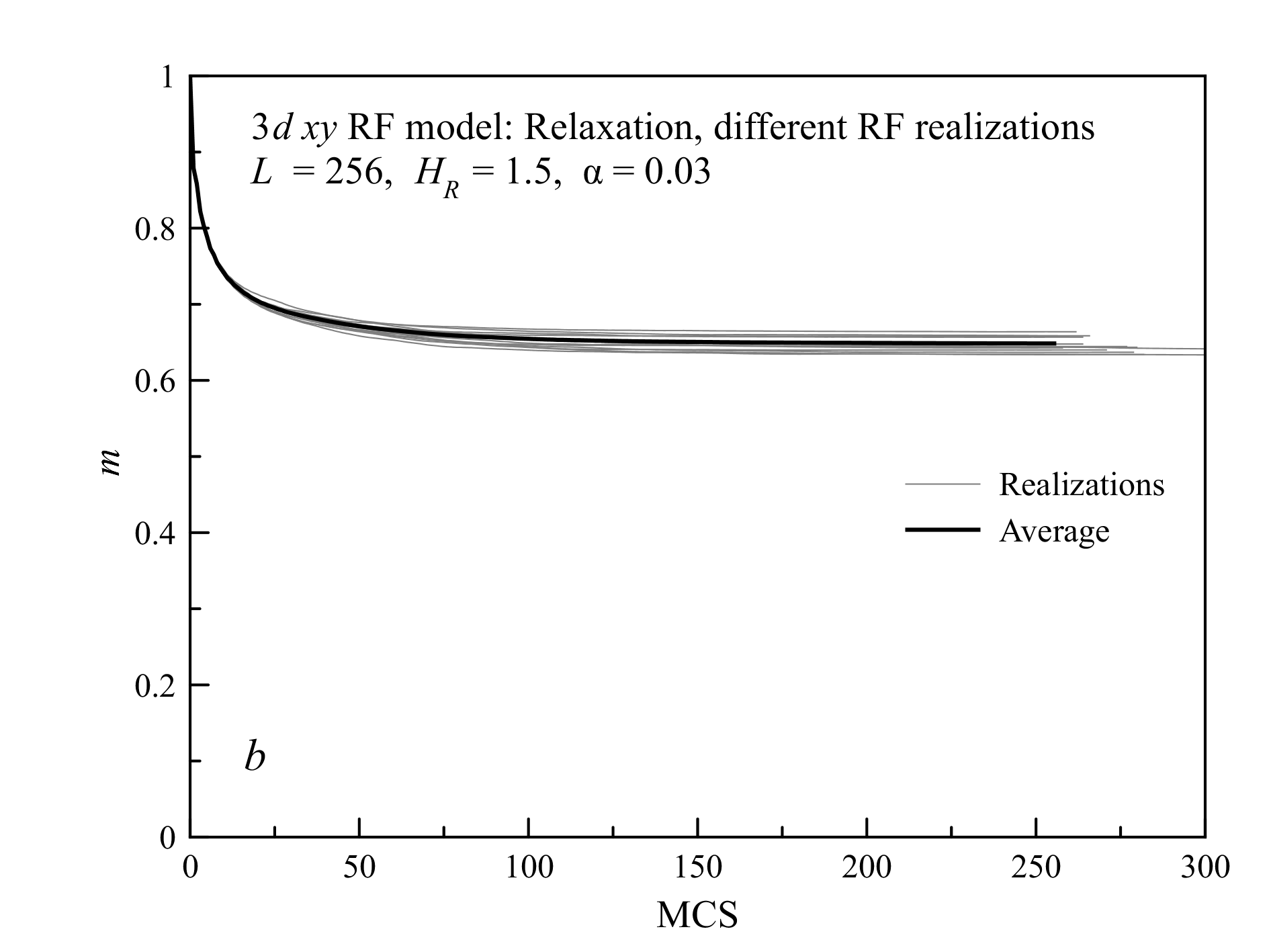}

\caption{Relaxation from collinear states for different realization of the random field.
Statistical scatter decreases with the system size due to self-averaging. a) $L=128$; b) $L=256$.
\label{Fig-m_vs_MCS_many_realizations_Nalp=2_Nx=Ny=Nz=128_HR=1.5}}
\end{figure}

During relaxation from any of the initial states we have tried, the
system's energy decreases. Starting from the collinear initial condition,
$m$ decreases until it reaches a constant value for $H_{R}\lesssim2$
and goes to zero at $H_{R}\gtrsim2$. Direction of the magnetization
vector $\mathbf{m}$ practically does not deviate from the initial
direction. Fig. \ref{Fig-m_vs_MCS_L=00003D216_HR=00003D1.5_pbc_coll_IC_alp=00003D1_0.03}
shows relaxation curves for $H_{R}=1.5$ with $m$ approaching a nonzero
constant and for $H_{R}=3$ with $m$ going to zero. One MCS (Monte
Carlo Step) means one complete spin update of the system. We use this
standard notation although we are not using Monte Carlo. One can see
that the pure finite rotation method ($\alpha=1$) is slow for our
problem in comparison to the combined method predominantly using over-relaxation
($\alpha=0.03$). For $H_{R}=3$, the system gets stuck in a metastable
state with $m\approx0.3$ and $\Delta E=-0.668$. However, the combined
method finds the state with a very small $m$ and the lower energy
$\Delta E=-0.671$. Here $\Delta E\equiv E-E_{0}$, where $E_{0}$is
the exchange energy of the collinear state, $-3J$ for the $3d$ model
with periodic boundary conditions (pbc). These results are in accordance
with the mechanism of relaxation sketched in Fig. \ref{Fig:Energy_minimization}.

In fact, already the pure over-relaxation method ($\alpha=0$) provides
a fast relaxation of $m$ at a constant energy. For this reason in
some computations we used the two-stage method, as shown in Fig. \ref{Fig-m_vs_MCS_L=00003D800_1000_HR=00003D1.5_0.5_pbc_coll_IC_alp=00003D0_0.03}.
The idea is that the conservative over-relaxation method has a potential
for the maximal possible disordering since it leads to states that
can be interpreted as thermal states with a small $T$ (the over-relaxation
plateau in Fig. \ref{Fig-m_vs_MCS_L=00003D800_1000_HR=00003D1.5_0.5_pbc_coll_IC_alp=00003D0_0.03}).
As the energy-relaxation mechanism is switched on, this temperature
goes to zero and ordering in the system increases, as is seen in the
Fig. \ref{Fig-m_vs_MCS_L=00003D800_1000_HR=00003D1.5_0.5_pbc_coll_IC_alp=00003D0_0.03}.
The states obtained in these computations are vortex free.

Fig. \ref{Fig-m_vs_MCS_many_coll_IC_same_RF_Nalp=2_Nx=Ny=Nz=128_HR=1.5} obtained by multiple relaxation events of a system with {\em the same} realization of the RF from differently oriented collinear states
shows different local energy minima achieved in different cases. This confirms glassy nature of a random-field magnet.
All these states are vortex-free, as above.

Fig. \ref{Fig-m_vs_MCS_many_realizations_Nalp=2_Nx=Ny=Nz=128_HR=1.5} shows similar computations with different realization of the RF.
One can see that Fig. \ref{Fig-m_vs_MCS_many_realizations_Nalp=2_Nx=Ny=Nz=128_HR=1.5}$a$ is similar to
Fig. \ref{Fig-m_vs_MCS_many_coll_IC_same_RF_Nalp=2_Nx=Ny=Nz=128_HR=1.5}.
Comparison of the two panels of Fig. \ref{Fig-m_vs_MCS_many_realizations_Nalp=2_Nx=Ny=Nz=128_HR=1.5} shows that the statistical scatter
decreases with the system size because of self-averaging.
For the standard deviation $\Delta m$ of the magnetization in the final state one has $\Delta m \approx 0.025$ for $L=128$
and $\Delta m \approx 0.0097$ for $L=256$.
On the other hand, $\Delta m L^{3/2} \approx 36$ for $L=128$ and $\Delta m L^{3/2} \approx 39$ for $L=256$ that are nearly the same.
This is in accord with the picture of correlated regions of linear size $R_f$ that are oriented independently from each other, leading to
\begin{equation}
\Delta m \propto \left(\frac{R_f}{L}\right)^{3/2}.\label{eq:Deltam-Def}
\end{equation}
Using $R_f$ of Eq. (\ref{eq:Rf-Def}) and the numerical factor from the computational results above, one can estimate the statistical scatter in all other cases.

The structure of the ferromagnetic state shown in Fig. \ref{Fig-spins_collinear_IC_L=00003D64_HR=00003D1}
has no singularities.

\begin{figure}
\centering\includegraphics[width=8cm]{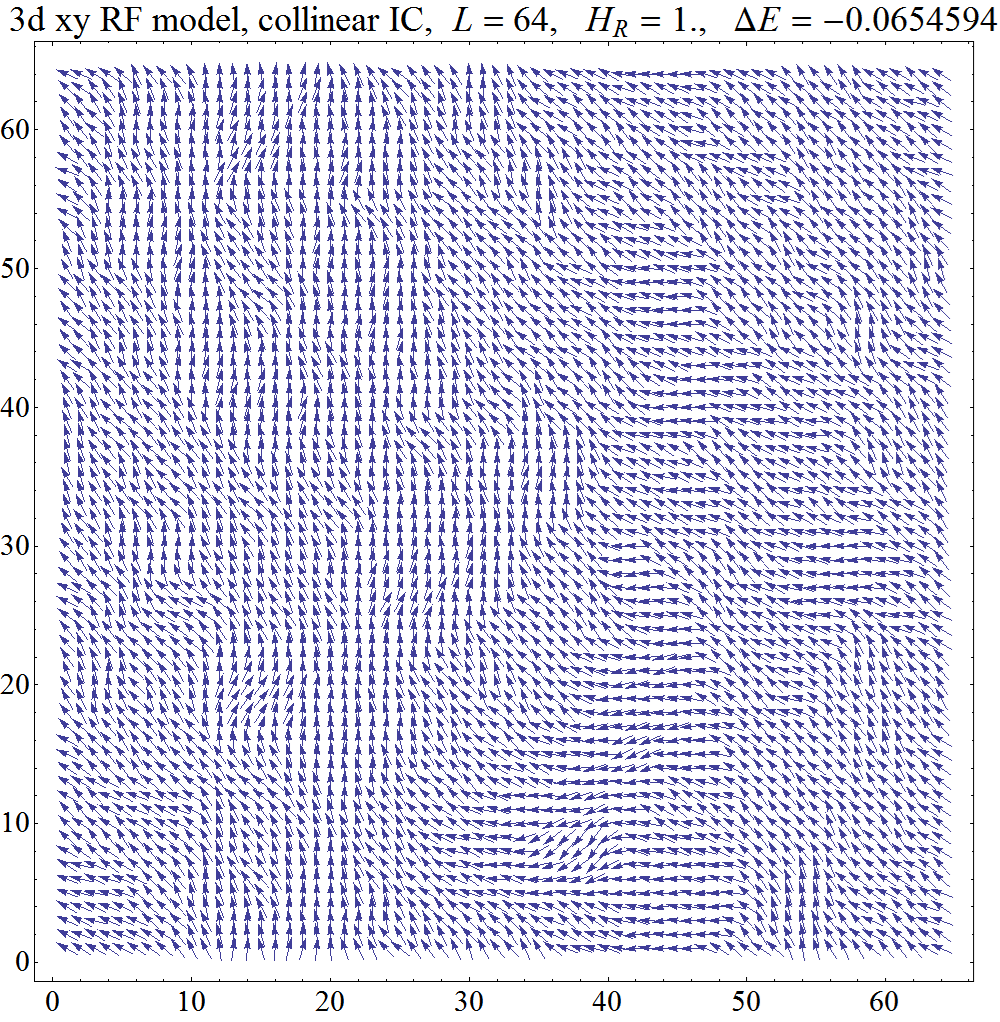}

\caption{Spin configuration obtained for $H_{R}=1$ from the collinear initial
condition}

\label{Fig-spins_collinear_IC_L=00003D64_HR=00003D1}
\end{figure}

\begin{figure}
\centering\includegraphics[width=8cm]{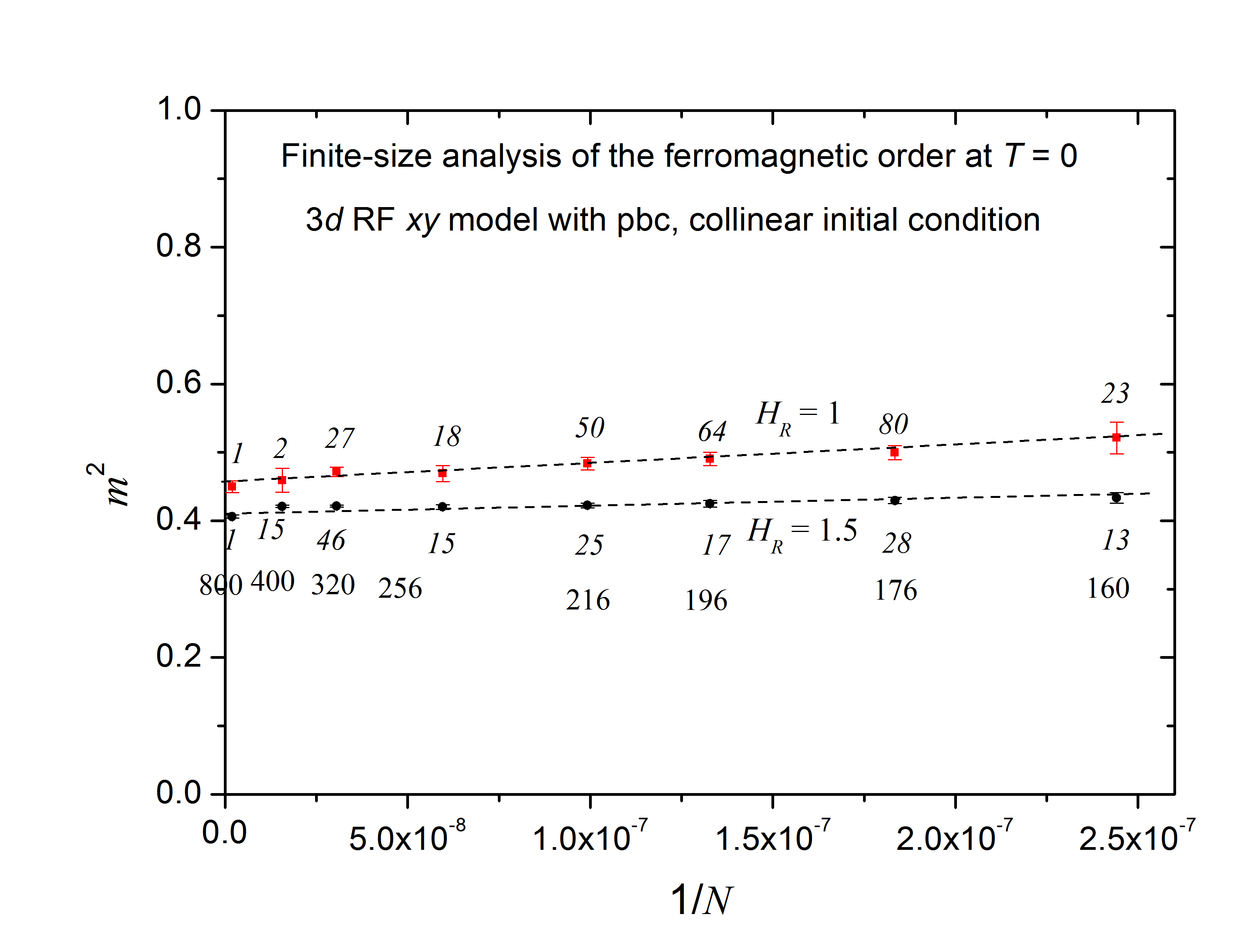}

\caption{Magnetization squared in the ferromagnetic state vs system volume
$V=N$. Italicized numbers are those of RF realizations used to compute the averages of $m$.
Upright numbers below points indicate the systems' linear size $L$.
Straight dashed lines are guides for the eye.}

\label{Fig-m^2_vs_size_HR=00003D1_and_1.5}
\end{figure}

With increasing the system size the numerically found $m$ does not
decrease to zero, as one can see by comparing Figs. \ref{Fig-m_vs_MCS_L=00003D216_HR=00003D1.5_pbc_coll_IC_alp=00003D1_0.03}
and \ref{Fig-m_vs_MCS_L=00003D800_1000_HR=00003D1.5_0.5_pbc_coll_IC_alp=00003D0_0.03}.
The stability of the ferromagnetic state is clearly seen from the
finite-size analysis shown in Fig. \ref{Fig-m^2_vs_size_HR=00003D1_and_1.5}.
Here all points except for $L=800$ have been obtained by averaging
over realizations of the random field, the number of realizations indicated by the italicized numbers.
Although there is self-averaging in the system, averaging over realizations allows to further reduce data scatter.
One can see that the points
fall on straight lines with a finite offset, in accordance with Eq.
(\ref{eq:m2_via_CF}). The error bars are the uncertainties of the average values computed as $\Delta m /\sqrt{n}$, where
$\Delta m$ is the standard deviation defined by Eq. (\ref{eq:Deltam-Def}) and $n$ is the number of realizations.

\subsection{Relaxation from the wavy state}

\label{sub:relaxation}

\begin{figure}
\centering\includegraphics[width=8cm]{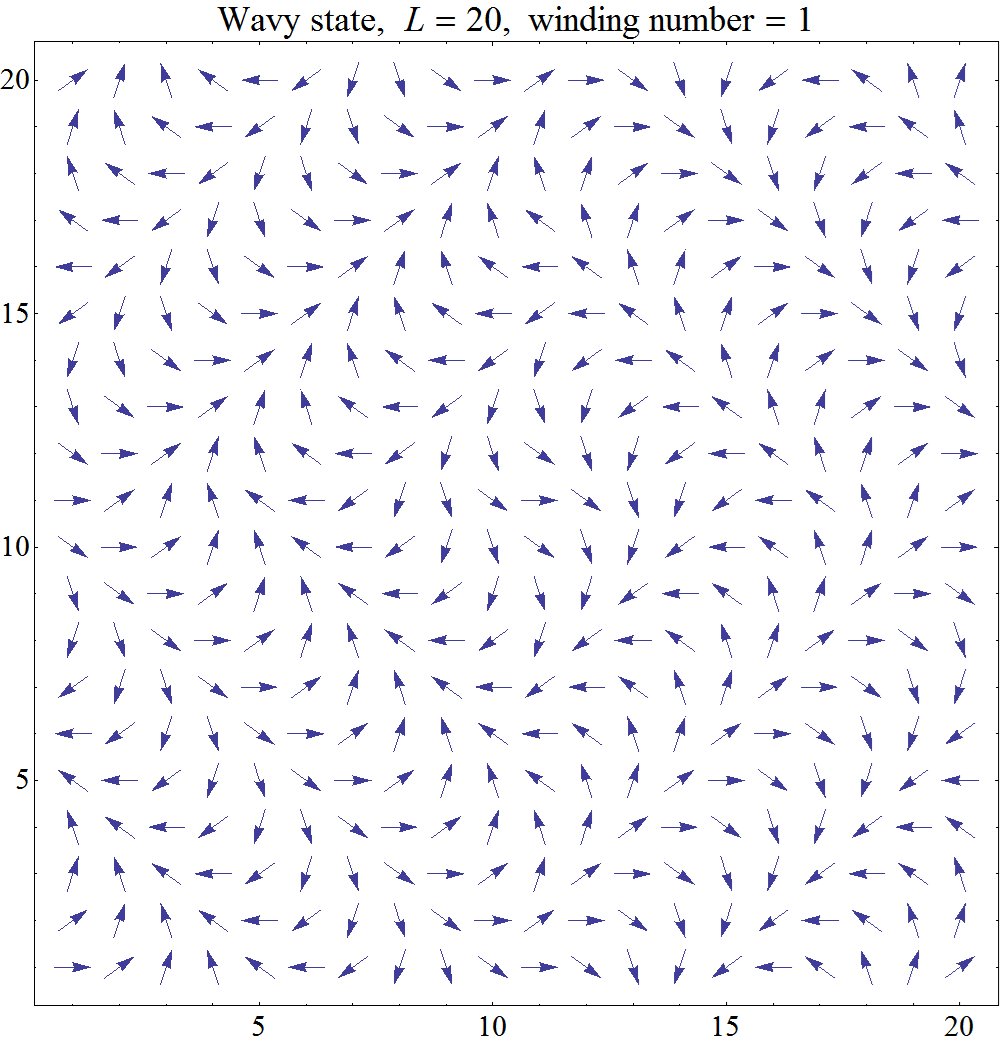}

\caption{Wavy state of spins in the $xy$ plane}

\label{Fig-wavy_state}
\end{figure}

\begin{figure}
\centering\includegraphics[width=8cm]{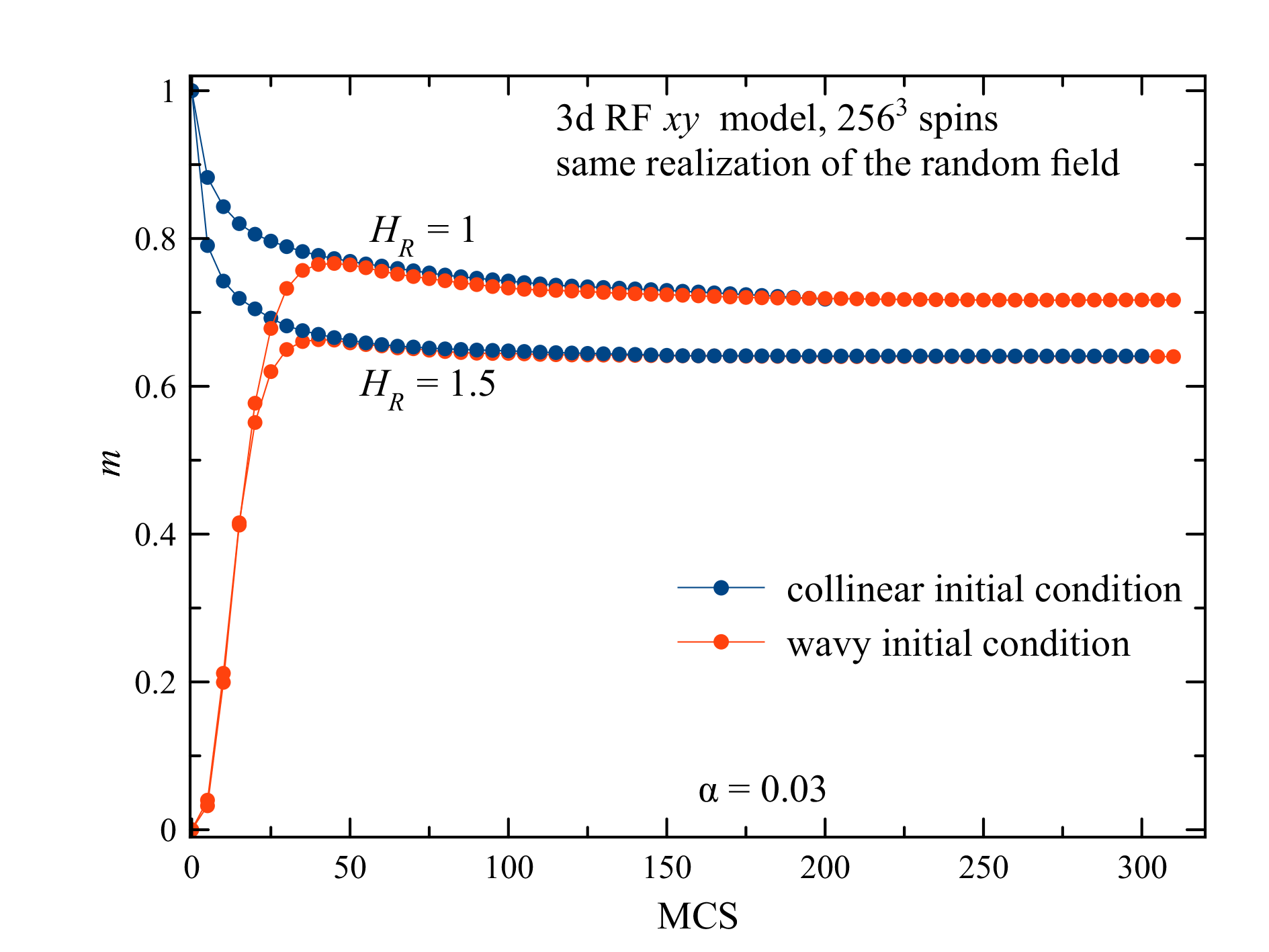}

\caption{Magnetization relaxation from the collinear and wavy initial states.}

\label{Fig-m_vs_MCS_coll_and_wavy_Nalp=00003D2_L=00003D256_HR=00003D1_1.5_pbc_alp=00003D0.03}
\end{figure}

\begin{figure}
\centering\includegraphics[width=8cm]{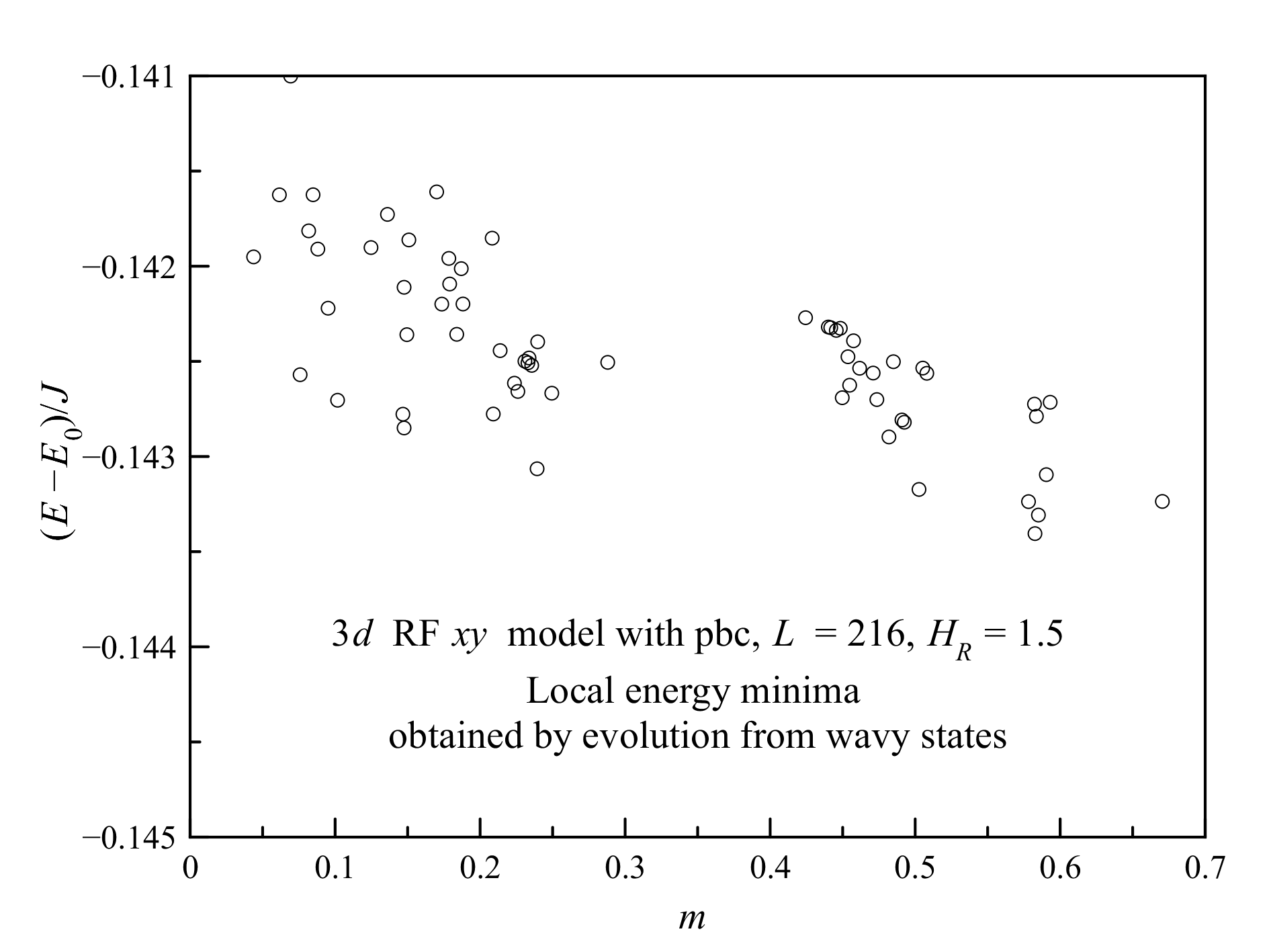}

\caption{Local energy minima, labelled by the corresponding magnetization values,
obtained by evolution from wavy states, Eq. (\ref{eq:WavyState-Def}),
with $k_{x,y,z}=0,1,2,3$.}

\label{Fig-dE_vs_m_wxyz_L=00003D216_HR=00003D1.5_pbc_alp=00003D0.03_eta=00003D0.02}
\end{figure}

One can argue that the ferromagnetic state obtained from the collinear
initial state is an artefact and ferromagnetism here is preselected.
An argument in support of ferromagnetic state can be obtained by starting
from a special kind of initial state that has $m=0$ and no vortices
or helicity. In this state, which we call a wavy state, spins rotate
in one direction and then in the opposite direction when the observer
is moving in any of the three directions in the cubic lattice. It
is defined by
\begin{equation}
\left(s_{x},s_{y}\right)=\left(\cos\left(\Phi\right),\sin\left(\Phi\right)\right),\label{eq:WavyState-Def}
\end{equation}
where
\begin{equation}
\Phi=\frac{2\pi k_{x}n_{x}}{N_{x}}\left(-1\right)^{\left[k_{x}n_{x}/N_{x}\right]}+(x\Rightarrow y)+\left(x\Rightarrow z\right).\label{eq:WavyState-Phase}
\end{equation}
Here $N_{x,y,z}$ are lattice sizes, $n_{x,y,z}=1,\ldots,N_{x,y,z}$
are lattice positions, $k_{x,y,z}$ are corresponding wave vectors
and $\left[x\right]$ means integer part. The wavy state is topologically
equivalent to the collinear state because it can be transformed into
the latter by continuous deformations without changing the topology.
This state resembles a spring that tends to straighten when released.
Its energy is $\sim J(a/L)^{2}$ above that of the collinear state.
An example of a wavy state is shown in Fig. \ref{Fig-wavy_state}.
Fig. \ref{Fig-m_vs_MCS_coll_and_wavy_Nalp=00003D2_L=00003D256_HR=00003D1_1.5_pbc_alp=00003D0.03}
shows magnetization relaxation curves starting from the collinear
and wavy initial conditions that lead to the same final state with
a high $m$. It must be noted that restoration of a large $m$ out
of the wavy state does not always take place. For $H_{R}\gtrsim2$
vortices are generated spontaneously out of any vortex-free state,
including the wavy state, so that the final state is a vortex glass
with $m$ close to zero. Even for $H_{R}=1.5$ the system randomly
lands in (vortex-free) states with small and large $m$, see Fig.
\ref{Fig-dE_vs_m_wxyz_L=00003D216_HR=00003D1.5_pbc_alp=00003D0.03_eta=00003D0.02}.
Note that states with higher $m$ in Fig. \ref{Fig-dE_vs_m_wxyz_L=00003D216_HR=00003D1.5_pbc_alp=00003D0.03_eta=00003D0.02}
typically have a lower energy.

\subsection{Vortex-glass state}

\label{sub:spin structures}

\begin{figure}
\centering

\centering\includegraphics[width=8cm]{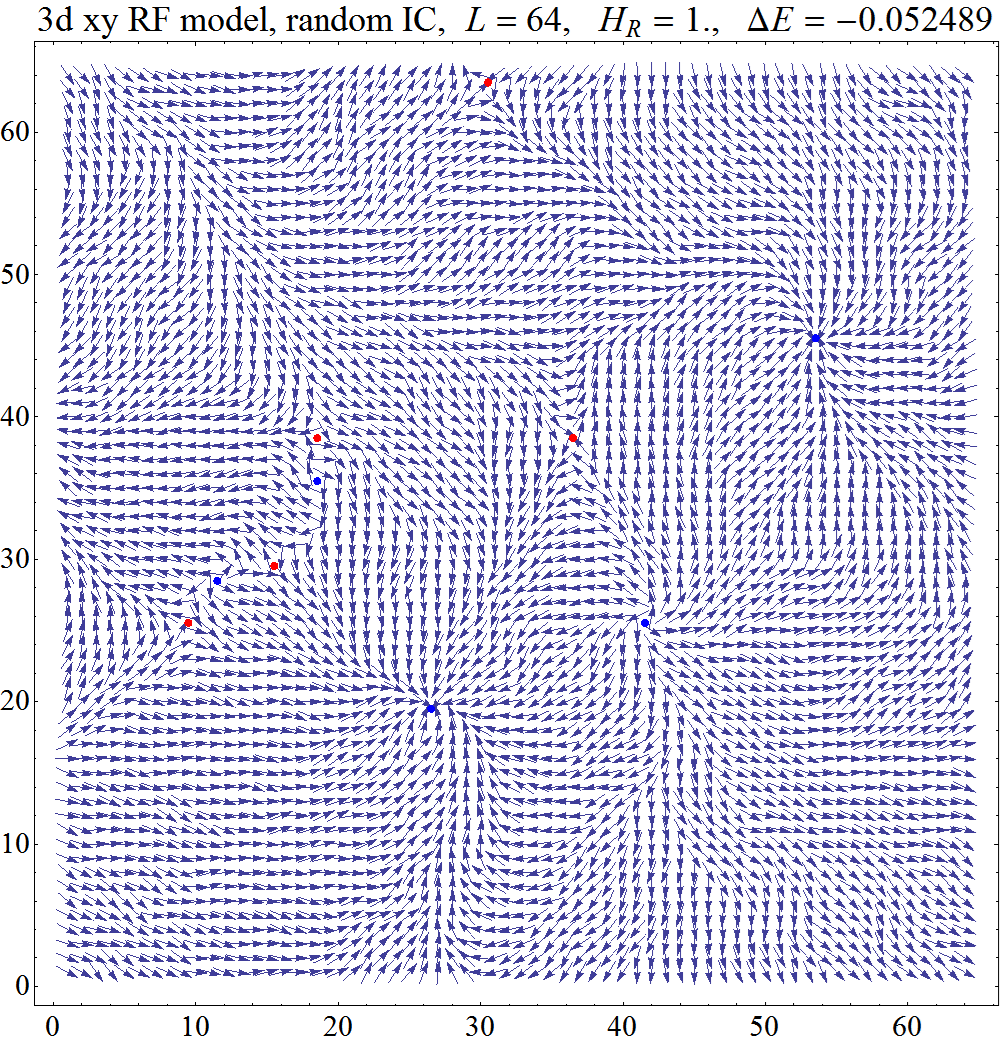}

\caption{Spin configuration obtained for $H_{R}=1$ from a random initial conditions.
Vortices/antivortices are shown by blue/red circles. }

\label{Fig-spins_random_IC_L=00003D64_HR=00003D1}
\end{figure}

The vortex-glass state contain singularities, vortices and antivortices,
shown in Fig. \ref{Fig-spins_random_IC_L=00003D64_HR=00003D1}. In
our $3d$ case these are vortex lines going through the sample and
vortex loops.

For larger $H_{R}$ vortex loops are created by the random field even
starting from the collinear initial condition. For any system size,
there is a critical value $H_{R,c}\approx2$ above which vortex loops
emerge. Slightly above $H_{R,c}$ these vortex loops are short, as
shown in Fig. \ref{Fig-vortex_loops} (top). With increasing $H_{R}$,
vortices quickly proliferate into the system and the number and length
of vortex loops increase. It is difficult to prove whether there exists
a size-independent threshold value $H_{R,c}$. Computations show that
$H_{R,c}$ slowly decreases with the size. However, this question
seems to be not very important because the vorticity increases with
a very small slope above the threshold. It may be that in the bulk
there are vortex loops at any finite $H_{R}$ but the vorticity for
small $H_{R}$ is extremely low.

Meanwhile, starting from random initial conditions one arrives at
states with long vortex lines that typically do not close into loops
but cross the system's boundaries, see Fig. \ref{Fig-vortex_loops}(bottom).
As vortices and antivortices can exist in all three available planes,
different singularities may exist at nearly the same point, e.g.,
a vortex in the $xy$ plane may occupy the same point as an antivortex
in the $yz$ plane. For this reason, some points in the figure may
contain both black and red paints.

Obtaining VG states with our algorithm amounts to slow cooling the
system. We have checked with Monte Carlo simulations that slow lowering
the temperature leads to the same effect: the system does not order
ferromagnetically but rather freezes into the VG state that for $H_{R}<H_{R,c}$
has a higher energy than the ferromagnetic state.

\begin{figure}
\centering\includegraphics[width=8cm]{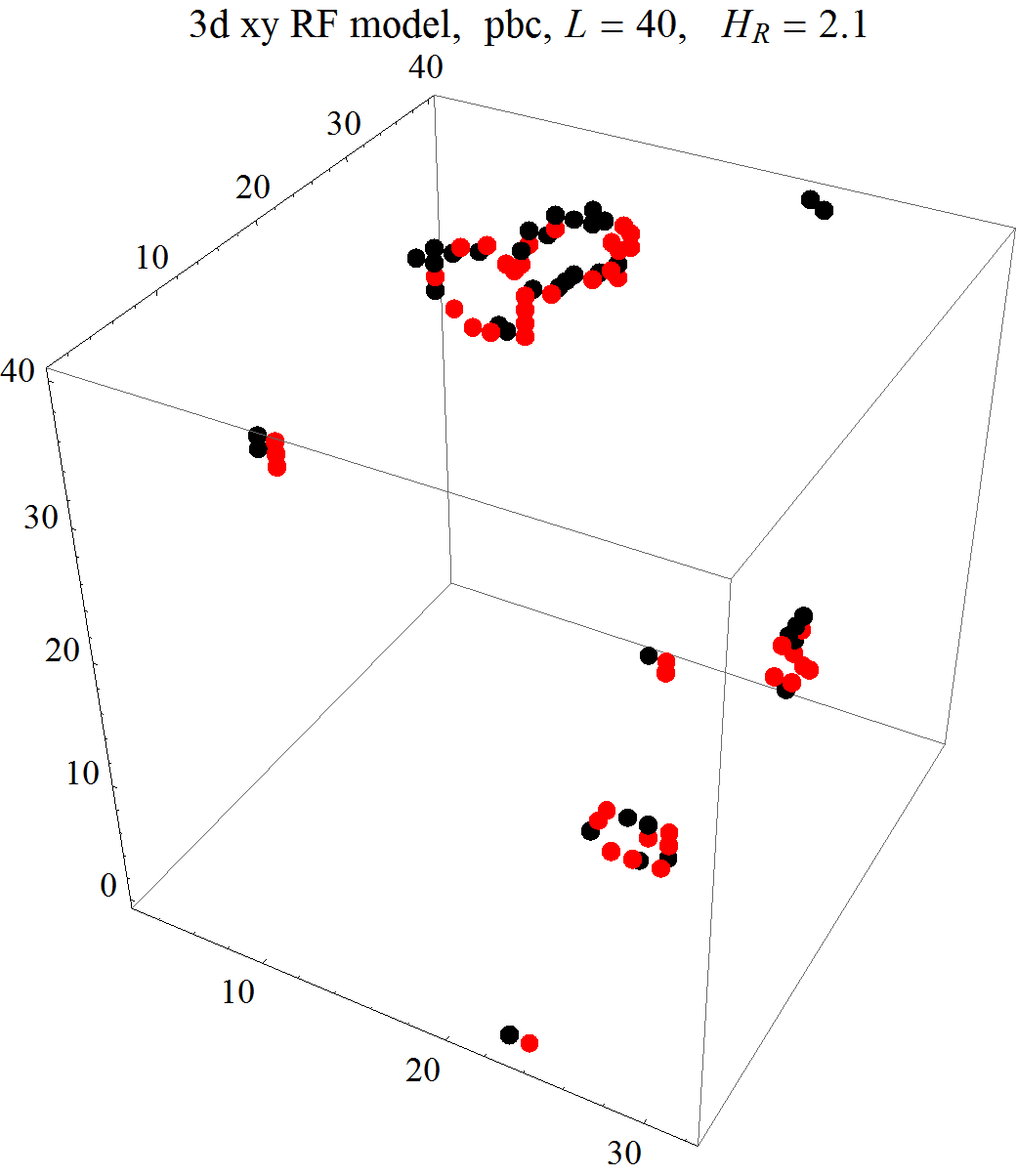}

\centering\includegraphics[width=8cm]{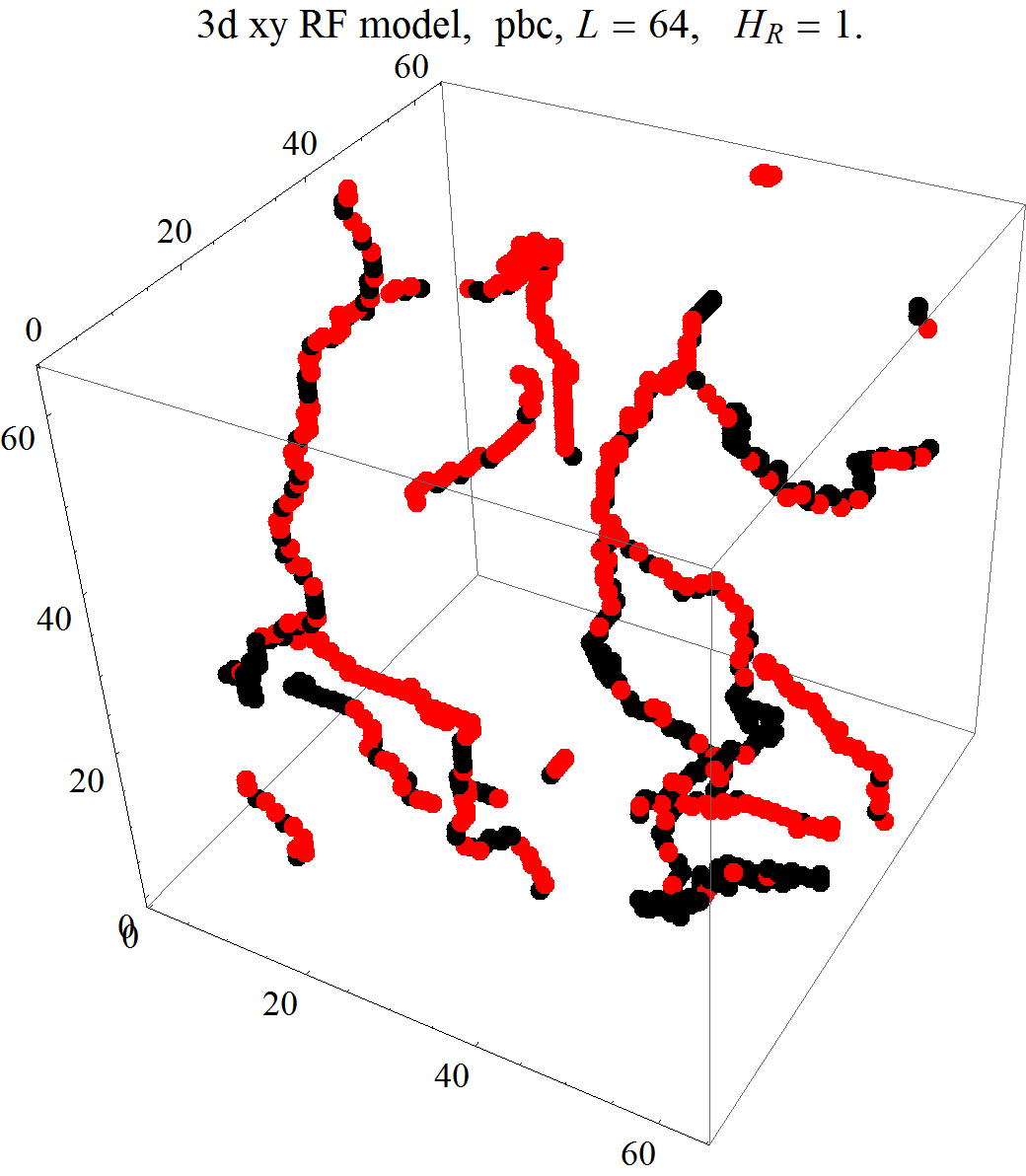}\caption{Vortex loops in $3d$ xy RF model. Collinear (top) and random (bottom)
initial conditions. Vortices/antivortices are shown by black/red.}

\label{Fig-vortex_loops}
\end{figure}

\subsection{Magnetization and vorticity}

\label{sub:mag-vorticity}

\begin{figure}
\centering\includegraphics[width=8cm]{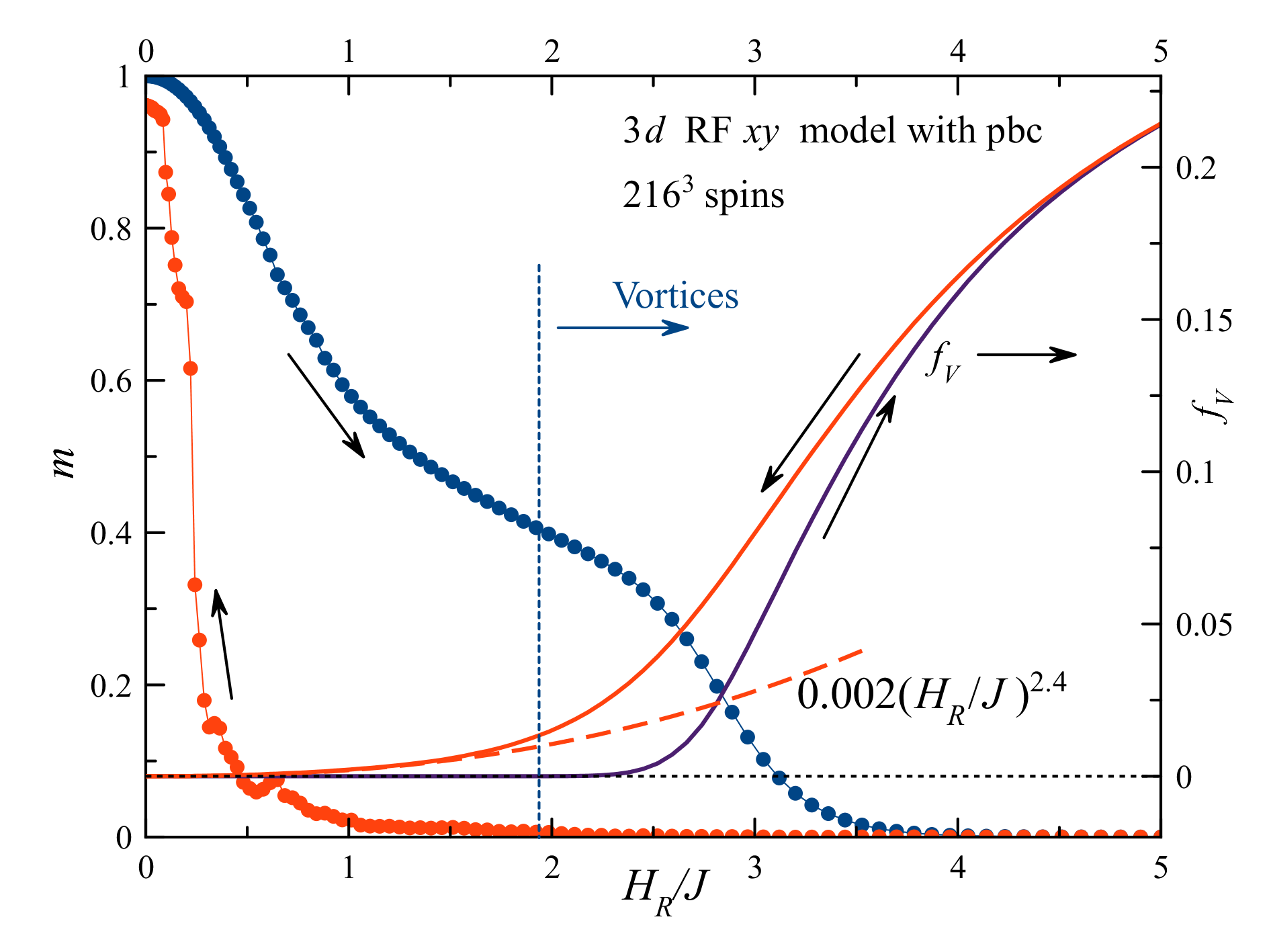}

\caption{Magnetization vs the random field strength $H_{R}$ for the model
with pbc of the size $L=216$.}

\label{Fig-m_vs_HR_L=00003D216_rc=00003D0_alpha=00003D0.05_pbc}
\end{figure}

\begin{figure}
\centering\includegraphics[width=8cm]{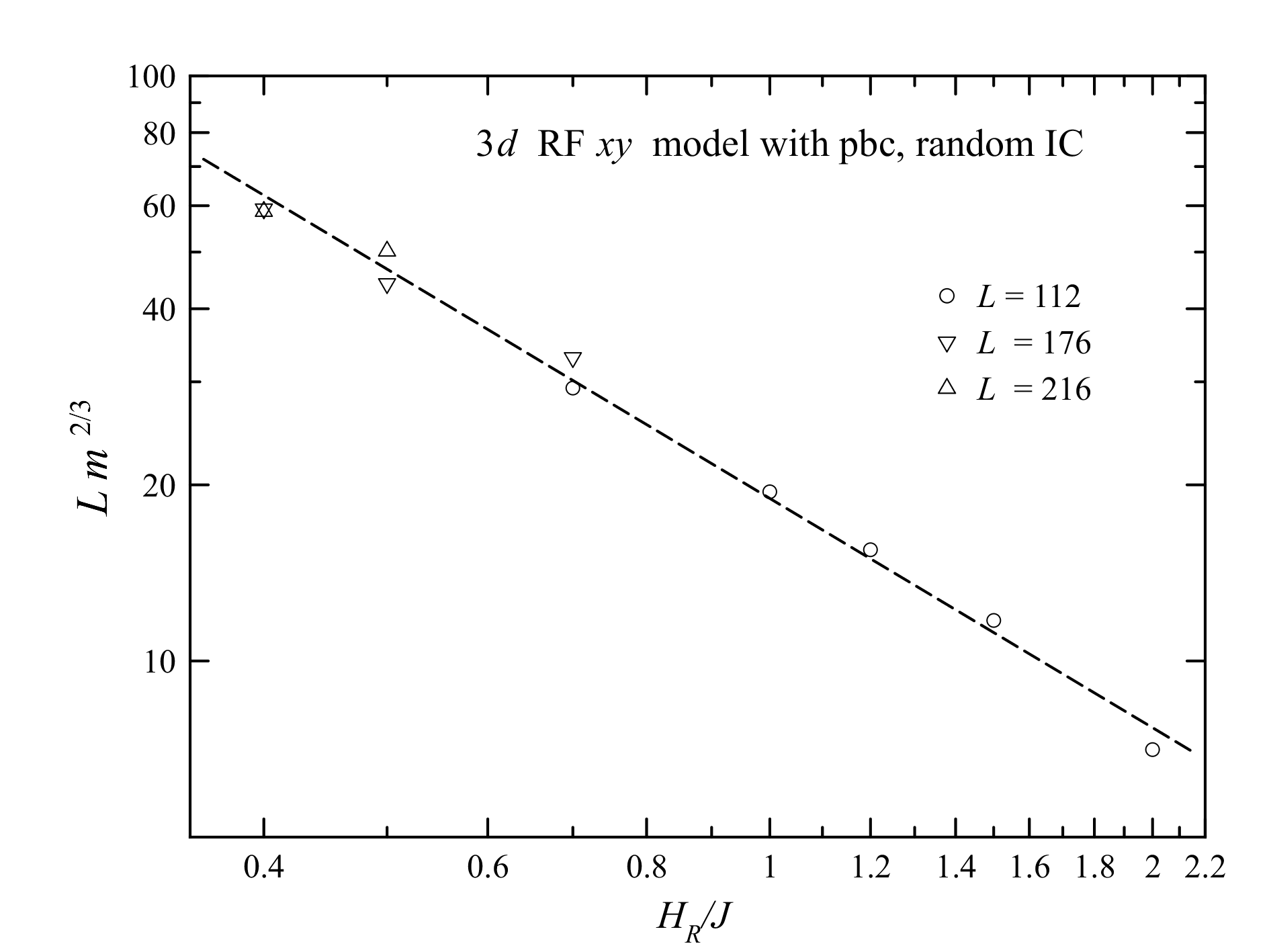}

\caption{Finite-size analysis of the magnetization in the vortex-glass phase}

\label{Fig-RV_vs_HR_from_m}
\end{figure}

The magnetization $m$ and vorticity $f_{V}$ as functions of $H_{R}$
are shown in Fig. \ref{Fig-m_vs_HR_L=00003D216_rc=00003D0_alpha=00003D0.05_pbc}.
Here the same realization of the random field was used and only its
strength $H_{R}$ was changed in small steps, using final states for
a given $H_{R}$ as initial conditions for the next value of $H_{R}$.
Different random-field realizations result in slightly different curves.

At $H_{R}=5$ vorticity is very high and it decreases upon lowering
$H_{R}$. The magnetization remains very small as the system enters
the vortex glass phase with a small but nonzero $f_{V}$. For $L=216$
the number of vortex lines in the system becomes small below $H_{R}=1$
{[}see Fig. \ref{Fig-vortex_loops}(bottom){]} and $m$ starts to
increase. For larger $L$, this happens at smaller $H_{R}.$ In some
cases the system reaches a collinear state with $m=1$ at $H_{R}=0$.
In other cases, as in Fig. \ref{Fig-m_vs_HR_L=00003D216_rc=00003D0_alpha=00003D0.05_pbc},
the system ends up in a topologically stable state with nonzero helicity
(for pbc) and $m<1$.

The magnetization in the ferromagnetic state decreases with $H_{R}$
as shown in Fig. \ref{Fig-m_vs_HR_L=00003D216_rc=00003D0_alpha=00003D0.05_pbc},
starting from the pure limit $m=1$ at $H_{R}=0$. For $H_{R}<1.8$
this state is vortex-free. Proliferation of vortices for $H_{R}>1.8$
results in the shoulder of this curve and full destruction of the
order at $H_{R}>4$.

The magnetization of the VG state for $H_{R}<H_{R,c}$ is small and
it scales as $m\propto1/L^{3/2}$, in accordance with Eq. (\ref{m_fluct}),
as shown in Fig. \ref{Fig-RV_vs_HR_from_m}. Putting together data
obtained for different $L$ and $H_{R}$ data, averaged over many
RF realizations, one obtains the dependence of the correlation radius
of the VG phase $R_{V}$ that replaces $R_{f}$ of Eq. (\ref{eq:Rf-Def}).
The result is
\begin{equation}
R_{V}\propto1/H_{R}^{1.2}\label{eq:Rv-Dependence}
\end{equation}
that is much shorter than $R_{f}$ at small $H_{R}$. The numerical
factor in this formula cannot be found by this method because the
form of the CF in the VG state is different from the simple exponential.
The precise form of $R_{V}$ will be found in the section on correlation
functions below.

On the other hand, vorticity data in the VG state in Fig. \ref{Fig-m_vs_HR_L=00003D216_rc=00003D0_alpha=00003D0.05_pbc}
can be roughly fitted to the form

\begin{equation}
f_{V}\approx0.002(H_{R}/J)^{2.4}.\label{eq:fV_fitted}
\end{equation}
Combining the two formulas above yields
\begin{equation}
R_{V}\propto1/f_{V}^{1/2}.\label{eq:RV_via_fV-propto}
\end{equation}
It is clear that vortices are the main reason for the decay of spin-spin
correlations in the vortex glass for $R_{V}\ll R_{f}$ . Thus there
must be a relation between $R_{V}$ and the vorticity $f_{V}$ defined
as the fraction of unit plaquettes with vortices or antivortices.
Naively one could expect that $R_{V}$ is proportional to the distance
between the singularities, so that in $3d$ one has $R_{V}\propto1/f_{V}^{1/3}$.
As vortex lines are linear objects, $R_{V}$ is proportional to the
average distance between vortex lines. This makes the situation effectively
two dimensional.

\subsection{Energy}

\label{sub:energy}

\begin{figure}
\centering\includegraphics[width=8cm]{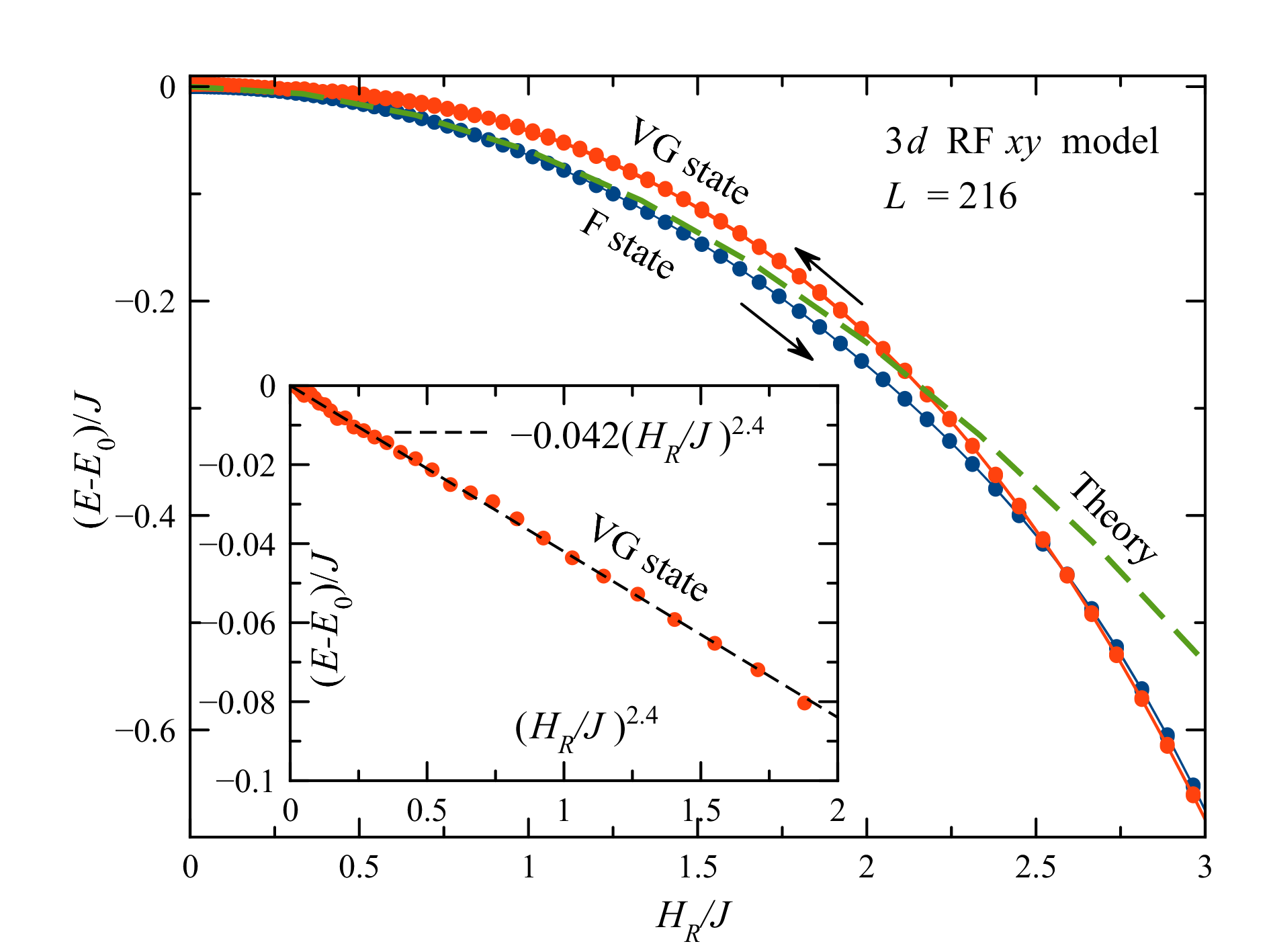}

\caption{Energy vs the random field strength $H_{R}$ in the ferromagnetic
and vortex-glass states. The dashed green line labelled ``theory''
is Eq. (\ref{total-SR}). Inset: Fitting the energy in the vortex-glass
state.}

\label{Fig-dE_vs_HR_L=00003D216_rc=00003D0_alpha=00003D0.05_pbc}
\end{figure}

\begin{figure}
\centering\includegraphics[width=8cm]{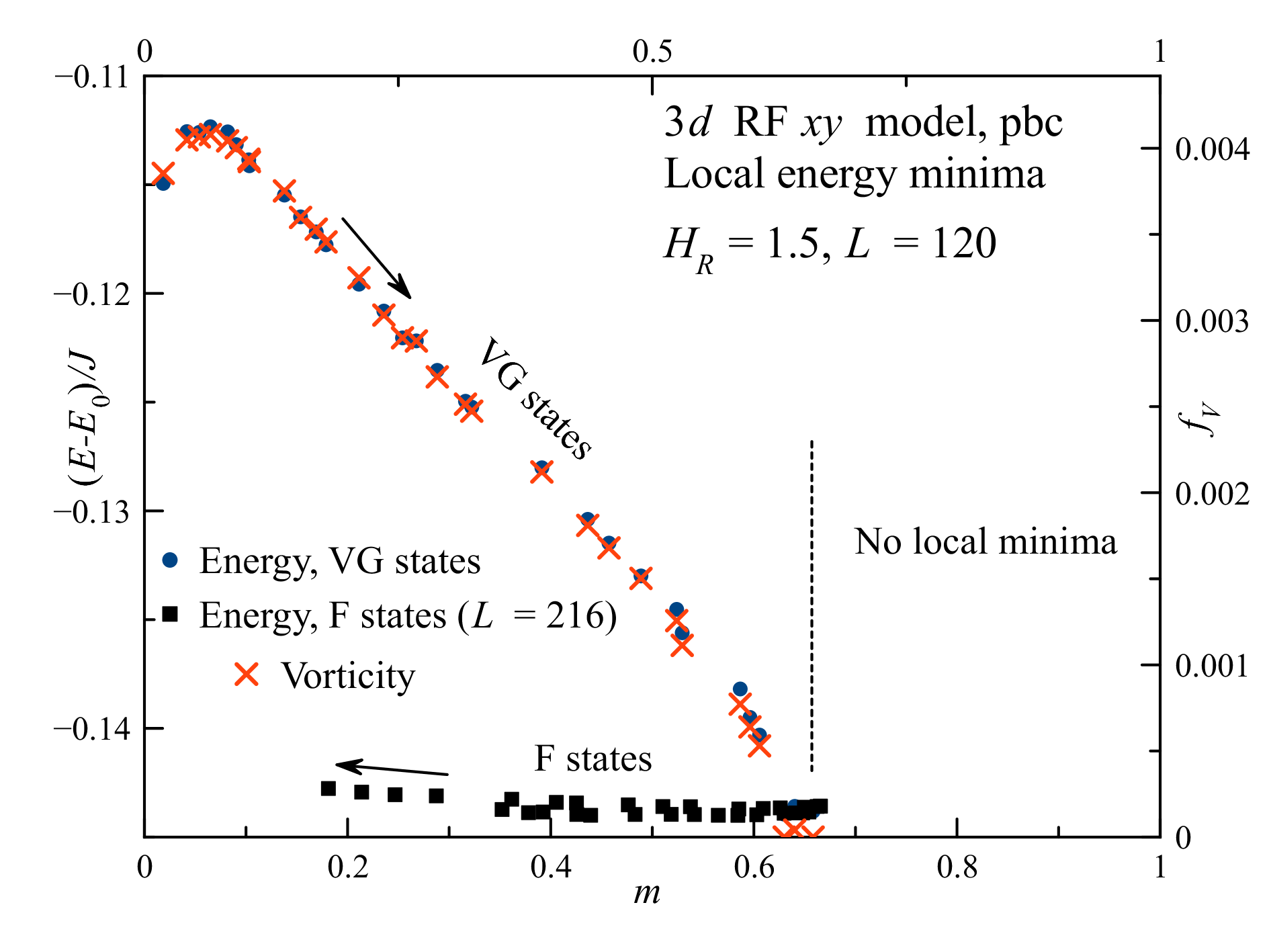}

\caption{Energies of metastable vortex-glass states sampled vs their magnetization.
Energy values in the vortex-glass state show a perfect correlation
with their vorticities. Energies of vortex-free ferromagnetic states
are comparable to those in Fig. \ref{Fig-dE_vs_m_F-states_L=00003D480_HR=00003D1.5}. }

\label{Fig-dE_vs_m_Nalp=00003D2_L=00003D120_HR=00003D1.5_rc=00003D0_pbc_alp=00003D0.1}
\end{figure}

Fig. \ref{Fig-dE_vs_HR_L=00003D216_rc=00003D0_alpha=00003D0.05_pbc}
obtained from the same computation as Fig. \ref{Fig-m_vs_HR_L=00003D216_rc=00003D0_alpha=00003D0.05_pbc}
shows that the energy of the vortex-glass state is higher than the
energy of the ferromagnetic state everywhere except for $H_{R}>2.5$
where creation of vortices becomes energetically favorable. However,
at these large values of $H_{R}$ the destruction of the ferromagnetic
state begins, see Fig. \ref{Fig-m_vs_HR_L=00003D216_rc=00003D0_alpha=00003D0.05_pbc}.
Thus the vortex-glass state is metastable in the most interesting
region of small $H_{R}$. The energy of the ferromagnetic state follows
Eq. (\ref{total-SR}) in the range $H_{R}\lesssim J$. The energy
per spin in the vortex-glass state can be fitted to
\begin{equation}
E-E_{0}\approx-0.042J(H_{R}/J)^{2.4}\approx-21f_{V}J,\label{eq:Energy_VG_via_HR}
\end{equation}
where Eq. (\ref{eq:fV_fitted}) was used. Note that by forming vortices
the system is lowering its energy with respect to the energy of the
collinear state. At the same time, creating vortices in the ferromagnetic
state costs energy.

We have studied the correlation between the energies of metastable
states and their magnetizations and vorticities. For this purpose,
for $H_{R}=1.5$ and $L=120$, we first allowed the system to relax
towards states with a preset value of $m_{z}$ by applying a self-adjusting
field $H$ as a Lagrange multiplier. Doing so, we moved from $m_{z}=0$
to $m_{z}=1$ starting from the random state at $m_{z}=0$ and using
the state with the preceding value of $m_{z}$ as the initial condition
for finding the state with the next value of $m_{z}$. In another
computation, we moved from $m_{z}=1$ to $m_{z}=0$ starting from
the collinear state at $m_{z}=1$. For each of these states with preset
$m_{z}$, we set $H$ to zero so that the system falls into the nearest
local energy minimum, using the larger-than-usual relaxation constant
$\alpha=0.1$. The energies and vorticities of the found metastable
states are plotted in Fig. \ref{Fig-dE_vs_m_Nalp=00003D2_L=00003D120_HR=00003D1.5_rc=00003D0_pbc_alp=00003D0.1}
vs $m$. While increasing preset $m_{z}$, we obtain VG states the
vorticity of which perfectly correlates with their energy. While decreasing
preset $m_{z}$, we obtain vortex-free ferromagnetic states with lower
energies. An interesting finding is that there are no local energy
minima for $m\gtrsim0.65$ in this plot, so that for all preset $m_{z}$
above this value the system typically slides into the deepest energy
minimum with $m\approx0.65$.

These results seem to be in contradiction with the theorem of Aisenman
and Wehr \cite{Aizenman-Wehr-PRL,Aizenman-Wehr-CMP} stating that
the system must have $m=0$ in the ground state. One possibility to
reconcile our findings with that theorem is this. Starting from a
vortex-free state, such as the states with $m\approx0.65$, one can
argue that there can be very rare configurations of the random field
that would energetically favor the formation of vortices. The vorticity
in these states is very small and locally they are very close to the
vortex-free states. However, even a very small but finite vorticity
could destroy spin correlations at large distances and render $m=0$.
Such states are not found if one starts with the collinear initial
condition because they require surmounting energy barriers. On the
contrary, starting from random initial conditions one ends up in the
states with a much larger vorticity and higher energy.

This argument is quite plausible in $2d$, where vortices are point
objects. However, it is less transparent in $3d$, where there are
vortex loops and vortex lines traversing the entire system. Configurations
of the random field that favor long vortex lines should be statistically
very rare and there must be many more short vortex loops. However,
the concentration of such vortex loops should be very small so that
they would not disturb the magnetic order at large distances as the
vortex lines do. Thus it is not clear whether a very diluted gas of
vortex loops in an infinite sample destroys the long-range order.
If it does it would be more along the lines of the Bragg glass theory.

\subsection{Approach to saturation, hysteresis and memory}

\label{sub:saturation}

Fig. \ref{Fig-mz_vs_H_L=00003D112_HR=00003D1.5_pbc-large_H}, which
shows approach to saturation for large $H$, is in accord with Eq.
(\ref{sqrt}).

\begin{figure}
\centering\includegraphics[width=8cm]{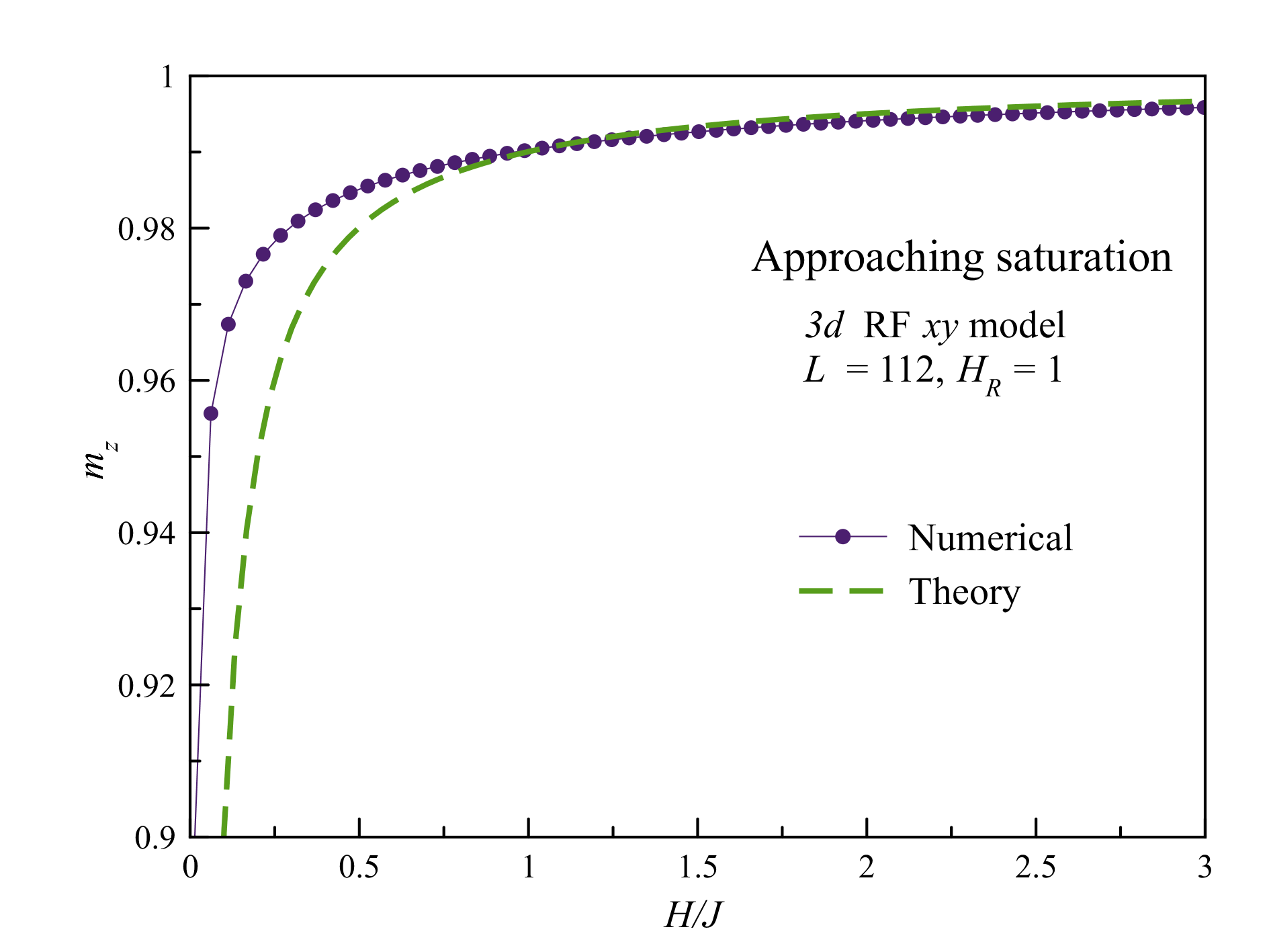}\caption{Approaching saturation in the $3d$ RF $xy$ model. Dashed line is
Eq. (\ref{sqrt}).}

\label{Fig-mz_vs_H_L=00003D112_HR=00003D1.5_pbc-large_H}
\end{figure}

\begin{figure}
\centering\includegraphics[width=8cm]{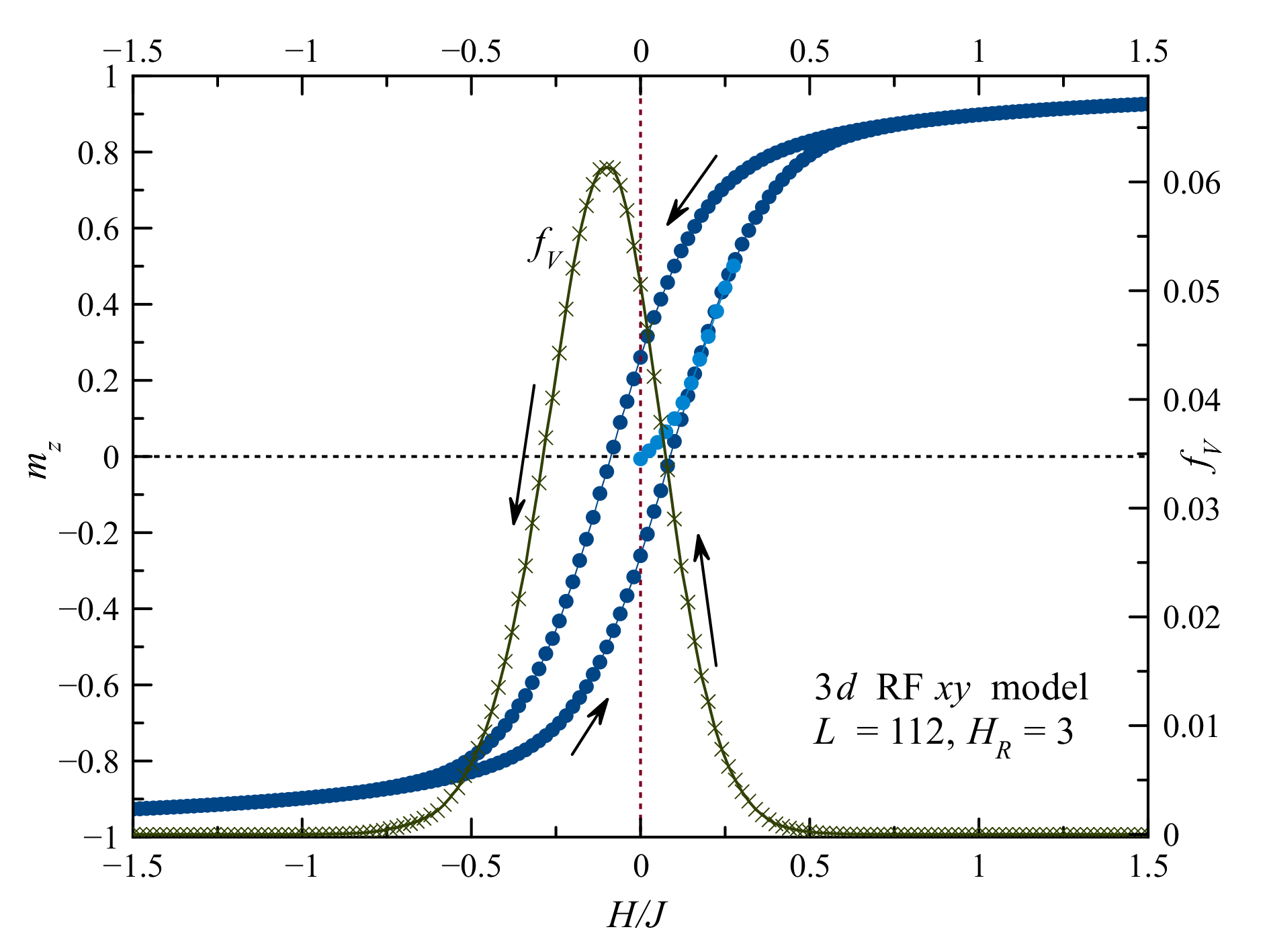}\caption{Hysteresis curves for $3d$ RF $xy$ model for $H_{R}=3$. Irreversibility
is clearly related to vorticity.}

\label{Fig-mz_vs_H_Nalp=00003D2_Nx=00003DNy=00003DNz=00003D112_HR=00003D3._rc=00003D0_pbc_alp=00003D0.03}
\end{figure}

\begin{figure}
\centering\includegraphics[width=8cm]{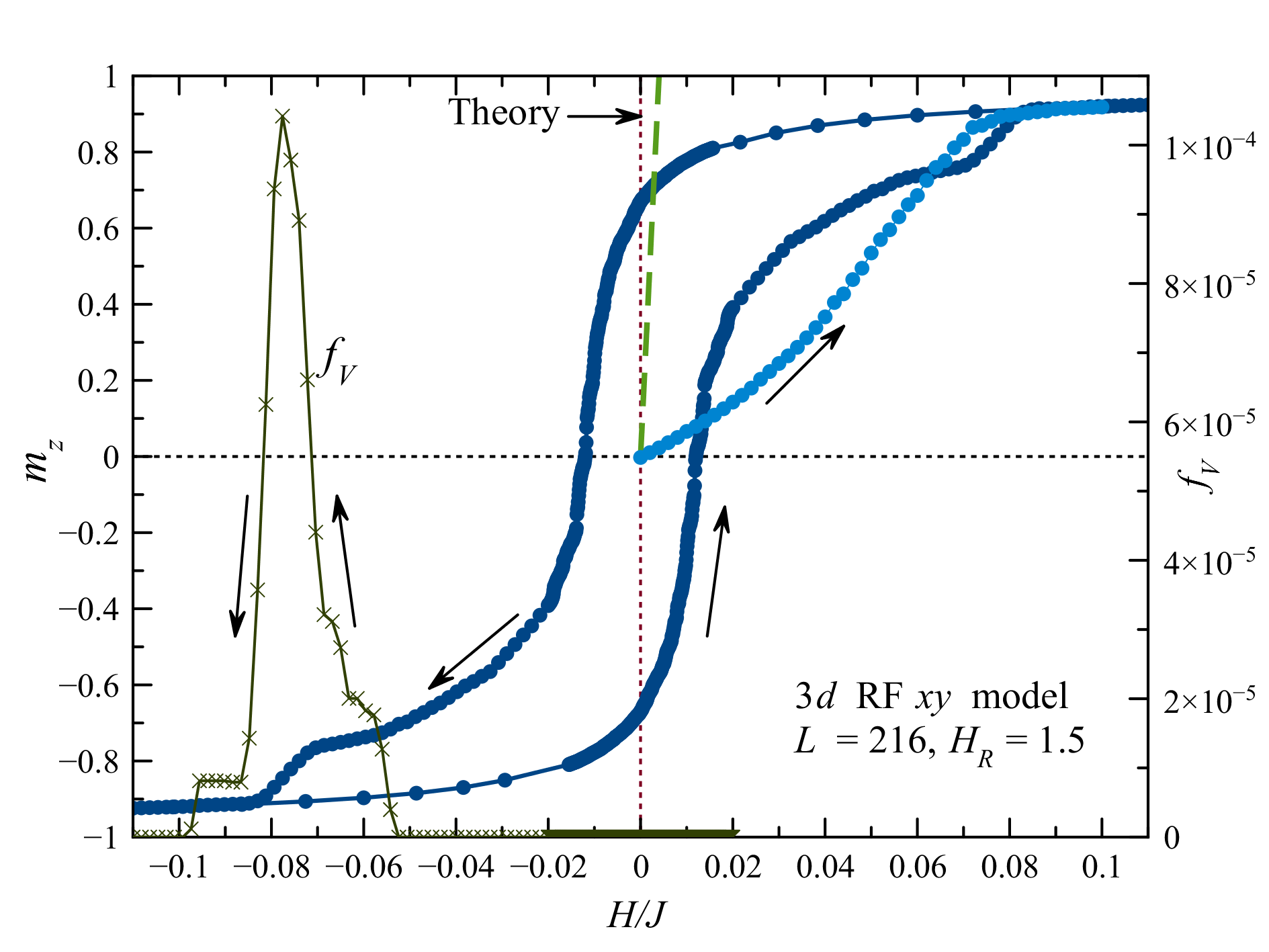}
\caption{Hysteresis curves for $3d$ RF $xy$ model for $H_{R}=1.5$. The straight
dashed line labeled ``Theory'' is based on Eq. (\ref{chi}).
Dense and rarified points are results for different realizations of the random field.
They overlap because of a sufficient self-averaging in the system.
}

\label{Fig-mz_vs_H_Nalp=00003D2_L=00003D216_HR=00003D1.5._rc=00003D0_pbc_alp=00003D0.03}
\end{figure}

\begin{figure}
\begin{centering}
\centering\includegraphics[width=8cm]{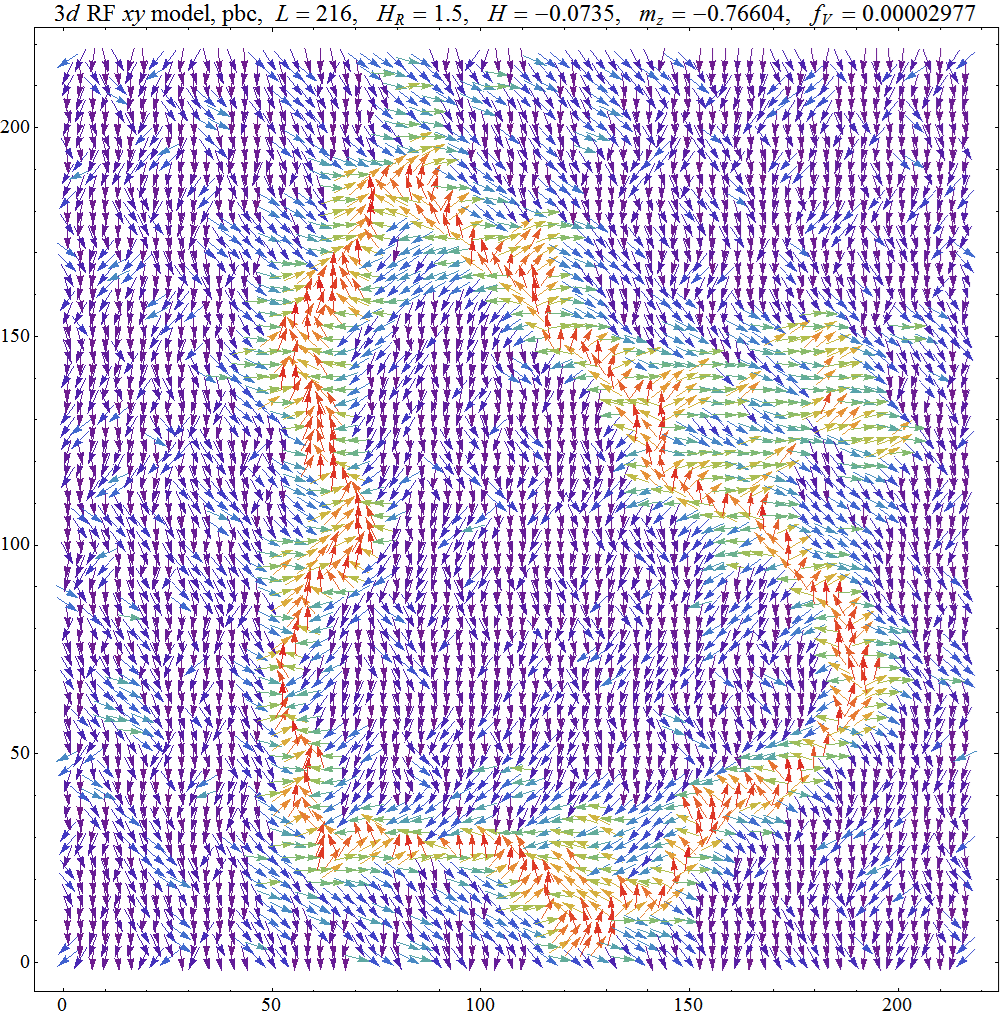}
\par\end{centering}

\caption{Walls of spins opposite to the field, pinned by the random field.
\label{fig:walls}}
\end{figure}

\begin{figure}
\centering\includegraphics[width=8cm]{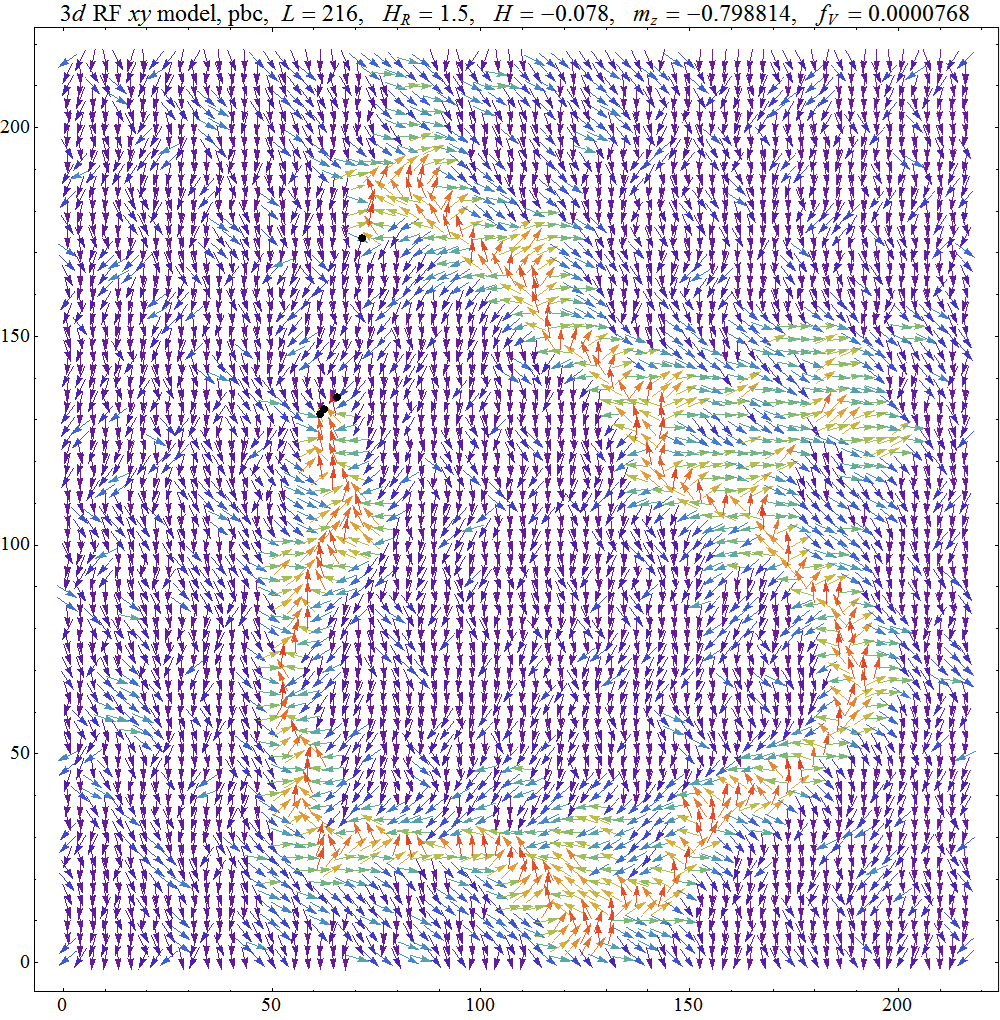}\caption{Walls of spins raptured by vortices (black points)}

\label{fig:walls-raptured}
\end{figure}

For a strong random field, such as $H_{R}=3$ in Fig. \ref{Fig-mz_vs_H_Nalp=00003D2_Nx=00003DNy=00003DNz=00003D112_HR=00003D3._rc=00003D0_pbc_alp=00003D0.03},
hysteresis curves have a standard form. The irreversibility is related
to the energy barriers at the atomic scale that changes the systems'
vorticity. The relation between vorticity and hysteresis is clearly
seen in the Fig. \ref{Fig-mz_vs_H_Nalp=00003D2_Nx=00003DNy=00003DNz=00003D112_HR=00003D3._rc=00003D0_pbc_alp=00003D0.03}.
In the course of the reversal the magnetization $m$ decreases down
to zero and then grows in magnitude again (not shown).

Fig. \ref{Fig-mz_vs_H_Nalp=00003D2_L=00003D216_HR=00003D1.5._rc=00003D0_pbc_alp=00003D0.03}
shows hysteresis curves of a random-field $xy$ magnet for $H_{R}=1.5$
and $L=216$. The initial magnetization curve that begins with $m_{z}=0$
at $H=0$ has a rather small slope, in a striking disagreement with
large zero-field susceptibility that follows from the Green-function
method, Eq.\ (\ref{chi}). This high rigidity of the vortex-glass
state is due to the pinning of vortices that Imry-Ma scenario does
not account for.

There is a large $m$ at $H=0$ along the $H$-down branch in Fig.
\ref{Fig-mz_vs_H_Nalp=00003D2_L=00003D216_HR=00003D1.5._rc=00003D0_pbc_alp=00003D0.03}
that does not depend strongly on the system size, which is in accord
with Fig. \ref{Fig-m^2_vs_size_HR=00003D1_and_1.5}. While the dependence
of $m_{z}$ on $H$ along the hysteresis curve is rather steep at
small fields, it is nearly smooth and has only small Barkhausen jumps
(not seen in the figure), with the slope in the ball park of that
given by Eq.\ (\ref{chi}). The magnetization of the sample does
not rotate as a whole from positive to negative values of $m_{z}$.
Instead, on average, the deviations of spins to the right and to the
left from the positive $z$ direction in different regions of space
increase smoothly as $H$ grows negative. In the process of spin reversal
the regions with right and left spin deviations occupy rather large
volumes separated by transient regions where spins are still directed
in the positive $z$ direction. Such transient regions form walls
of topological origin, see the cross-section of the sample in Fig.
(\ref{fig:walls}). They are pinned by the random field.

As the magnetization reversal proceeds along the hysteresis curve,
the walls rapture, with the raptured area bounded by the vortex loop,
as shown in Fig. \ref{fig:walls-raptured}. The loops then grow and
eat the walls away, completing the reversal. This happens at $H=-H_{V}\approx0.075$
in Fig. \ref{Fig-mz_vs_H_Nalp=00003D2_L=00003D216_HR=00003D1.5._rc=00003D0_pbc_alp=00003D0.03},
where $m_{z}$ has a shoulder and vorticity has a peak. Such a behavior
is typical for the $xy$ random magnet of size large compared to the
ferromagnetic correlation length. Systems of smaller sizes typically
switch their magnetization via rotation as a whole that leads to a
jump from positive to negative values at a coercive field. This behavior
is similar to that of a single-domain magnetic particle.

For $H>-H_{V}$, the upper hysteresis branch is quasi-reversible:
Removing the field leads to partial restoration a large magnetization
of the ferromagnetic state in $H=0$, which can interpreted as a memory
effect. The simulated relaxation curves are shown in Fig. \ref{Fig-mz_vs_MCS_recovery_L=00003D216_HR=00003D1.5_alp=00003D0.02}.
The recovery happens because the ferromagnetic state with spin walls
exhibits elasticity. As the field is reversed, it stores energy and
tends to return to the initial state when the stress due to the opposite
field is removed. This behavior is a good evidence of the stability
of the ferromagnetic state. The incomplete restoration of the magnetization
in this experiment should be due to energy barriers not related to
vortices.

\begin{figure}
\centering\includegraphics[width=8cm]{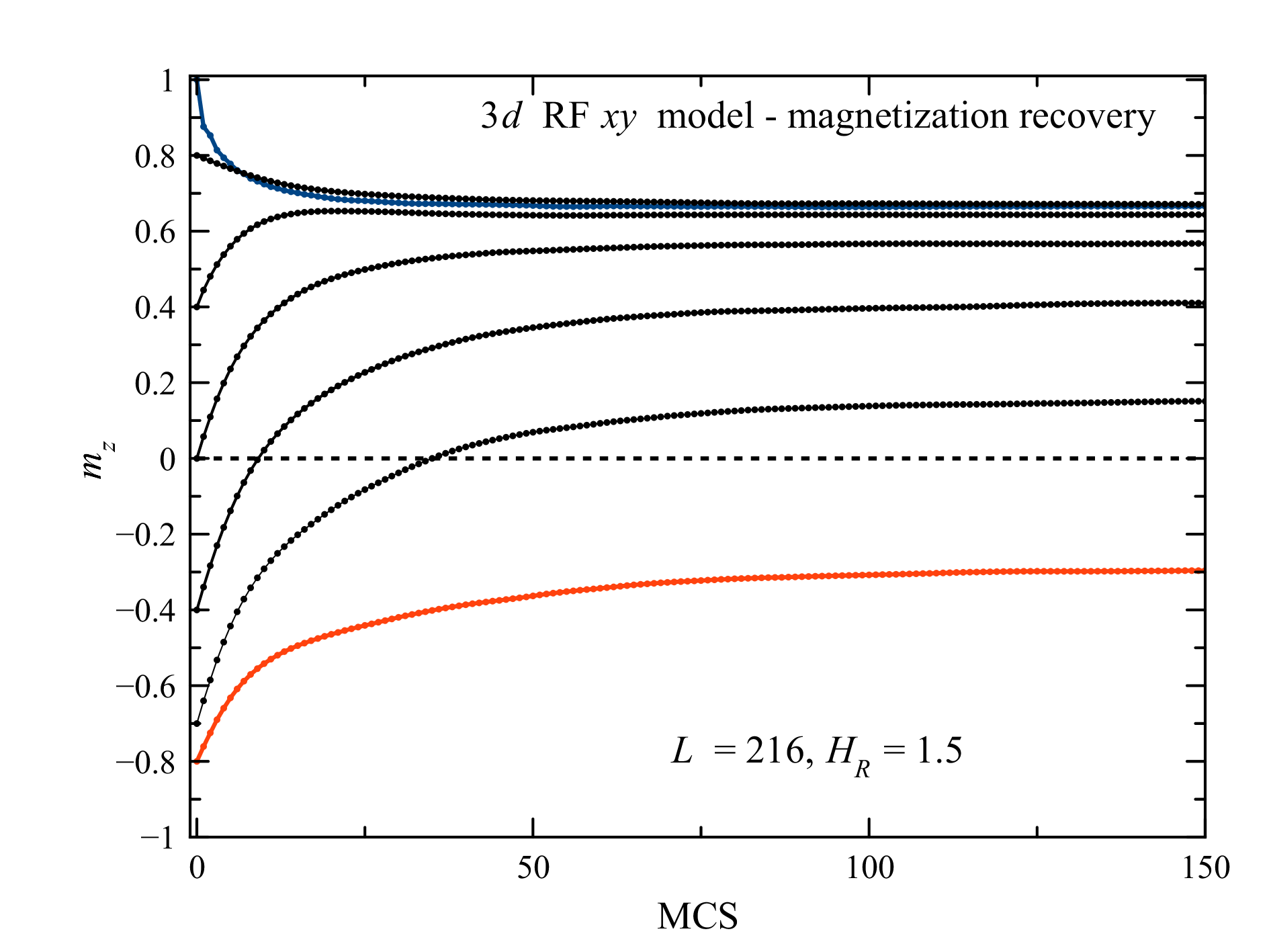}

\caption{Magnetization recovery from the quasi-reversible branch of the hysteresis
curve ($H>-H_{V}$ in Fig. \ref{Fig-mz_vs_H_Nalp=00003D2_L=00003D216_HR=00003D1.5._rc=00003D0_pbc_alp=00003D0.03}),
computed after setting $H=0$. The red curve corresponding to the
initial value $m_{z}=-0.8$ does not go into the positive region because
this initial state is beyond the quasi-reversible branch and has a
large vorticity.}

\label{Fig-mz_vs_MCS_recovery_L=00003D216_HR=00003D1.5_alp=00003D0.02}
\end{figure}

\begin{figure}
\centering\includegraphics[width=8cm]{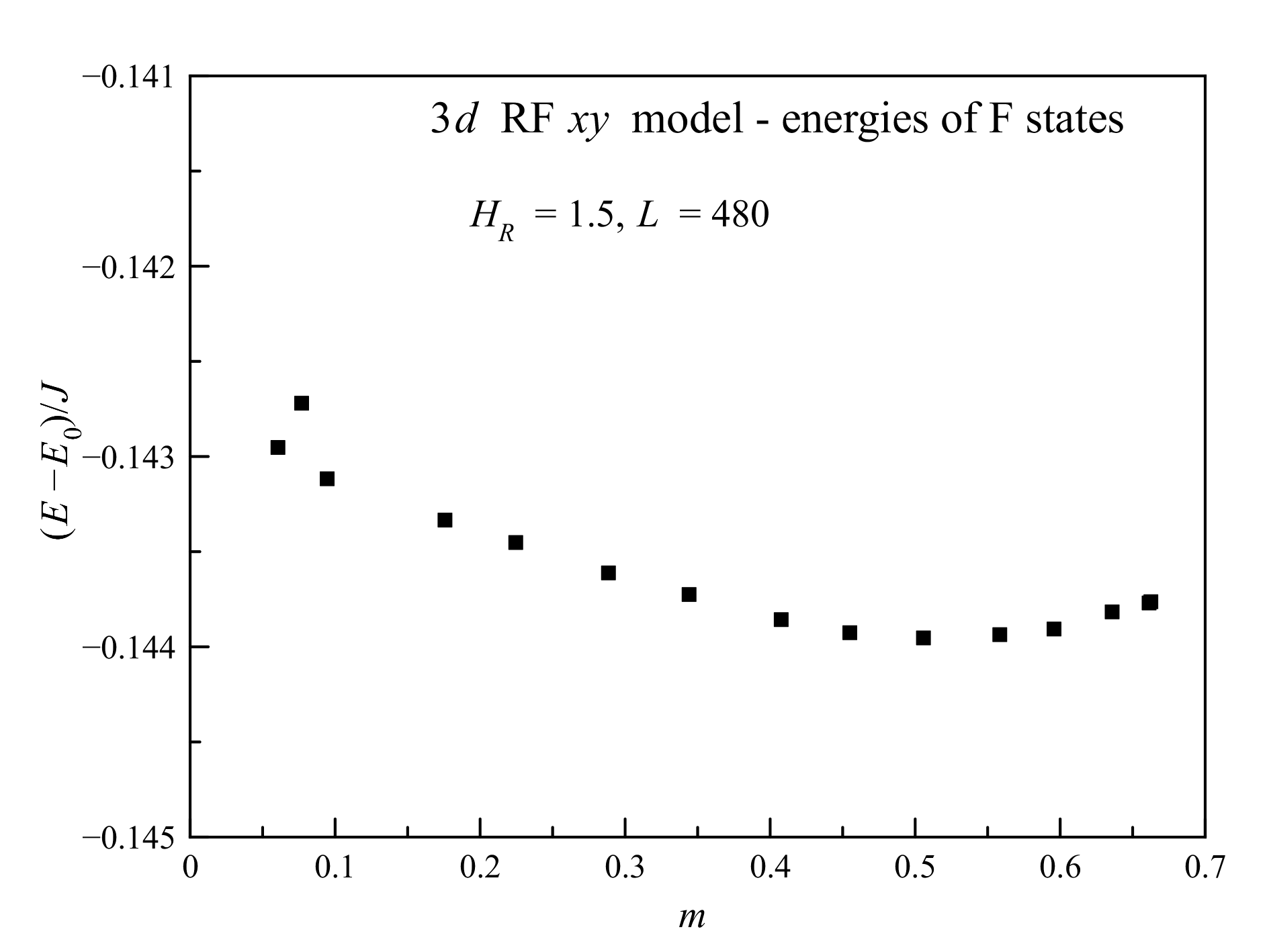}

\caption{Energies of vortex-free ferromagnetic states (local energy minima)
obtained by magnetization recovery of the type shown in Fig. \ref{Fig-mz_vs_MCS_recovery_L=00003D216_HR=00003D1.5_alp=00003D0.02}.
The rightmost state is obtained by relaxation from any state with
$m\gtrsim0.7$.}

\label{Fig-dE_vs_m_F-states_L=00003D480_HR=00003D1.5}
\end{figure}

\begin{figure}
\centering\includegraphics[width=8cm]{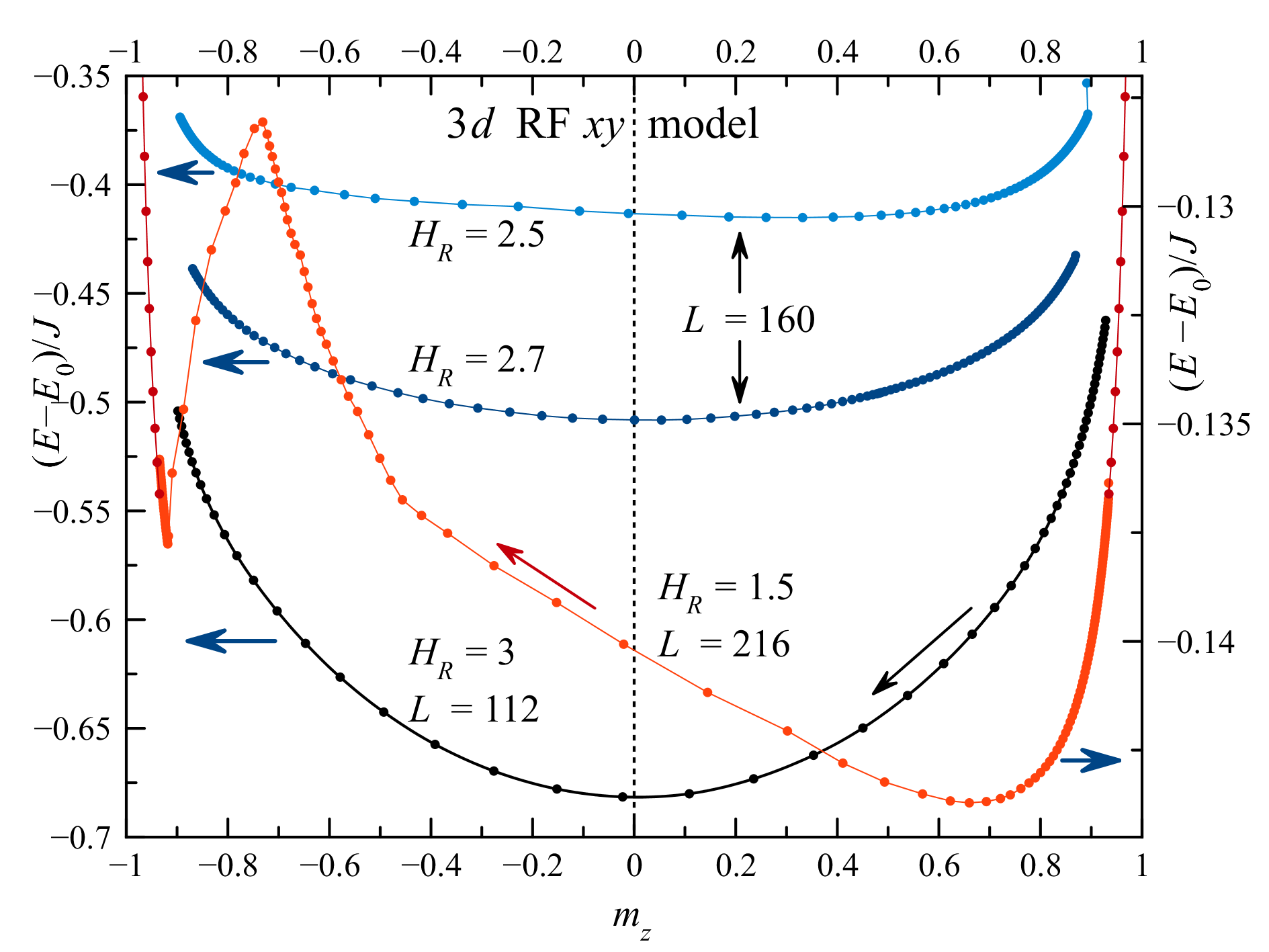}

\caption{Energies of the states created by the external field $H$ vs $m_{z}$
(with the energy due to $H$ subtracted). The lowest-energy state
corresponds to $m_{z}=0$ for $H_{R}\gtrsim2.6$ and to $m_{z}>0$
for $H_{R}\lesssim2.6$. }

\label{Fig-dEnoH_vs_mz_L=00003D112_216_HR=00003D3_1.5_pbc_alp=00003D0.03_eta=00003D0.02}
\end{figure}

The magnetization-recovery experiment provides an access to more ferromagnetic
states than just relaxation from a collinear state. Because of small
barriers there is a big number of metastable ferromagnetic states
that differ by energy and magnetization $m$, shown in Fig. \ref{Fig-dE_vs_m_F-states_L=00003D480_HR=00003D1.5}.
States with smaller $m$ occur due to relaxation from states with
smaller $m_{z}$ in the upper hysteresis branch in Fig. \ref{Fig-mz_vs_H_Nalp=00003D2_L=00003D216_HR=00003D1.5._rc=00003D0_pbc_alp=00003D0.03}.
This is also seen in Fig. \ref{Fig-mz_vs_MCS_recovery_L=00003D216_HR=00003D1.5_alp=00003D0.02}.
The rightmost state in Fig. \ref{Fig-dE_vs_m_F-states_L=00003D480_HR=00003D1.5}
is obtained by relaxation from any state with $m\gtrsim0.7$ because
there are no local energy minima in this range. There is a significant
interval of $m$ values in the ferromagnetic states in Fig. \ref{Fig-dE_vs_m_F-states_L=00003D480_HR=00003D1.5},
all having very close energies, in contrast with much larger energy
differences between vortex-glass states in Fig. \ref{Fig-dE_vs_m_Nalp=00003D2_L=00003D120_HR=00003D1.5_rc=00003D0_pbc_alp=00003D0.1}.
One can clearly see that the lowest-energy state is at $m\approx0.5$,
a value that varies a bit depending on the random field realization.
The energy values in Fig. \ref{Fig-dE_vs_m_F-states_L=00003D480_HR=00003D1.5}
are comparable to those of the ferromagnetic states for $L=216$ in
Fig. \ref{Fig-dE_vs_m_Nalp=00003D2_L=00003D120_HR=00003D1.5_rc=00003D0_pbc_alp=00003D0.1}
and the states in Fig. \ref{Fig-dE_vs_m_wxyz_L=00003D216_HR=00003D1.5_pbc_alp=00003D0.03_eta=00003D0.02}.

Another method of accessing the energies of the states vs their magnetization
is to plot the energy obtained in the computation of the hysteresis
(with the energy due to the external magnetic field $H$ subtracted)
vs $m_{z}$. In this way one can access not only local energy minima,
as in Fig. \ref{Fig-dE_vs_m_F-states_L=00003D480_HR=00003D1.5} but
also the energies of all unstable states supported by the external
field. Fig. \ref{Fig-dEnoH_vs_mz_L=00003D112_216_HR=00003D3_1.5_pbc_alp=00003D0.03_eta=00003D0.02}
shows the computed energies for different values of $H_{R}$. A striking
feature is the transition between the energy minimum at $m=0$ to
an energy munimum at $m>0$ on $H_{R}$ that occurs at $H_{R}\approx2.6$.
One can see that for $H_{R}=1.5$ the results are very close to those
for the local energy minima in Fig. \ref{Fig-dE_vs_m_F-states_L=00003D480_HR=00003D1.5}
but also contain unstable states with $m\gtrsim0.7$. Suppression
of ferromagnetic states at large $H_{R}$ was already seen in Fig.
\ref{Fig-m_vs_HR_L=00003D216_rc=00003D0_alpha=00003D0.05_pbc}. The
energy maximum at $m_{z}\approx-0.8$ corresponds to the shoulder
at this $m_{z}$ in Fig. \ref{Fig-mz_vs_H_Nalp=00003D2_L=00003D216_HR=00003D1.5._rc=00003D0_pbc_alp=00003D0.03}.
On decreasing $H_{R}$, its increasing part is due to the energy input
into compressed spin walls while its decreasing part and it is due
to rapture of spin walls by vortex loops.

\subsection{Ordering by decreasing rotating field}

\label{sub:decreasing_rotating_field}

\begin{figure}
\centering\includegraphics[width=8cm]{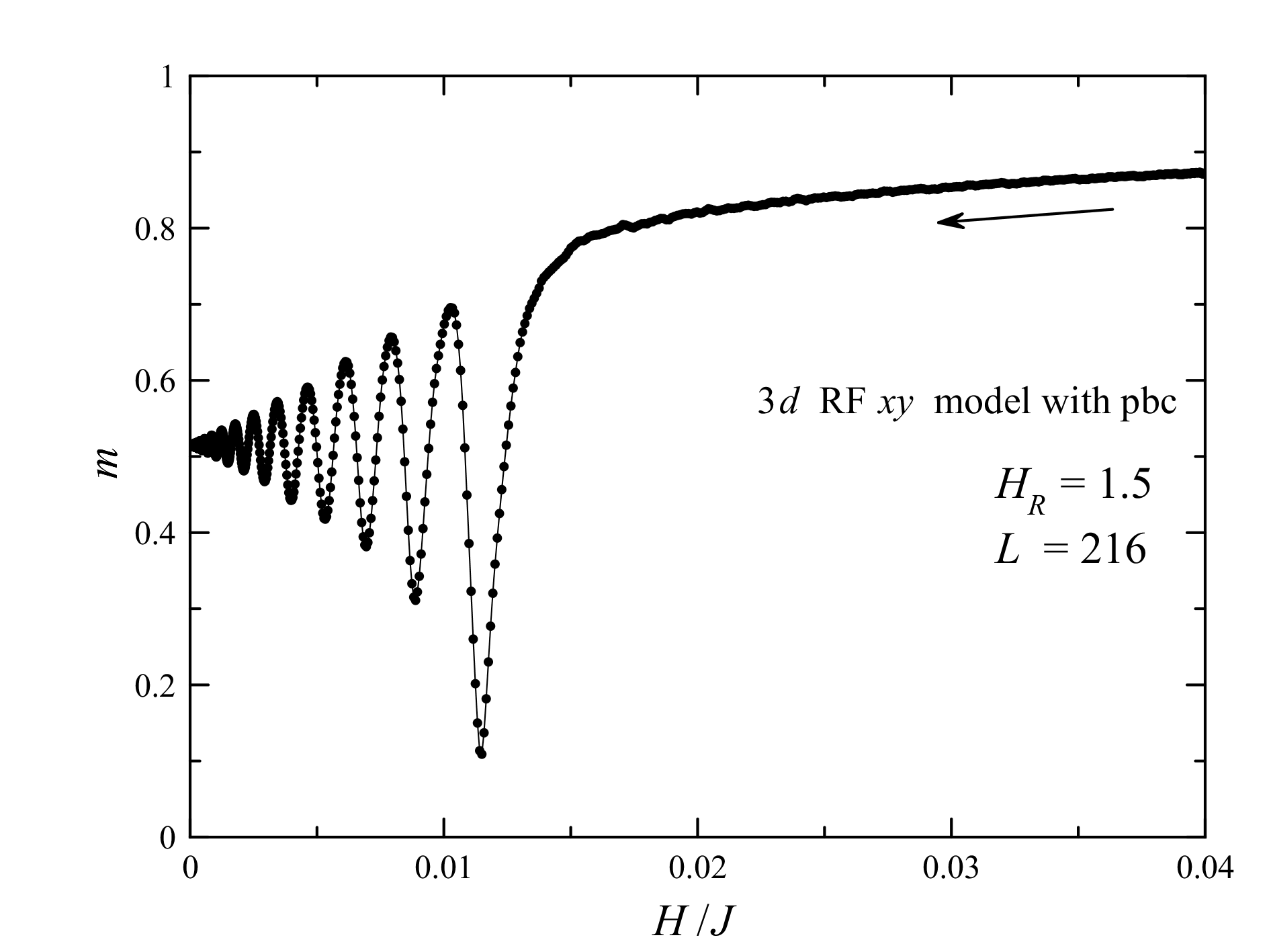}

\caption{Magnetization vs decreasing magnitude of a rotating field $\mathbf{H}$}

\label{Fig-m_vs_rotating_H_Nalp=00003D2_Nx=00003DNy=00003DNz=00003D216_HR=00003D1.5_pbc_alp=00003D0.03-subtr}
\end{figure}

\begin{figure}
\centering\includegraphics[width=8cm]{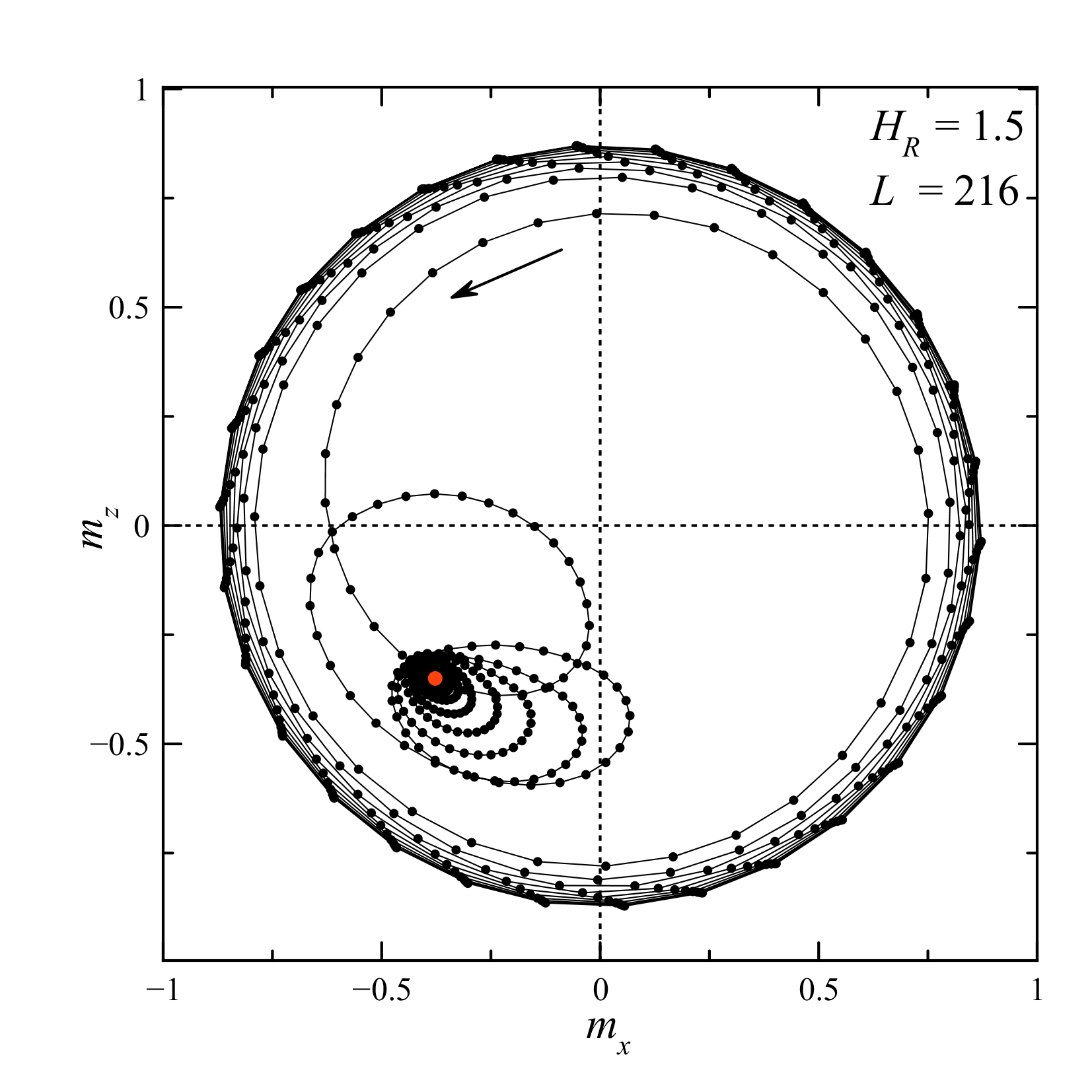}\caption{Components of the magnetization vector $\mathbf{m}$ in the rotating-field
experiment}

\label{Fig-mx_my_Nalp=00003D2_Nx=00003DNy=00003DNz=00003D216_HR=00003D1.5_pbc_alp=00003D0.03_eta=00003D0.02_1_StAn-subtr}
\end{figure}

Another type of numeric experiment showing ferromagnetic ordering
is relaxation in the presence of a rotating external field $\mathbf{H}$
with the magnitude slowly decreasing to zero. This is a version of
the method of \textit{stimulated annealing} that helps the system
to overcome barriers preventing it from relaxing to states with a
lower energy. If there were states with a small or zero magnetization
having a lower energy that in our other numerical experiments, these
states were likely to be reached by this method.

Numerical results shown in Figs. \ref{Fig-m_vs_rotating_H_Nalp=00003D2_Nx=00003DNy=00003DNz=00003D216_HR=00003D1.5_pbc_alp=00003D0.03-subtr}
and \ref{Fig-mx_my_Nalp=00003D2_Nx=00003DNy=00003DNz=00003D216_HR=00003D1.5_pbc_alp=00003D0.03_eta=00003D0.02_1_StAn-subtr}
show that also in the decreasing rotating field experiment ferromagnetically
ordered states are reached. For the field magnitude $H$ large enough,
the direction of $\mathbf{m}$ follows that of $\mathbf{H}$, while
both $H$ and $m$ are decreasing. As $H$ goes below 0.015 (see Fig.
\ref{Fig-m_vs_rotating_H_Nalp=00003D2_Nx=00003DNy=00003DNz=00003D216_HR=00003D1.5_pbc_alp=00003D0.03-subtr}),
direction of $\mathbf{m}$ decouples from that of $\mathbf{H}$ and,
after oscillations around an energy minimum corresponding to a significant
value of $m$, the system reaches this energy minimum. In the above
numerical experiment, the final magnetization value is $m=0.5148$.
It turns out that our method leads to the energy values very close
to those of Fig. \ref{Fig-dE_vs_m_F-states_L=00003D480_HR=00003D1.5}.
Thus, no states with a smaller $m$ and lower energy have been found,
that again proves robustness of the ferromagnetically ordered state.

In another type of numerical experiment, a field slowly oscillating
parallel to a fixed direction with the amplitude slowly decreasing
to zero had been applied. Here one could obtain states with a small
magnetization. However, the energy of such states was higher than
the energy of the F state because of the vorticity generated by rapturing
spin walls, see Sec. \ref{sub:saturation}.

\subsection{Correlation functions}

\label{sub:corr-functions}

\begin{figure}
\centering\includegraphics[width=8cm]{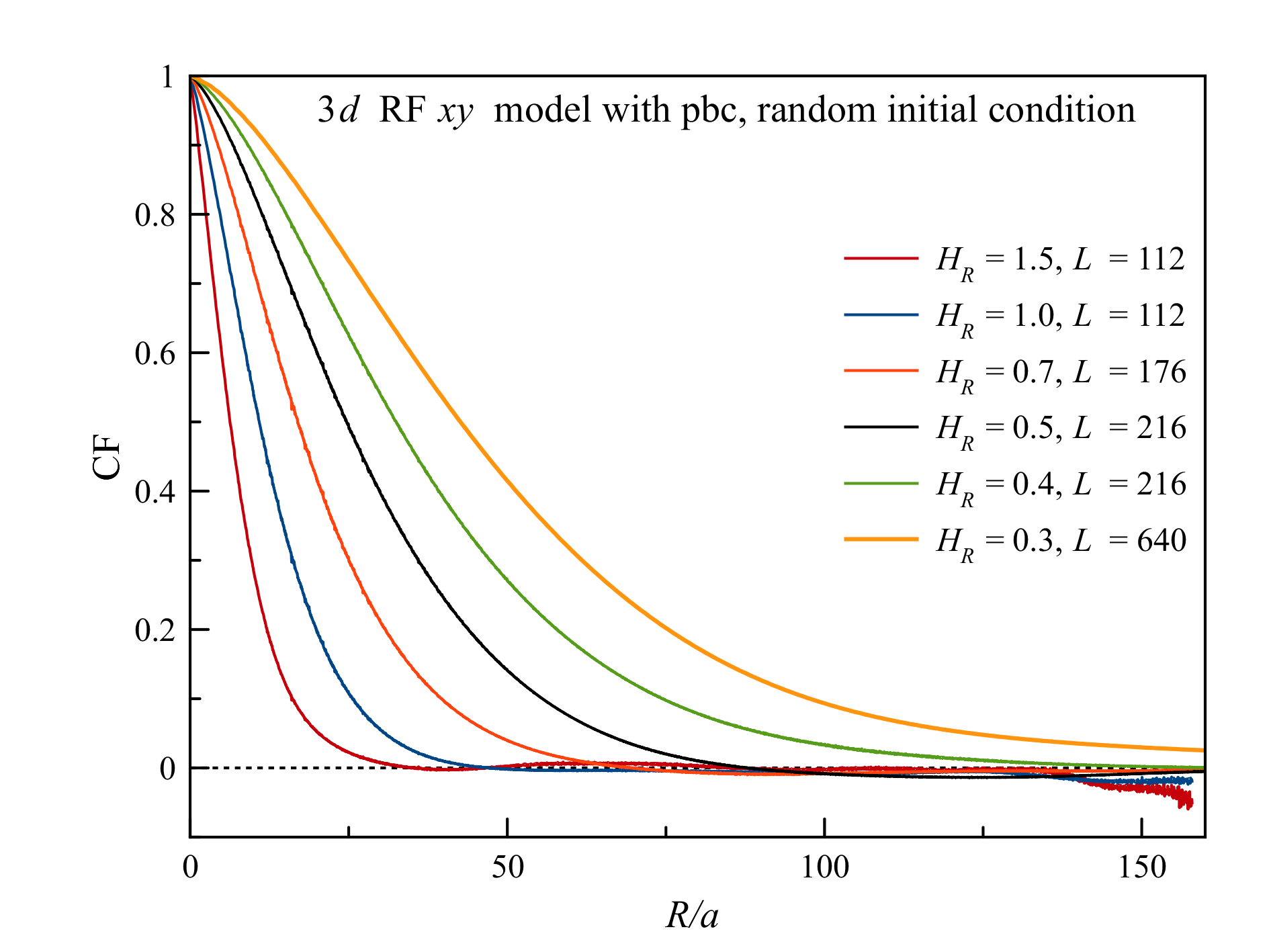}

\centering\includegraphics[width=8cm]{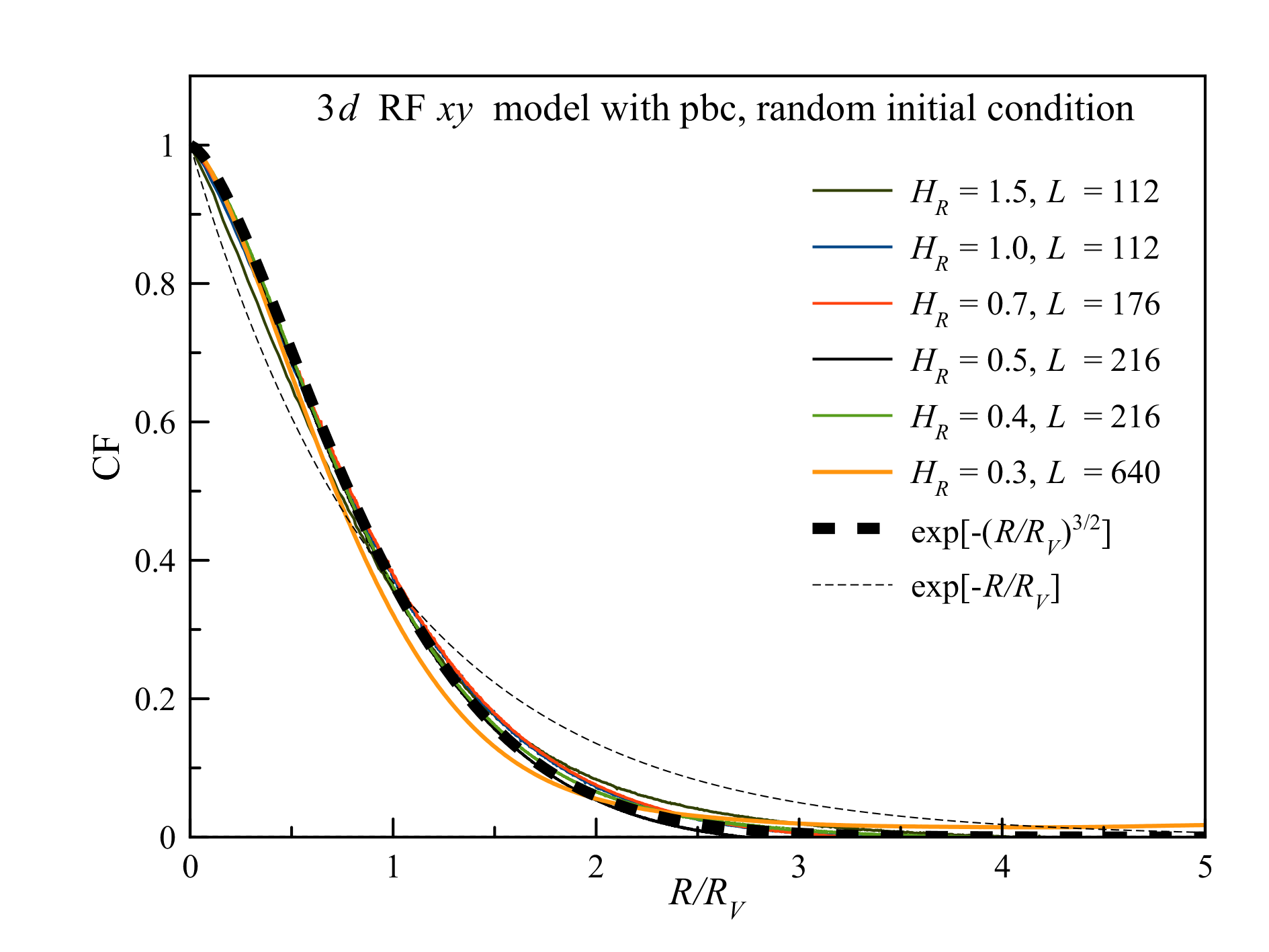}

\caption{Correlation functions of the $3d$ RF $xy$ model in the vortex glass
state obtained starting from random initial conditions. Natural (top)
and scaled (bottom) presentations. $R_{V}$ is given by Eq. (\ref{eq:RV_fitted}).
\label{fig:CF-VG}}
\end{figure}

\begin{figure}
\centering\includegraphics[width=8cm]{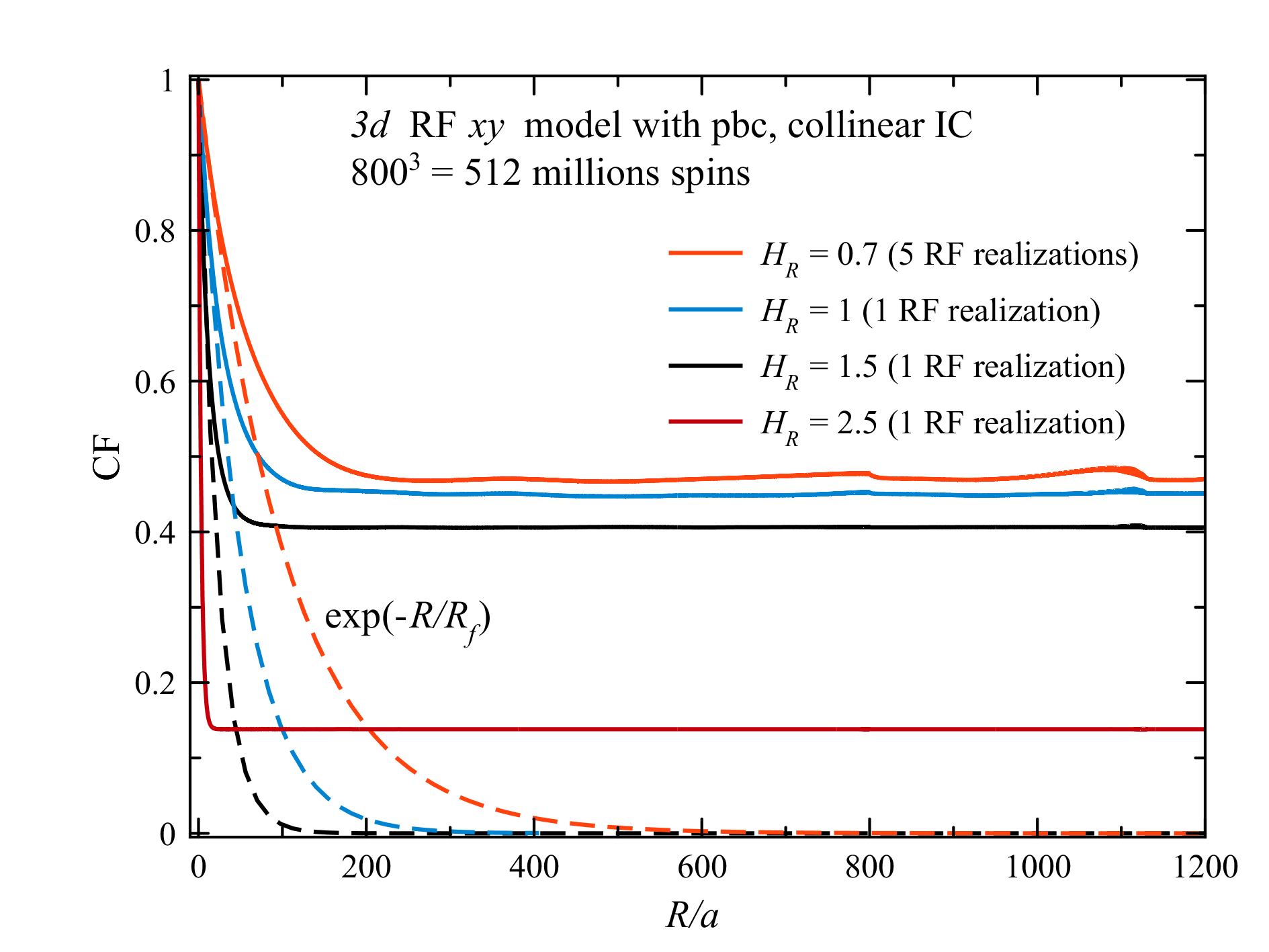}

\caption{Correlation functions of the $3d$ RF $xy$ model in the ferromagnetic
state obtained starting from collinear initial conditions. }

\label{Fig-CF_vs_n_Nx=00003DNy=00003DNz=00003D800_HR=00003D0.7,_1,_1.5_rc=00003D0_pbc_coll_IC}
\end{figure}

We have computed correlation functions in the energy minima of our
system that we have found by our relaxation algorithm. After computing
CFs we averaged them over directions of $\mathbf{R}\equiv{\bf r}_{1}-{\bf r}_{2}$.

In the vortex-glass state obtained from random initial conditions,
correlation functions shown in Fig. \ref{fig:CF-VG} decay to zero
but their form and correlation radius is different from Eqs. (\ref{Exp-CF})
and (\ref{eq:Rf-Def}). The results can be fitted by the stretched
exponential

\begin{equation}
\langle{\bf s}({\bf r}_{1})\cdot{\bf s}({\bf r}_{2})\rangle=s^{2}e^{-(|{\bf r}_{1}-{\bf r}_{2}|/R_{V})^{3/2}},\quad R_{V}\simeq14\left(Js/H_{R}\right)^{1.2}.\label{eq:RV_fitted}
\end{equation}
Note that the dependence of $R_{V}$ on $H_{R}$ is much weaker than
$R_{f}\propto1/H_{R}^{2}$ of Eq. (\ref{eq:Rf-Def}) and thus $R_{V}\ll R_{f}$
at small $H_{R}$. Using the vorticity dependence of Eq. (\ref{eq:fV_fitted}),
one can express the VG correlation length $R_{V}$ as
\begin{equation}
R_{V}\simeq0.6/f_{V}^{1/2},\label{eq:RV_via_fV}
\end{equation}
c.f. Eq. (\ref{eq:RV_via_fV-propto}). This dependence is in agreement
with the $2d$ nature of vortices discussed below Eq. (\ref{eq:RV_via_fV-propto}).

If the initial state is collinear and $H_{R}$ is not too large, the
correlation functions have plateaus at large distances. At $R\lesssim R_{f}$
they exactly follow Eq. (\ref{Exp-CF}). The results for our largest
size $L=800$ are presented in Fig. \ref{Fig-CF_vs_n_Nx=00003DNy=00003DNz=00003D800_HR=00003D0.7,_1,_1.5_rc=00003D0_pbc_coll_IC}.
For $H_{R}=1$ and 1.5 there is enough self-averaging and we show
correlation functions obtained for only one random-field realization.
They have well-defined plateaus with small fluctuations. For $H_{R}=0.7$,
correlation functions obtained with one random-field realization are
too bumpy and averaging over realizations is needed. The bumps at
$R=800$ and $\sqrt{2}\times800$ are artifacts of periodic boundary
conditions. The length of the plateaus show that the large magnetization
in the ferromagnetic state is not a fluctuational magnetization.

The perfect plateau for $H_{R}=2.5$ shows that the appreciable vorticity
$f_{V}=0.01766$ in this state does not yet disrupt ferromagnetic
order at long distances. This should be the consequence of vortices
forming small closed loops such as in Fig. \ref{Fig-vortex_loops}(top).
Meanwhile, one can expect that even a small concentration of vortex
lines that go through the whole sample, as is the case in the vortex-glass
state, see Fig. \ref{Fig-vortex_loops}(bottom), will destroy the
long-range order.

\section{The Imry-Ma argument and vortices}

\label{sec:IM-vortices}

Surprising robustness of the ferromagnetic state found in our different
calculation requires an explanation. According to the Imry-Ma scenario,
starting from collinear state, spins would relax towards directions
of the random field averaged over correlated regions of linear size
$R_{f}$, so that the magnetization would go to zero if $R_{f}$ is
small compared to the size of the system. In our computations we indeed
observe a fast initial disordering but then the magnetization stops
to decrease at an appreciable value, Figs. \ref{Fig-m_vs_MCS_L=00003D216_HR=00003D1.5_pbc_coll_IC_alp=00003D1_0.03}
and \ref{Fig-m_vs_MCS_L=00003D800_1000_HR=00003D1.5_0.5_pbc_coll_IC_alp=00003D0_0.03}.
What could be the factor that prevents it from relaxing to zero?

\begin{figure}
\centering\includegraphics[width=8cm]{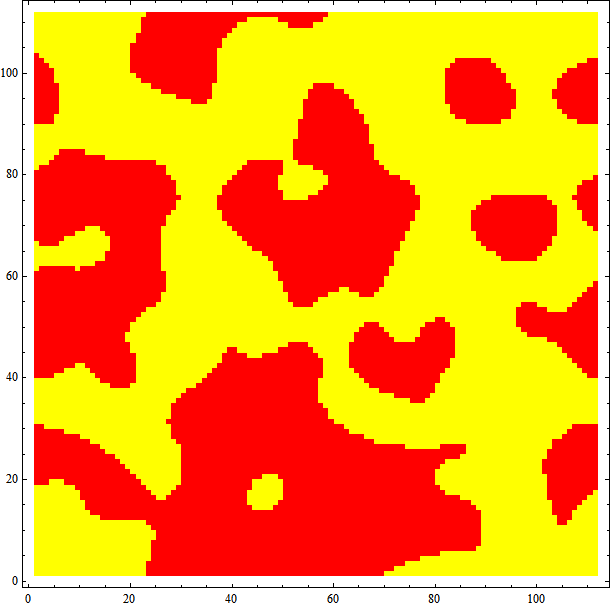}

\centering\includegraphics[width=8cm]{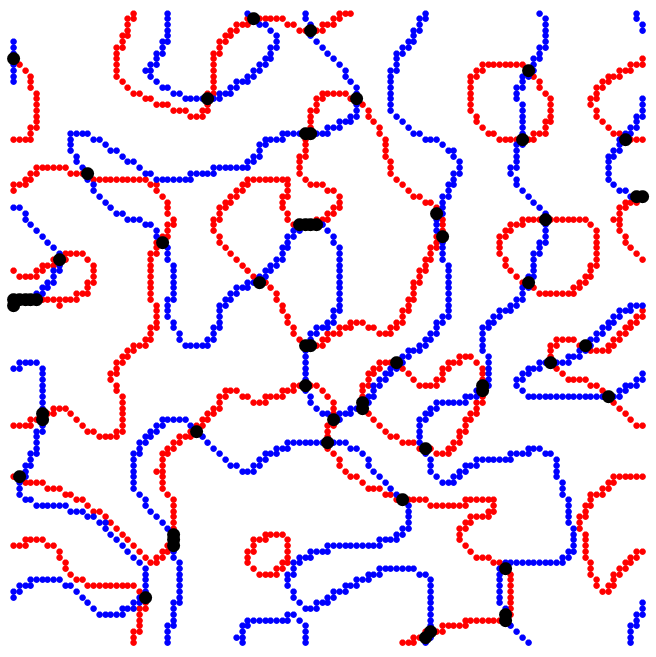}

\caption{(Top) Domains of positive and negative $h_{x}(\mathbf{r})$ in a $xy$
plane. (Bottom) Singularities at the crossings of domain boundaries
for $h_{x}(\mathbf{r})$ and $h_{y}(\mathbf{r})$ in a $xy$ plane.}

\label{Fig-IM-domains}
\end{figure}

\begin{figure}
\centering\includegraphics[width=8cm]{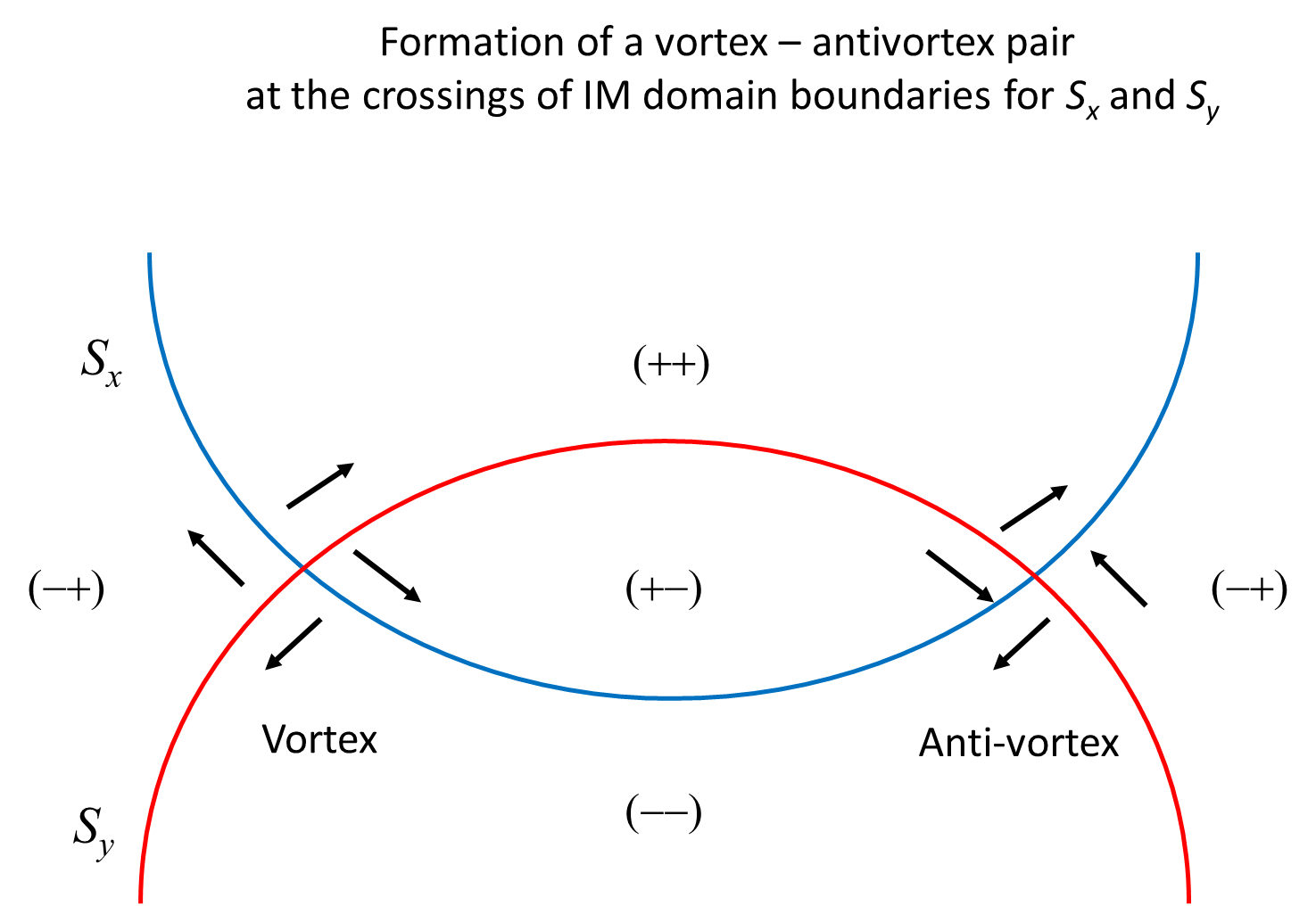}

\caption{Formation of vortices and antivortices at the crossings of domain
boundaries for $h_{x}(\mathbf{r})$ and $h_{y}(\mathbf{r})$.}

\label{Fig-vortex-antivortex_pair}
\end{figure}

The answer to this question seems to be that the magnetization cannot
smoothly follow the average random field without the formation of
vortices in $2d$ and vortex loops in $3d$. The latter cost energy
that prevents relaxation towards a completely disordered state. Thus
ferromagnetically ordered state is topologically protected.

This can be demonstrated by considering the average of the random
field over the correlated region around each point $\mathbf{r}$,
the so-called moving average, for instance,
\begin{equation}
\mathbf{\bar{h}}(\mathbf{r})=\frac{1}{V_{f}}\intop_{|\mathbf{r}'|\leq R_{f}}d^{d}\mathbf{r}'\mathbf{h}(\mathbf{r}'+\mathbf{r}),\label{eq:RF-moving_average}
\end{equation}
where $V_{f}$ is the correlated volume, $V_{f}=\left(4\pi/3\right)R_{f}^{3}$
in $3d$. This is exactly a mathematical implementation of the original
Imry-Ma argument. The averaged random field $\mathbf{\bar{h}}(\mathbf{r})$
describes a disorder correlated at length $R_{f}$. Since its components
$\bar{h}_{x}(\mathbf{r})$ and $\bar{h}_{y}(\mathbf{r})$ are sums
of many random variables, they have a Gaussian distribution at any
point $\mathbf{r}$ and are statistically independent. Spin field
in the Imry-Ma state, aligned with $\mathbf{\bar{h}}(\mathbf{r})$,
should be of the form
\begin{equation}
\mathbf{s}_{\mathrm{IM}}(\mathbf{r})=\frac{\mathbf{\bar{h}}(\mathbf{r})}{\left|\mathbf{\bar{h}}(\mathbf{r})\right|}.\label{eq:s_IM-aligned}
\end{equation}

Now, it can be shown that such defined spin field has singularities.
This happens when $\left|\mathbf{\bar{h}}(\mathbf{r})\right|=0$,
that is, both components of $\mathbf{\bar{h}}(\mathbf{r})$ turn to
zero. Regions of positive and negative $s_{x}(\mathbf{r})$ in a $xy$
plane, generated by Eq. (\ref{eq:s_IM-aligned}) are shown on Fig.
\ref{Fig-IM-domains}. The areas of positive and negative $s_{x}(\mathbf{r})$
are on average the same and the boundaries between domains are random
lines shown in Fig. \ref{Fig-IM-domains}(top). Domain boundaries
for $s_{y}(\mathbf{r})$ are also random lines statistically independent
from the former. Thus domain boundaries for $s_{x}(\mathbf{r})$ and
$s_{y}(\mathbf{r})$ will cross at some points, as shown in Fig. \ref{Fig-IM-domains}(bottom).
At these points vortices or antivortices will be generated because
of the denominator in Eq. (\ref{eq:s_IM-aligned}), as illustrated
in Fig. \ref{Fig-vortex-antivortex_pair}. In $3d$ there will be
vortex loops that cost much more energy than a vortex in $2d$.

Let us now estimate the energy gain in the IM state with vortices.
There is about one vortex per IM domain with size $R_{f}$, having
the energy
\begin{equation}
E_{V}\sim Js^{2}\left(\frac{R_{f}}{a}\right)\ln\left(\frac{R_{f}}{a}\right).
\end{equation}
The corresponding exchange energy per spin is
\begin{equation}
E_{\mathrm{ex}-V}\sim Js^{2}\left(\frac{a}{R_{f}}\right)^{2}\ln\left(\frac{R_{f}}{a}\right)
\end{equation}
that should replace the second term in Eq. (\ref{eq:Imry-Ma-1}).
Minimization with respect to $R_{f}$ in that expression gives
\begin{equation}
R_{f}\sim a\left(\frac{Js}{h}\right)^{2}\ln^{2}\left(\frac{Js}{h}\right)
\end{equation}
that is longer than the IM correlation radius because of the large
lorarithm. The corresponding energy gain
\begin{equation}
E-E_{0}\sim-Js^{2}\left(\frac{h}{Js}\right)^{4}\left[\ln\left(\frac{Js}{h}\right)\right]^{-3}\sim\frac{\Delta E_{IM}}{\ln^{3}(Js/h)}
\end{equation}
is the IM energy gain divided by a large logarithmic term.

On the other hand, the ferromagnetic state we have found numerically
can be understood as an incompletely disordered IM state, in which
the energy gain is $\Delta E_{IM}$ reduced by a numerical factor
of order one rather than by a large logarithmic term. The energy of
this ferromagnetic state should be lower than that of the IM state
with vortices, in accordance with our numerical results (see, e.g.,
Figs. \ref{Fig-dE_vs_HR_L=00003D216_rc=00003D0_alpha=00003D0.05_pbc}
and \ref{Fig-dE_vs_m_Nalp=00003D2_L=00003D120_HR=00003D1.5_rc=00003D0_pbc_alp=00003D0.1}).
The rapid relaxation out of the collinear state followed by a plateau
in Figs. \ref{Fig-m_vs_MCS_L=00003D216_HR=00003D1.5_pbc_coll_IC_alp=00003D1_0.03}
and \ref{Fig-m_vs_MCS_L=00003D800_1000_HR=00003D1.5_0.5_pbc_coll_IC_alp=00003D0_0.03}
can be explained as follows. Spins are readily relaxing in the direction
of the net RF in the regions of linear size $R_{f}$ until their further
rotation toward the totally disordered IM state requires creation
of vortices. As the latter costs energy, relaxation stops at this
point.

Of course, there is a non-zero probability that the random field at
the location of the vortex is vortex-like and almost parallel to the
spin field. In this case the energy gain from the vortex will be significantly
higher. However, since the Imry-Ma state in which the spin field follows
the direction of the average local random field is unique for every choice of the correlated volume $V_f$,
so should be the positions of the vortices. The fraction of the lucky vortices
mentioned above is determined by the probability of the corresponding
lucky configuration of the random field, which is small. Consequently,
it cannot affect the above argument .

\section{Discussion}

\label{discussion}

We have studied states of local energy minima of the random-field
$xy$ model focusing on weak random fields. The minimal
random-field value $h\equiv H_{R}$ in our work is defined by $h/J=0.3$,
see Fig. \ref{fig:CF-VG}. This should be considered weak for the following reason. In the
cubic lattice each spin has six nearest neighbors that are nearly
collinear for small $h$, the exchange field is $J_{0}\equiv6J$.
Thus, physically it makes more sense to consider the dimensionless parameter
$h/J_{0}$ that in our computations has the minimal value $h/J_{0}=0.05$
being manifestly small. In terms of $J_{0}$ formulas of the LIM theory
do not contain large numbers. For instance, $R_{f}$ in Eq. (\ref{eq:Rf-Def})
can be rewritten as $R_{f}/a=(4\pi/9)(J_{0}s/h)^{2}$.

Computations have been performed on lattices up to $1000\times1000\times1000$
spins. Our main finding is that completely disordered ($m\cong0$) states
are dominated by vortices and have higher energy than vortex-free
ferromagnetically ordered states. There
are unsurpassable energy barriers between different states even in
the case of a weak random field because switching between different
spin configurations involves large groups of correlated spins. This
makes the magnetic states depend strongly on the initial conditions.
At first glance this may appear conceptually similar to the behavior
of a conventional ferromagnet with pinning of domain walls. Prepared
with random orientations of spins, it would freeze in a state with
small magnetic domains and high energy due to many domain walls. In
a similar fashion, the random-field magnet freezes in a high-energy
state due to many vortices pinned by the random field. When prepared
with collinear spins, the conventional ferromagnet would remain in
a magnetized state because pinning prevents domain walls from proliferating
into the sample and achieving the ground state with zero total magnetization.
Similarly, the random-field magnet prepared with collinear spins relaxes
to a state with non-zero magnetic moment.

There is an essential difference between the two systems
though. While the conventional ferromagnet tends to relax toward an
$m=0$ state via diffusion of domain walls out of local energy minima,
the random-field magnet in our computations does not have this tendency
to relax to the zero-magnetization state out of the magnetized state.
In fact the energies of zero-magnetization states found in our various
types of computations are always higher than the energies of magnetized
states. One possibility is that the zero-magnetization state is not
the ground state. Another possibility is that there are energy barriers
to relaxation out of the magnetized state that involve collective
behavior of large volumes of spins and they are actually greater for a
weaker random field. This would be very different from shallow local energy
barriers for the diffusion of domain walls in conventional ferromagnets.

The bottom line of our analysis is that the Imry-Ma state in which
the system breaks into finite-size domains providing zero total magnetic
moment is impossible without formation of vortex loops. They become
very long and possess very large energy when the random field
becomes very small. This makes the barriers associated with the formation
of the zero-magnetization state unsurpassable at any temperature even in the limit of weak random field. The above argument is based upon $h$-dependence of $R_f$ and it stands as long as $R_f$ is small compared to the size of the system.
One can ask how close the vortex-glass state is to the Imry-Ma state.
To address this question, for $H_{R}=1.5$ we have created an Imry-Ma
state of Eq. (\ref{eq:s_IM-aligned}) and let it relax. As the result,
the vorticity decreased from $f_{V}\approx0.008$ in the Imry-Ma state
to $f_{V}\approx0.0006$ in the vortex-glass state. This means that
the system tries to annihilate vortices to reduce its energy but it
cannot do it completely because some vortices are pinned. Similar conclusion
regarding dislocations in two-dimensional pinned flux lattices has been reached in Ref. \onlinecite{Fisher-PRL99}.

On the other hand, it must be stressed that the ground state of the
system was not systematically searched for, and, moreover, it is of
little relevance in glassy systems. A single vortex loop going across
the whole sample will totally destroy magnetic order while its excess
energy, as well as its vorticity, will be vanishingly small. It cannot
be excluded that such type of states has the lowest possible energy.
However, these states are exotic and they were not studied here. Consequently,
we cannot rule out the existence a completely disordered
vortex-free ground state in our computations. However, finding such
a state may require a special initial condition or a more sophisticated
numerical argorithm anticipating the result.

It is generally believed, see, e.g., Refs.
\onlinecite{Fisher-PRL1997,Nat-review}, that in the presence of
quenched randomness the elastic interactions, like the ones in the
atomic or vortex lattices, or exchange in spin lattices, provide the
elastic-glass ground state that is characterized by the
power law decay of correlations at large distances. We have not found such a behavior
for the random-field $xy$ spin model in three dimensions.
The relation between that model and randomly pinned
flux lattices in superconductors has been discussed in some detail
in Ref. \onlinecite{Gingras-Huse-PRB1996}. The role of topological
defects in flux lattices is played by dislocations as compared to
vortices in spin models. Large areas of defect-free flux
lattices have been observed in experiment, see, e.g., Ref. \onlinecite{Klein-Nature2001}.
When analyzing such experiments, one should remember, however, that
for weak disorder the correlation length in $3d$ can be very large,
making it difficult to distinguish large defect-free slightly disordered
domains from the Bragg glass. While it is possible that some of the conclusions of this
paper apply to pinned flux lattices the latter requires a separate study because the two models
have different symmetry and different kinds of interaction with the random field.

\section{Acknowledgements}

The authors acknowledge useful discussions with Joseph Imry, Thomas
Nattermann, and Valerii Vinokur. This work has been supported by the
U.S. Department of Energy through Grant No. DE-FG02-93ER45487.

\end{document}